\documentclass[fleqn,10pt]{wlscirep}
\usepackage[utf8]{inputenc}
\usepackage[T1]{fontenc}

\usepackage{caption}
\usepackage{amsmath} 
\usepackage{url}
\usepackage{subfig}
\usepackage{graphicx}

\usepackage{alphalph}

\newcommand{\RNum}[1]{%
  \textup{\uppercase\expandafter{\romannumeral#1}}%
}


\newcommand{\vvv}{V$^3$~}
\newcommand{\vvvtail}{V$^3$}
\newcommand{\nvc}{NVC~}
\newcommand{\nvctail}{NVC}

\title{Quantifying the Economic Impact of COVID-19 in Mainland China Using Human Mobility Data}

\author[1,*]{Jizhou Huang}
\author[1,*]{Haifeng Wang}
\author[1]{Haoyi Xiong}
\author[1]{Miao Fan}
\author[1]{An Zhuo}
\author[1]{Ying Li}
\author[1]{Dejing Dou}

\affil[1]{Baidu Inc., Beijing, China}

\affil[*]{Corresponding author. Email: \{huangjizhou01, wanghaifeng\}@baidu.com}

\begin{abstract}
To contain the pandemic of coronavirus (COVID-19) in Mainland China, the authorities have put in place a series of measures, including quarantines, social distancing, and travel restrictions. While these strategies have effectively dealt with the critical situations of outbreaks, the combination of the pandemic and mobility controls has slowed China's economic growth, resulting in the first quarterly decline of Gross Domestic Product (GDP) since GDP began to be calculated, in 1992. To characterize the potential shrinkage of the domestic economy, from the perspective of mobility, we propose two new economic indicators: the \emph{New Venues Created} (NVC) and the \emph{Volumes of Visits to Venue} (V$^3$), as the complementary measures to domestic investments and consumption activities, using the data of Baidu Maps. The historical records of these two indicators demonstrated strong correlations with the past figures of Chinese GDP, while the status quo has dramatically changed this year, due to the pandemic. We hereby presented a quantitative analysis to project the impact of the pandemic on economies, using the recent trends of NVC and V$^3$. We found that the most affected sectors would be travel-dependent businesses, such as \textit{hotels}, \textit{educational institutes}, and \textit{public transportation}, while the sectors that are mandatory to human life, such as \textit{workplaces}, \textit{residential areas}, \emph{restaurants}, and \emph{shopping sites}, have been recovering rapidly. Analysis at the provincial level showed that the self-sufficient and self-sustainable economic regions, with internal supplies, production, and consumption, have recovered faster than those regions relying on global supply chains. 
\end{abstract}

\begin{document}

\flushbottom
\maketitle

\thispagestyle{empty}

\section*{Main Finding}
The World Health Organization (WHO) has declared a global pandemic as the COVID-19 disease rapidly spreads across the world\cite{sohrabi2020world}. 
The current countermeasures, carried out by the authorities of many countries around the world, include quarantines, travel restrictions, social distancing, etc. Some interventions, such as digital contact tracing\cite{Ferrettieabb6936}, suspending intra-city public transport\cite{Tianeabb6105}, shutting down the recreation venues\cite{Tianeabb6105}, and banning public gatherings\cite{Tianeabb6105}, have effectively curbed the spread of COVID-19\cite{Kraemereabb4218,Maiereabb4557}; but they have also crushed many national economies, globally, due to reduced mobility.

The combination of the pandemic and mobility controls has slowed the world's economic growth, resulting in the first quarterly decline of Gross Domestic Product (GDP). For example, the GDP of U.S., the largest economy in the world, decreased at an annual rate of $4.8\%$ in the first quarter of 2020, according to the statistics released by the Bureau of Economic Analysis on April 29, 2020\cite{usbeagpdq1}. 
In addition, the GDP of Mainland China, the second-largest economy in the world, shrank $6.8\%$ in the first quarter of 2020 compared to a year earlier, according to the statistics that Chinese authorities released on April 17, 2020\cite{chinagpdq1}. 

Therefore, this work proposed to quantify the impact of COVID-19 to the economy of Mainland China. As the real-time mobility data and the past records collected from Baidu Maps have been successfully leveraged to study the social impact of COVID-19\cite{Tianeabb6105,Kraemereabb4218,Chinazzieaba9757}, we also adopted the same data source for the analysis. Considering that Baidu Maps has over 340 million monthly active users by the end of December 2016\cite{baiduirrp}, who mainly live and travel in Mainland China, we believe the data should well characterize the economic performance of the country.

To carry out the quantitative research, our work consists of two steps, as follows. First, we would like to search some \emph{key economic performance indicators} from the historical records of Baidu Maps, as these indicators are expected to correlate with the core econometrics, such as the figures of GDP of Mainland China, in various sectors and in different geographical regions. Second, we hope to see \emph{how these indicators have changed during the pandemic in Mainland China}, especially to what degree these indicators have been drifted since the mobility restrictions were put in place, as well as the year-to-year variations of these indicators compared to the status quo of past years. 

\begin{figure*}[!htp]
\centering
\includegraphics[width=1.0\textwidth,trim={0.9cm 1.1cm 1.2cm 1.4cm},clip]{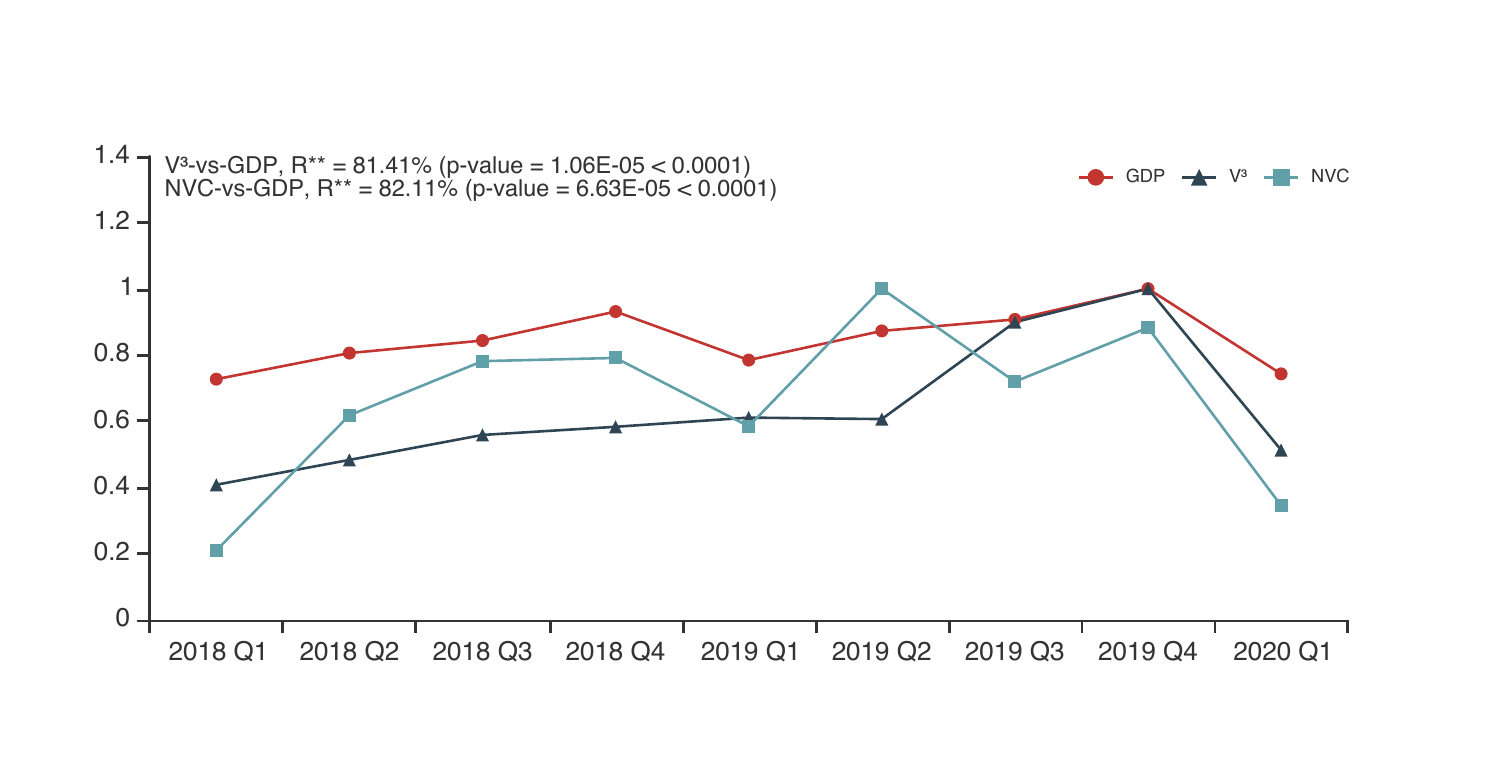}
\caption{The traces and trends of the GDP of Mainland China, nationwide \vvvtail, and nationwide \nvc in each quarter, from 2018 Q1 to 2020 Q1.}
\label{fig:gdp-rbe-v2p}
\end{figure*}

As was known, the macroeconomic performance of Mainland China has benefited majorly from the strength of investment, consumption, and import-export in past decades\cite{chinasyr2019}. To understand the behaviors related to the domestic economy, we especially focused on the ways to quantify the activities related to investment and consumption, especially for those related to small businesses that cannot be fully screened by nationwide statistics. Therefore, we first retrieved from Baidu Maps information for all venues located in Mainland China: namely, for points-of-interest including residential areas, workplaces, restaurants, shopping centers, scenic spots, public transportation hubs, etc.\cite{fan2019monopoly}. With the venue information, we collected the real-time and historical records on the creation of new venues, as well as the visits to the venues. We hereby proposed two important indicators, as follows.

\begin{itemize}
    \item \emph{New Venues Created (\nvctail): } To characterize the strength of investment into businesses, we performed statistical analysis on the numbers of New Venues Created (\nvctail) every week from January 1, 2018 to present. We further mapped such weekly \nvc traces to the provinces of Mainland China, and across different types of venues: 11 types in total, ranging from public transportation, to private housing, to restaurant and shopping, etc. With a greater \nvc in a venue type/province, we could expect a higher investment in the corresponding industrial sector/region.
    
    \item \emph{Volumes of Visits to Venues (\vvvtail): } To directly measure the on-site consumption in different industrial sectors or economic regions, we counted the weekly volumes of visits to (all types of) venues, nationwide, from January 1, 2018 to present. Similarly, we also mapped such weekly \vvv traces to different venue types and various provinces of Mainland China. Though, sometimes, each individual visit might not include a transaction of consumption, the overall volumes of \vvv might tell us the trends about on-site consumption.
\end{itemize}

To validate the effectiveness of the two new above indicators, we use the quarterly GDP figures from 2018 Q1 to 2020 Q1, published by the National Bureau of Statistics of China, to study the correlations between the indicators and GDP.
The statistics regarding the GDP of Mainland China can be freely downloaded from the official website of National Bureau of Statistics of China\cite{chinagdpbyq}.
Figure~\ref{fig:gdp-rbe-v2p} illustrates three curves, representing the GDP of Mainland China, the traces of nationwide \vvv aggregated by quarters, and the traces of nationwide \nvc aggregated by quarters, respectively. All the values are normalized proportionally by the maximum, accordingly (see Supplementary Materials). To figure out the strength of the associations, we apply Pearson's correlation coefficient\cite{ROUSSEAU201867} to every two groups of the data, i.e., to  \vvvtail-vs-GDP and to \nvctail-vs-GDP. It shows that the Pearson's correlation between the GDP of Mainland China and the \vvv is $81.41\%$ ($N=9$ and p-value $= 1.06 \times 10^{-5} \ll 0.0001$); and that the Pearson's correlation between the GDP of Mainland China and the \nvc is $82.11\%$ ($N=9$ and p-value $= 6.63 \times 10^{-5} \ll 0.0001$). We discover that the GDP has significant positive correlations with both the \vvv and the \nvctail, at the national level.
A detailed analysis was further carried out to explore the \vvvtail-vs-GDP and \nvctail-vs-GDP correlations at the level of the provinces in Mainland China (see Supplementary Materials). The results show that both the \vvv and the \nvc have significant correlations with the GDP in all $31$ provinces excluding Hong Kong, Macao, and Taiwan of China, supporting the main finding of this article.

So far, we have tested the hypotheses that correlate the \vvv and the \nvc with the quarterly GDP records of Mainland China in recent years. Given the strong correlations between the economic performance and these two indicators, in the rest of this work, we aim at projecting the impact of COVID-19 on Chinese economies using the recent trends of the two indicators. 

\begin{figure}[th!]
    \centering
    \subfloat[The \vvv in Mainland China]{\includegraphics[width=0.49\textwidth,trim={0.48cm 1.08cm 0.58cm 1.08cm},clip]{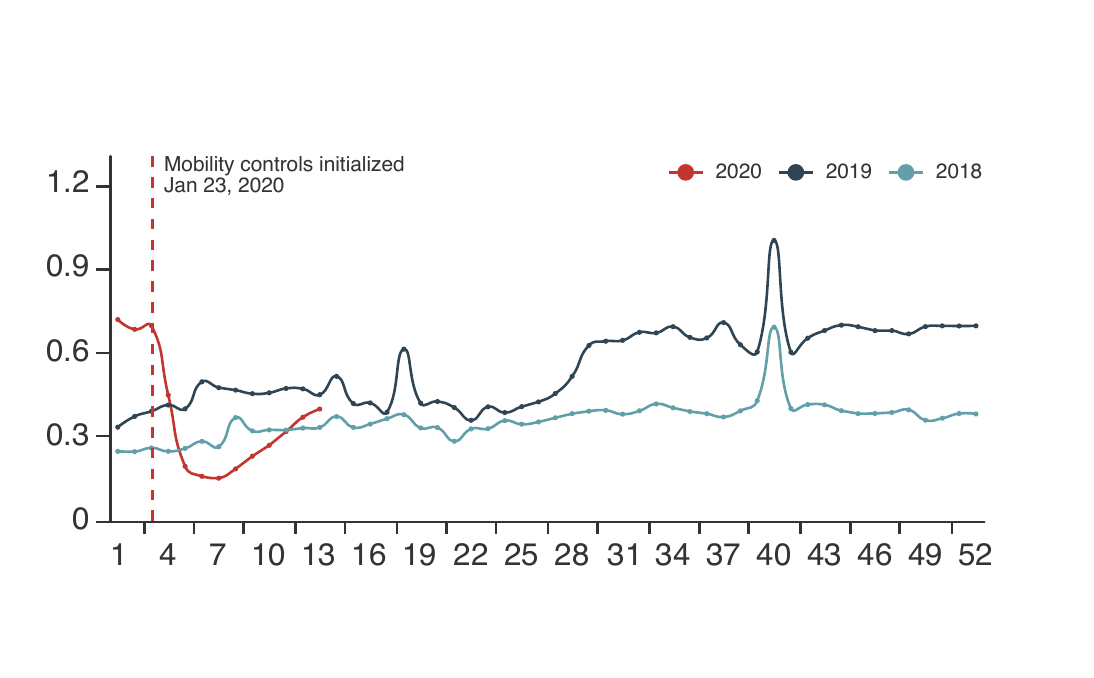}}\
    \subfloat[The \nvc for Mainland China]{\includegraphics[width=0.49\textwidth,trim={0.48cm 1.08cm 0.58cm 1.08cm},clip]{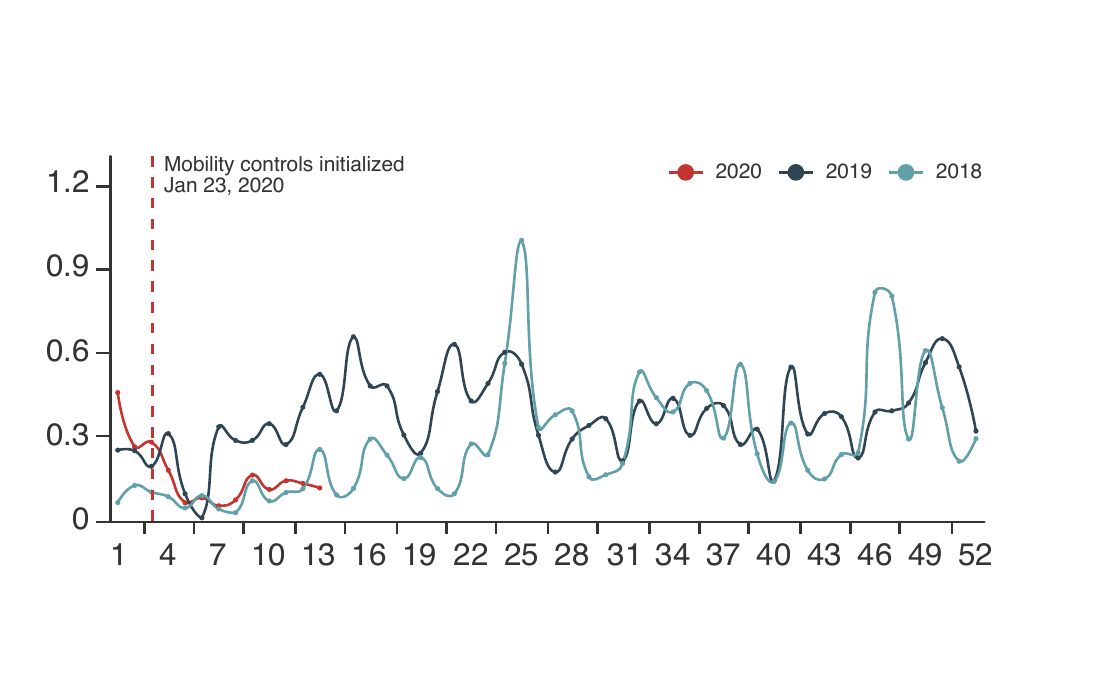}}
    \caption{The recent trends and historical records of the \vvv and the \nvc in Mainland China for 2020, 2019, and 2018.}
    \label{fig:two-factors-china}
\end{figure}

\section*{Projecting the impact of COVID-19 on Chinese economies}
\begin{figure}[ht!]
    \centering
    \subfloat[The \vvv in Hubei]{\label{fig:v-hubei-v3}\includegraphics[width=0.49\textwidth,trim={0.48cm 1.08cm 0.58cm 1.08cm},clip]{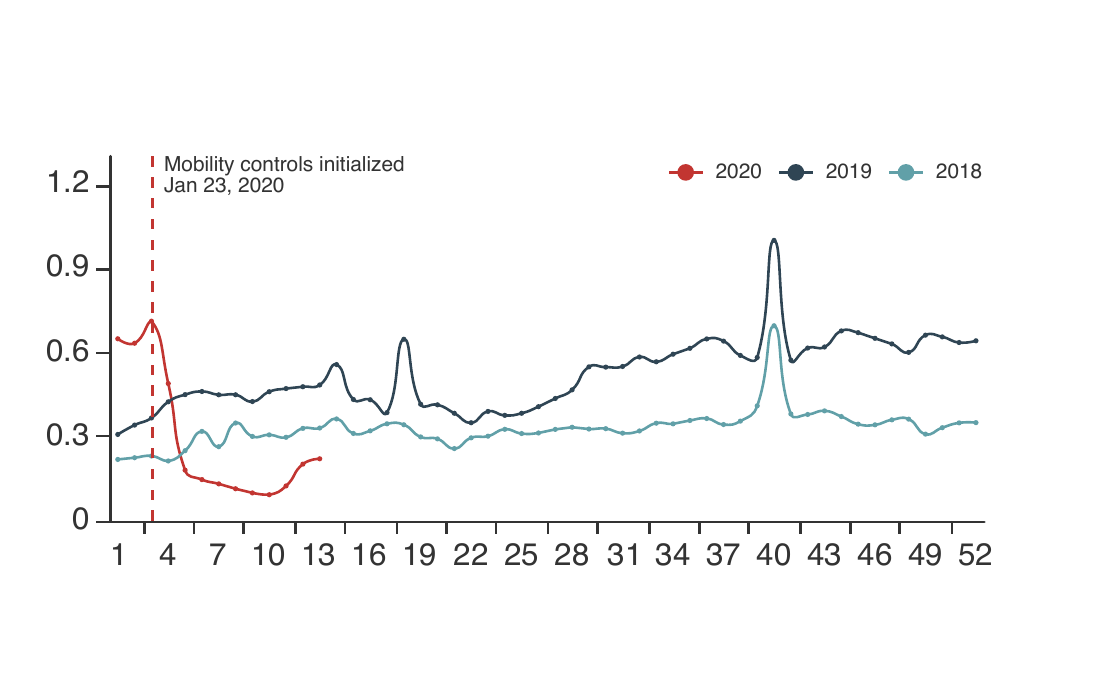}}
    \subfloat[The \nvc for Hubei]{\label{fig:v-hubei-nvc}\includegraphics[width=0.49\textwidth,trim={0.48cm 1.08cm 0.58cm 1.08cm},clip]{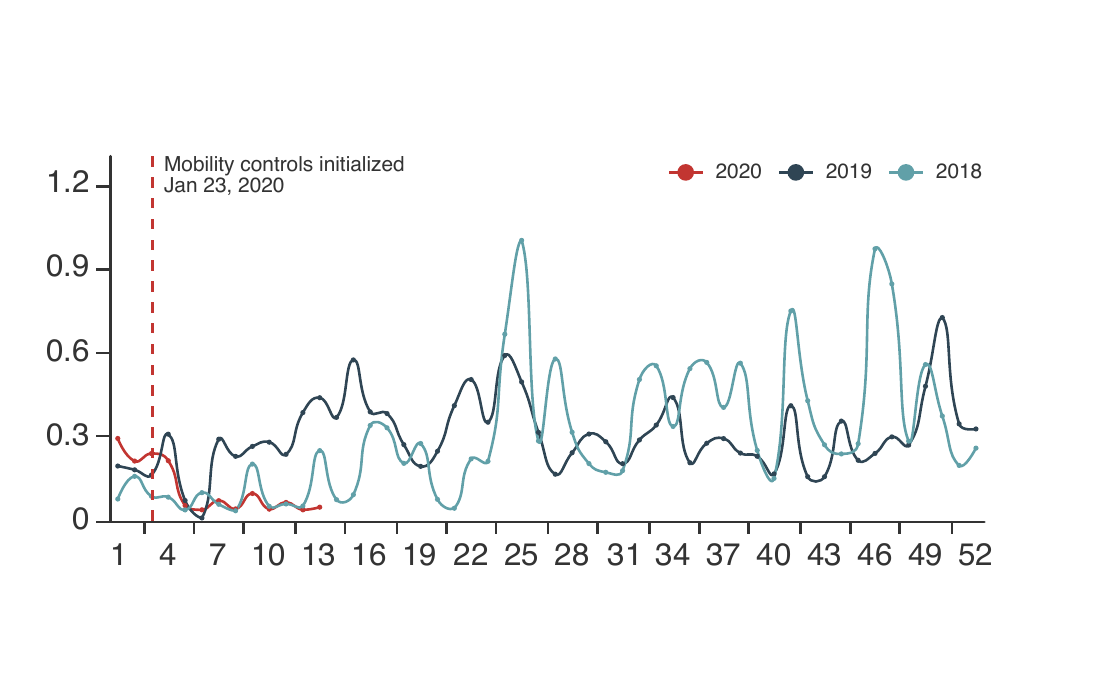}}\
    \subfloat[The \vvv in Beijing]{\label{fig:v-beijing-v3}\includegraphics[width=0.49\textwidth,trim={0.48cm 1.08cm 0.58cm 1.08cm},clip]{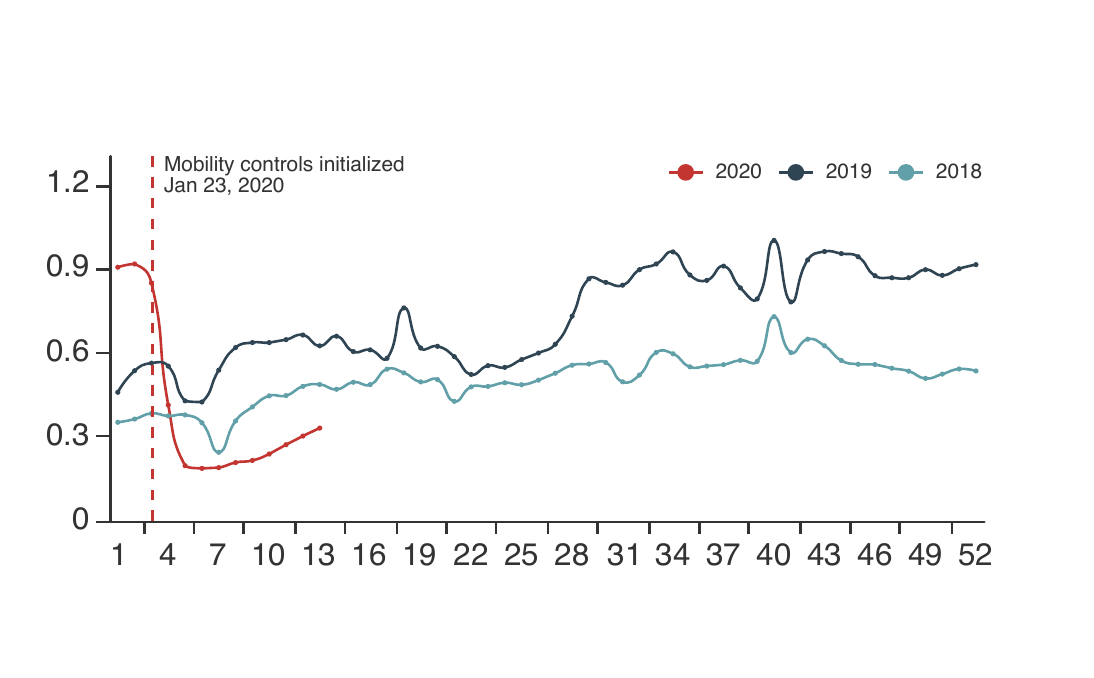}}
    \subfloat[The \nvc for Beijing]{\label{fig:v-beijing-nvc}\includegraphics[width=0.49\textwidth,trim={0.48cm 1.08cm 0.58cm 1.08cm},clip]{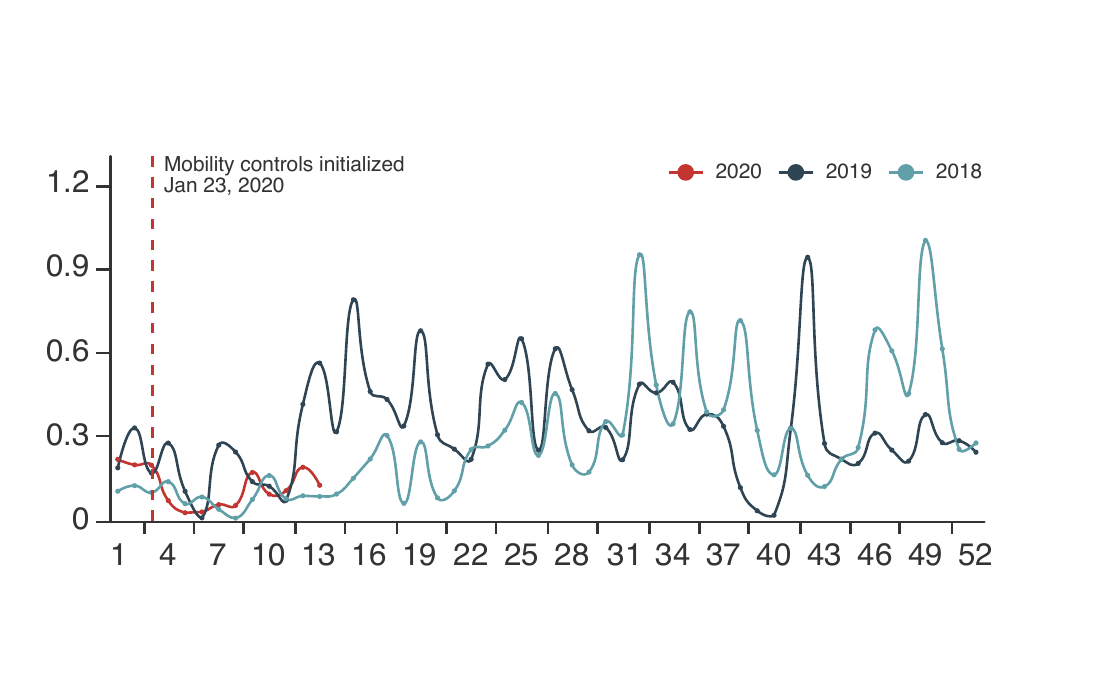}}\
    \subfloat[The \vvv in Tianjin]{\label{fig:v-tianjin-v3}\includegraphics[width=0.49\textwidth,trim={0.48cm 1.08cm 0.58cm 1.08cm},clip]{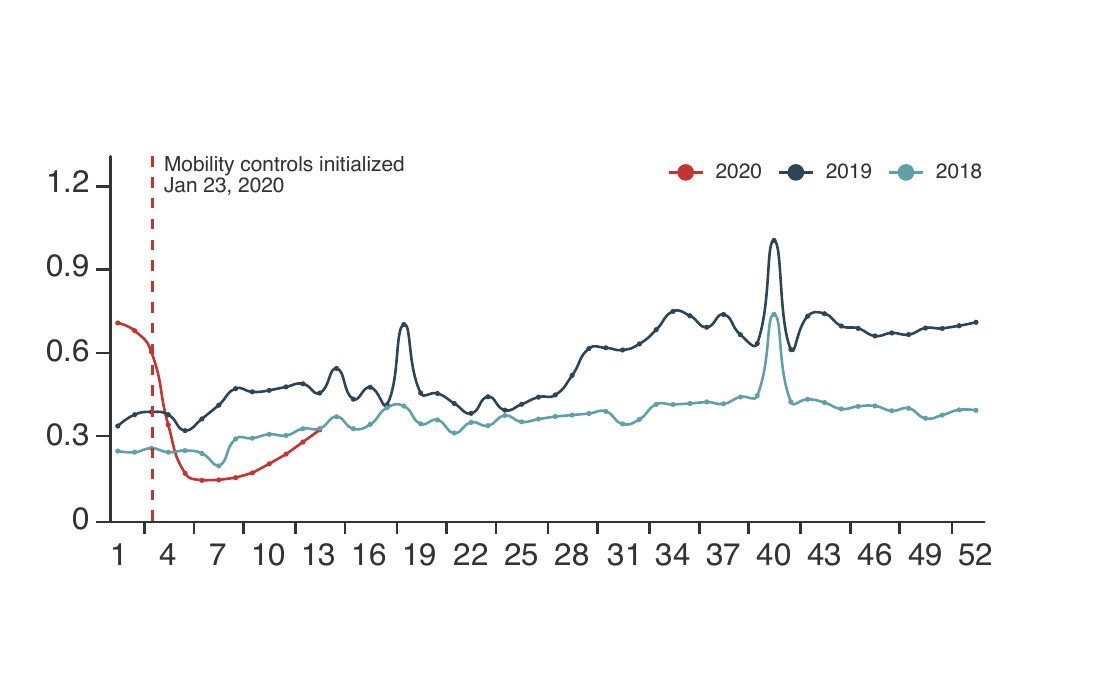}}\
    \subfloat[The \nvc for Tianjin]{\label{fig:v-tianjin-nvc}\includegraphics[width=0.49\textwidth,trim={0.48cm 1.08cm 0.58cm 1.08cm},clip]{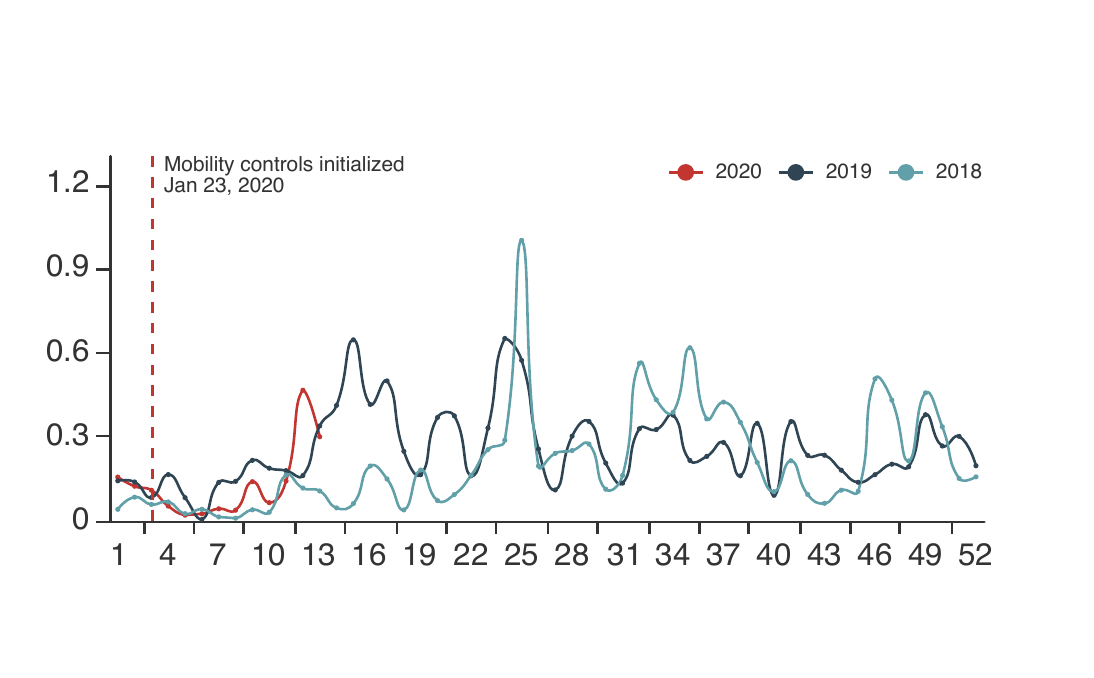}}
    \caption{The L-shaped recession of the \vvv in Hubei, Beijing, and Tianjin.}
    \label{fig:a-tale-of-three-cities}
\end{figure}

\begin{figure}[ht!]
    \centering
    \subfloat[The \vvv in the Big Northwestern region]{\label{fig:ck-f3-nwr-v3}\includegraphics[width=0.49\textwidth,trim={0.48cm 1.08cm 0.58cm 1.08cm},clip]{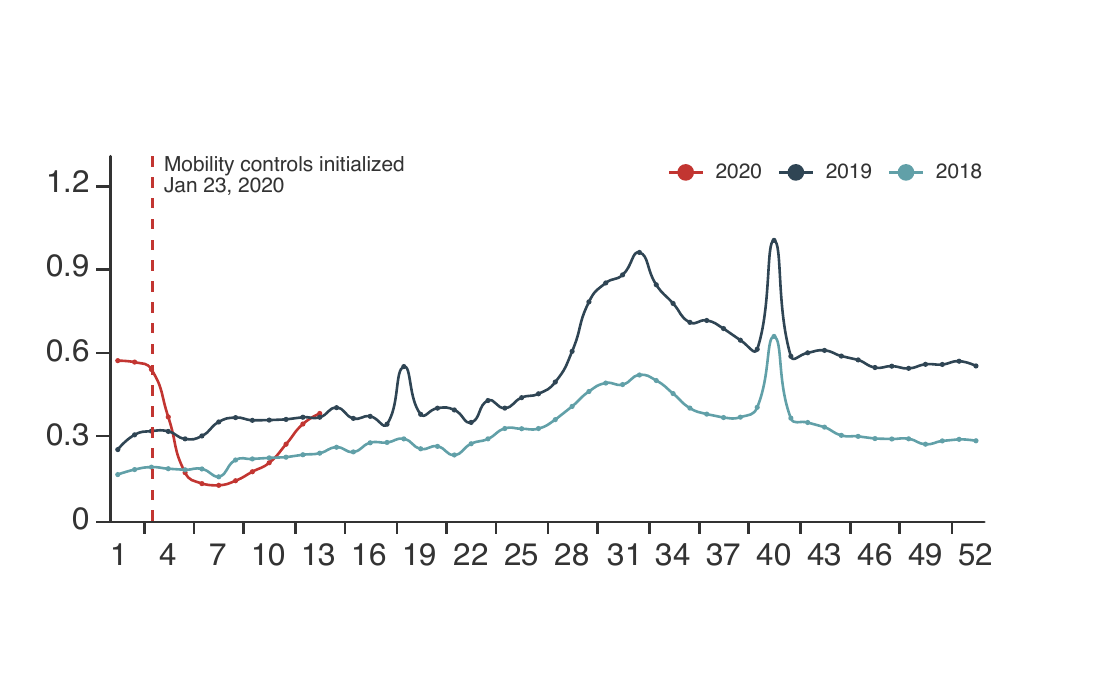}}\
    \subfloat[The \nvc for the Big Northwestern region]{\label{fig:ck-f3-nwr-nvc}\includegraphics[width=0.49\textwidth,trim={0.48cm 1.08cm 0.58cm 1.08cm},clip]{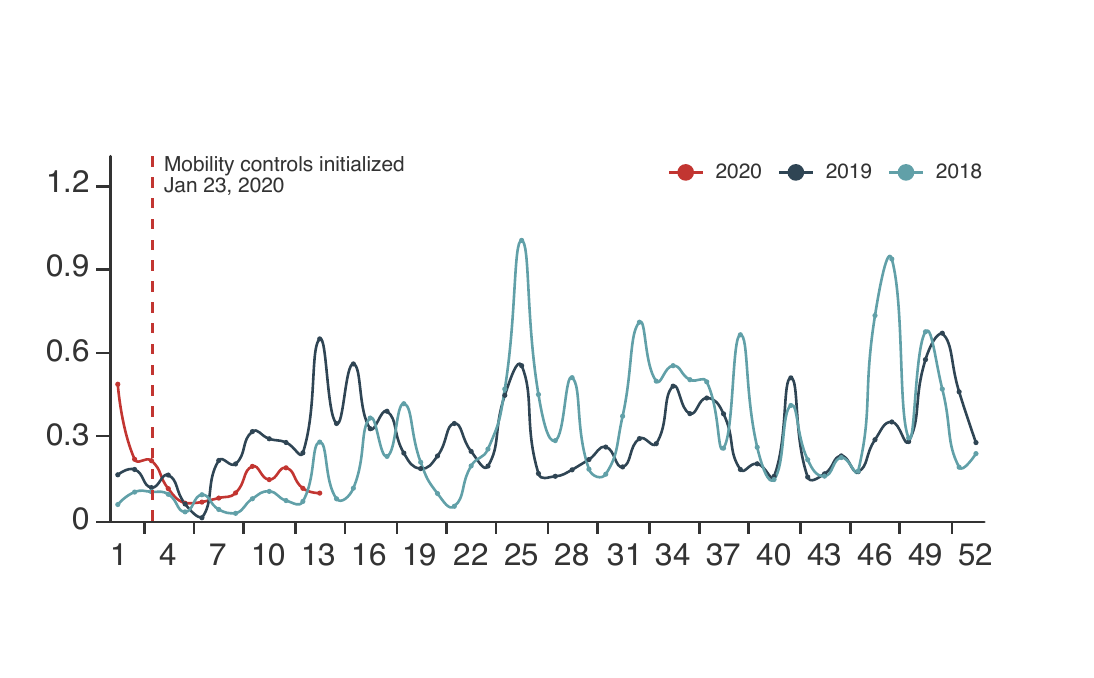}}\
    \subfloat[The \vvv in the Northeastern region]{\label{fig:ck-f3-ner-v3}\includegraphics[width=0.49\textwidth,trim={0.48cm 1.08cm 0.58cm 1.08cm},clip]{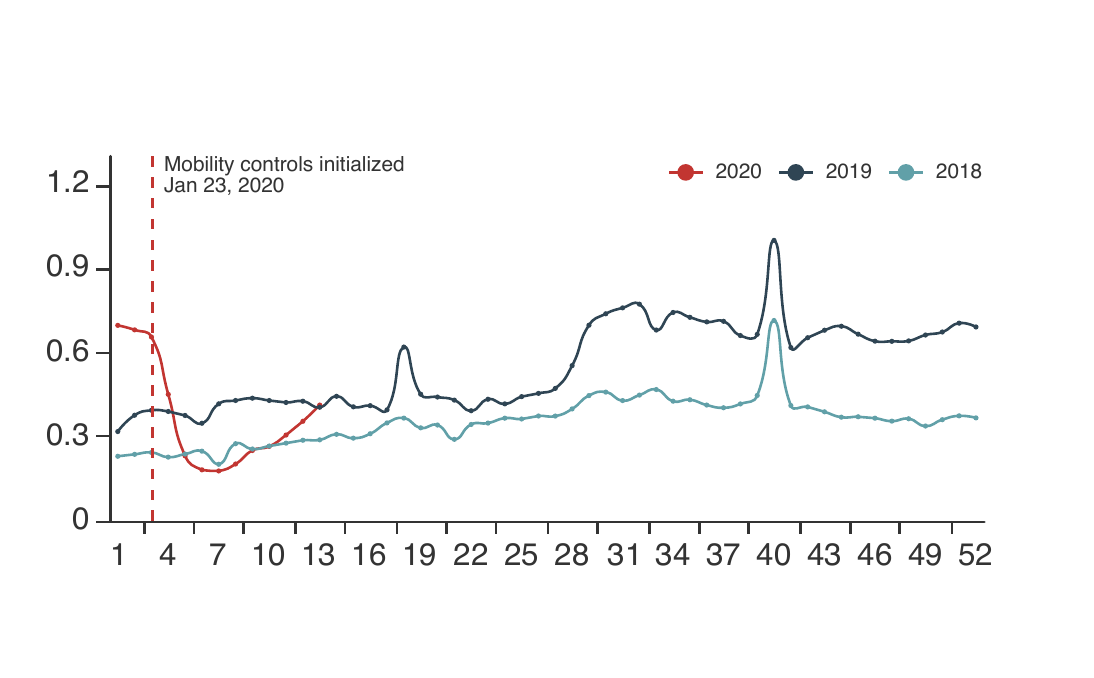}}\
    \subfloat[The \nvc for the Northeastern region]{\label{fig:ck-f3-ner-nvc}\includegraphics[width=0.49\textwidth,trim={0.48cm 1.08cm 0.58cm 1.08cm},clip]{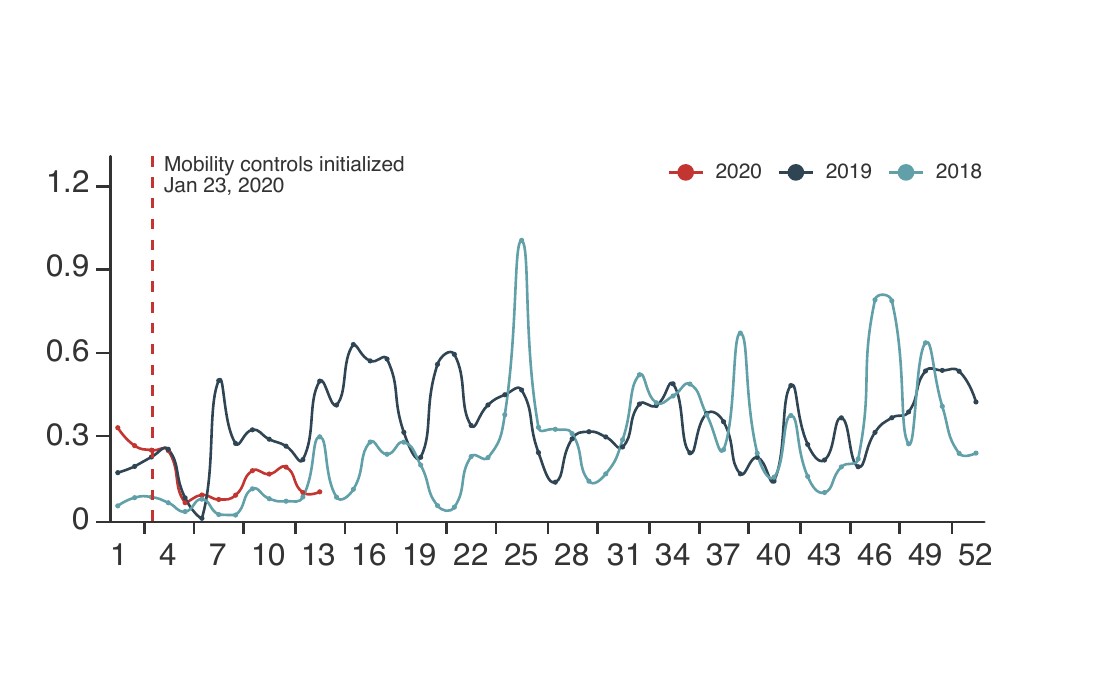}}\
    \subfloat[The \vvv in the Southwestern region]{\label{fig:ck-f3-swr-v3}\includegraphics[width=0.49\textwidth,trim={0.48cm 1.08cm 0.58cm 1.08cm},clip]{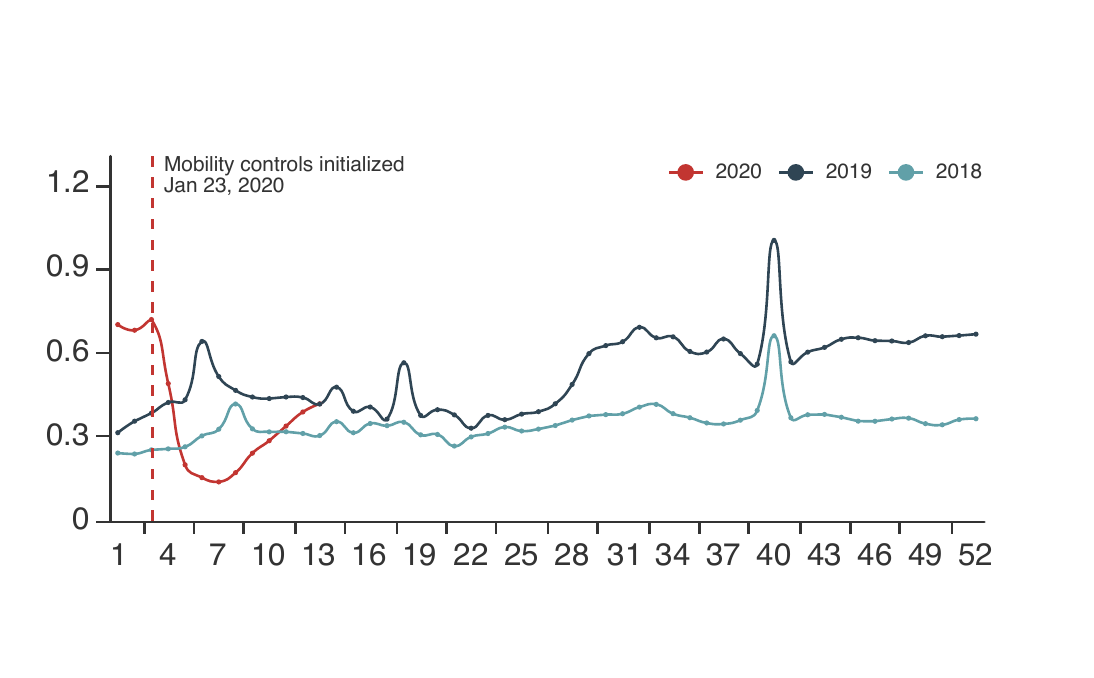}} \
    \subfloat[The \nvc for the Southwestern region]{\label{fig:ck-f3-swr-nvc}\includegraphics[width=0.49\textwidth,trim={0.48cm 1.08cm 0.58cm 1.08cm},clip]{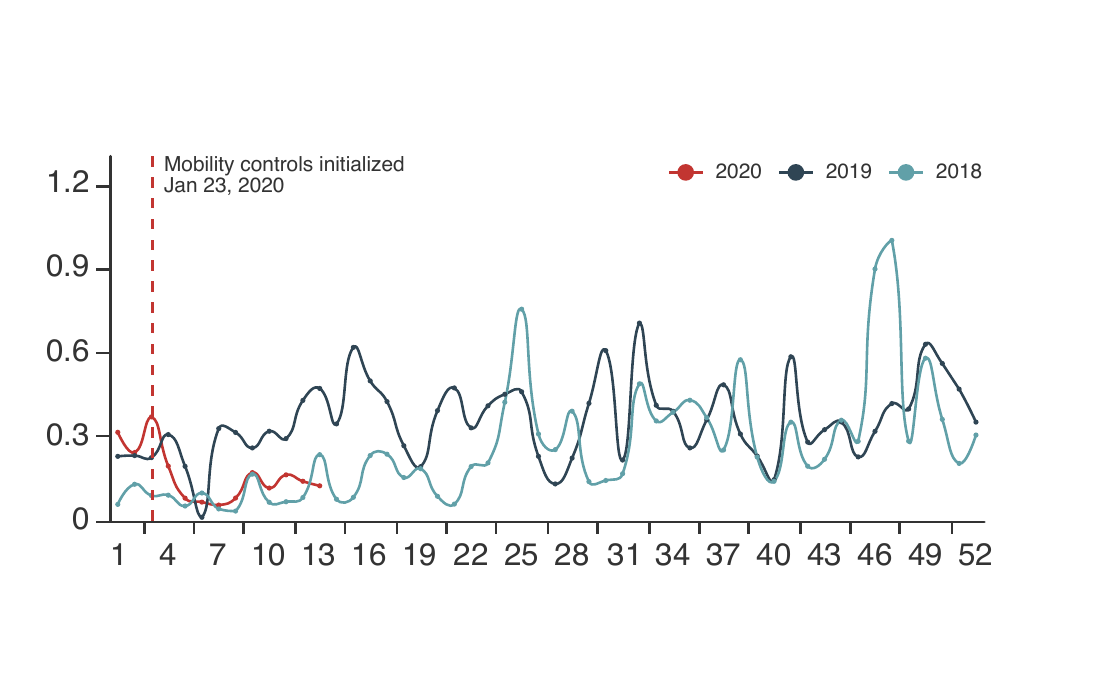}}\
    \caption{The emerging, V-shaped bounces of the \vvv in three economic regions of Mainland China.}
    \label{fig:v-shaped-areas}
\end{figure}

Recent trends and historical records of the two indicators are illustrated by Figure~\ref{fig:two-factors-china} where the X-axis represents 52 weeks in a year (The same is true from Figure~\ref{fig:l-shaped-cats} to Figure~\ref{fig:v3-nvc-31-provinces}).
We can observe the \emph{fall-and-rebound} in the recent trends of the \vvv in Mainland China, which might suggest the recovery from the disruption caused by COVID-19. 
Clearly, a huge shrinkage of both indicators for the year of 2020, especially from the 3$^{rd}$ week to 7$^{th}$ week of 2020, can be found, through comparison with the records of 2019 and 2018. Fortunately, it also shows a clear and strong bounce of the \vvv since the 8$^{th}$ week of 2020, which might suggest evidence of the recovery of on-site consumption activities. While we could roughly conclude that, from the perspectives of the \vvvtail, the on-site consumption activities have already been begun to return to the levels between 2018 and 2019, the domestic investment activities represented by the \nvc were still low. The recent trends of the \nvc were quite similar to the situations of 2018. All in all, we could expect a reasonable growth of both indicators, as well as the economic performance, regarding to the momentum of the \vvv recovery and the historical \nvc trajectories. 

In the following sections, we would like to present the dissections of the two indicators, to understand the details of economic performance, with respect to different industrial sectors in different parts of Mainland China.

\subsection*{Economic impact of COVID-19 on industrial sectors}
\begin{figure}[th!]
    \centering
    \subfloat[The \vvv in Airports]{\label{fig:l-airports-v3}\includegraphics[width=0.49\textwidth,trim={0.48cm 1.08cm 0.58cm 1.08cm},clip]{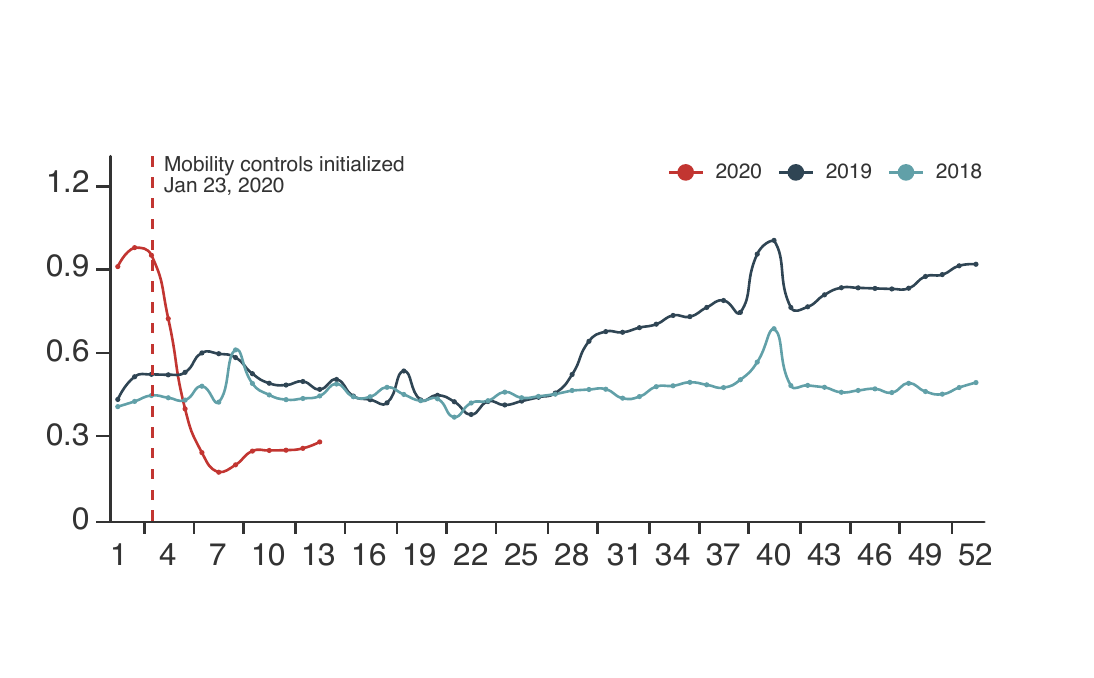}} \
    \subfloat[The \nvc for Airports]{\label{fig:l-airports-nvc}\includegraphics[width=0.49\textwidth,trim={0.48cm 1.08cm 0.58cm 1.08cm},clip]{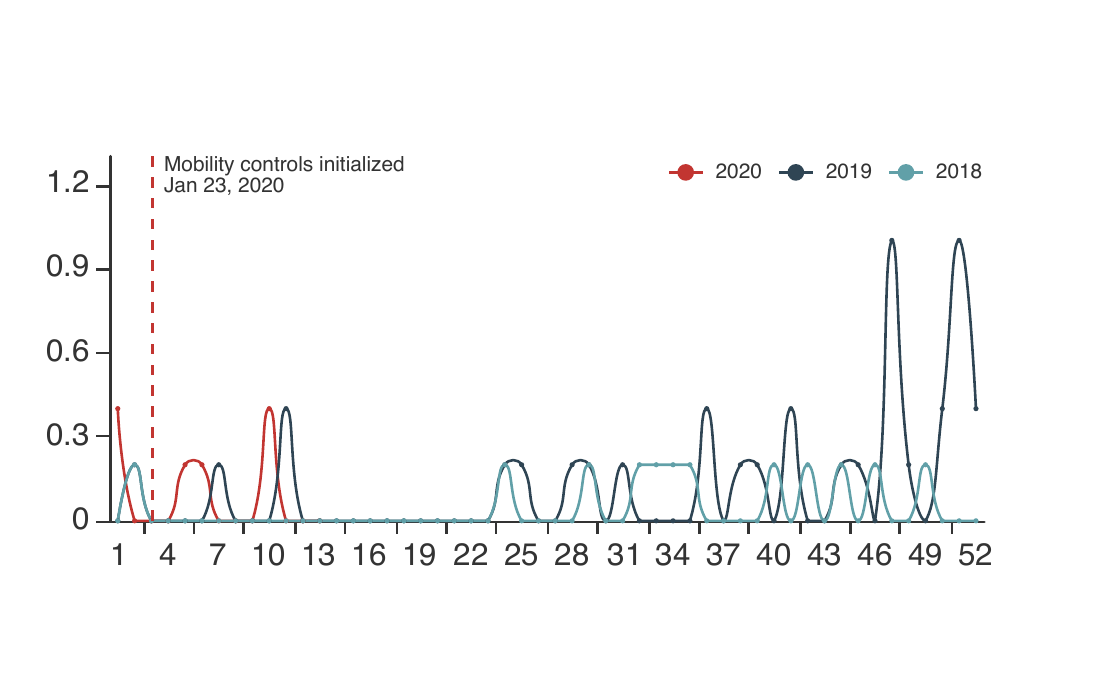}}\
    \subfloat[The \vvv in Train Stations]{\label{fig:l-trainst-v3}\includegraphics[width=0.49\textwidth,trim={0.48cm 1.08cm 0.58cm 1.08cm},clip]{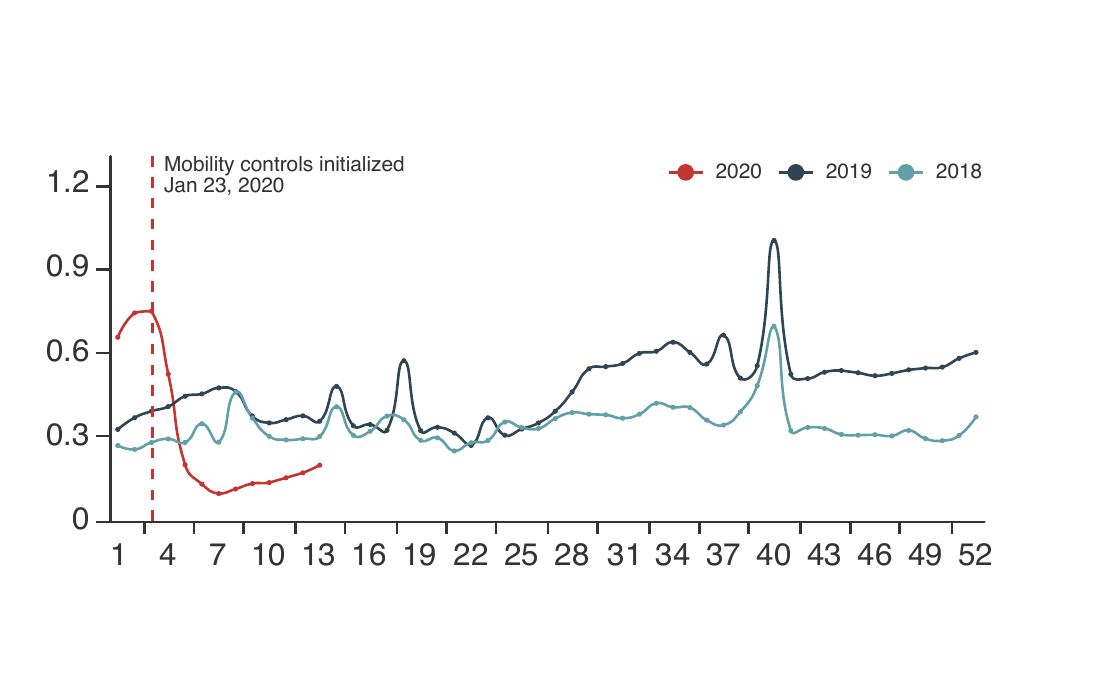}}\
    \subfloat[The \nvc for Train Stations]{\label{fig:l-trainst-nvc}\includegraphics[width=0.49\textwidth,trim={0.48cm 1.08cm 0.58cm 1.08cm},clip]{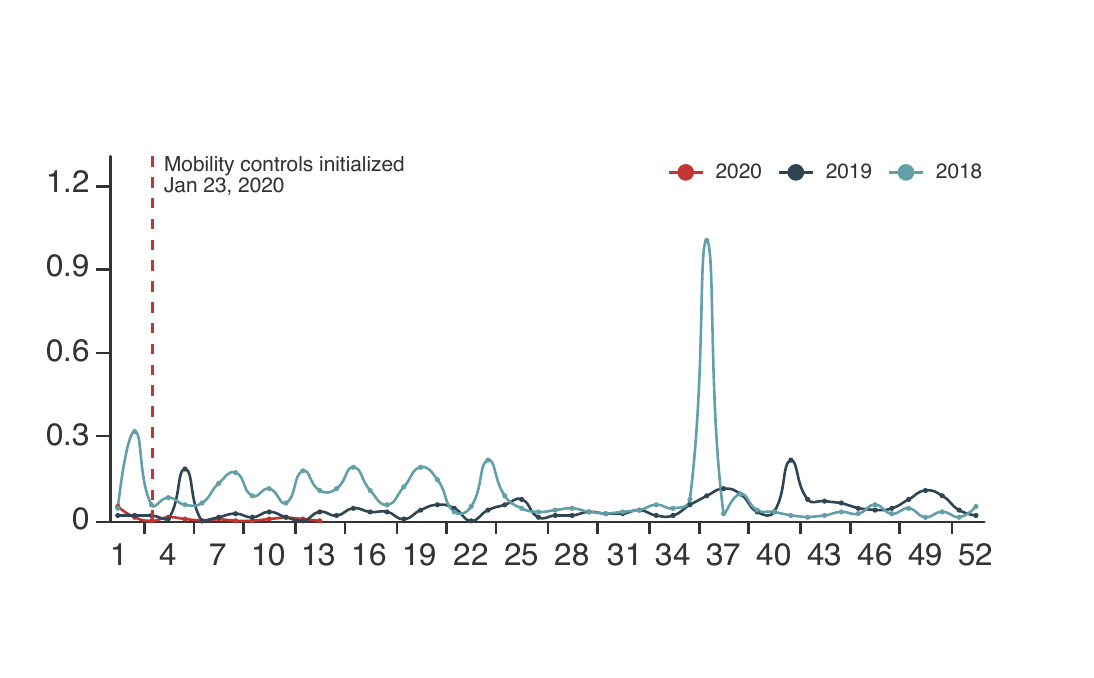}}\
    \subfloat[The \vvv in Educational Institutes]{\label{fig:l-eduins-v3}\includegraphics[width=0.49\textwidth,trim={0.48cm 1.08cm 0.58cm 1.08cm},clip]{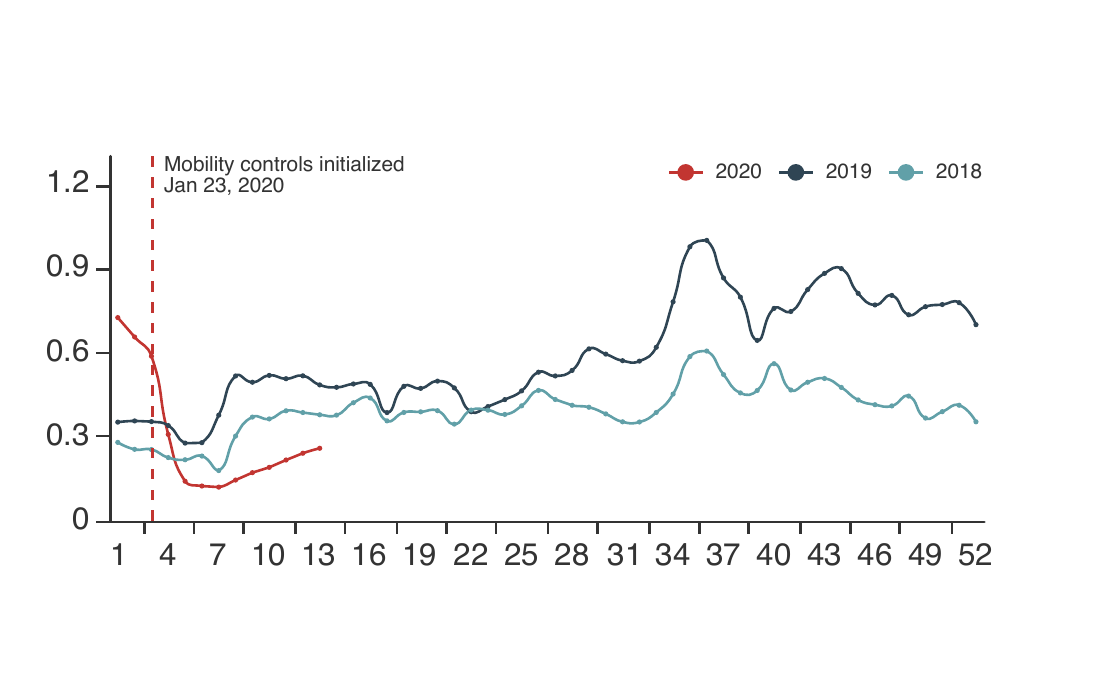}}\
    \subfloat[The \nvc for Educational Institutes]{\label{fig:l-eduins-nvc}\includegraphics[width=0.49\textwidth,trim={0.48cm 1.08cm 0.58cm 1.08cm},clip]{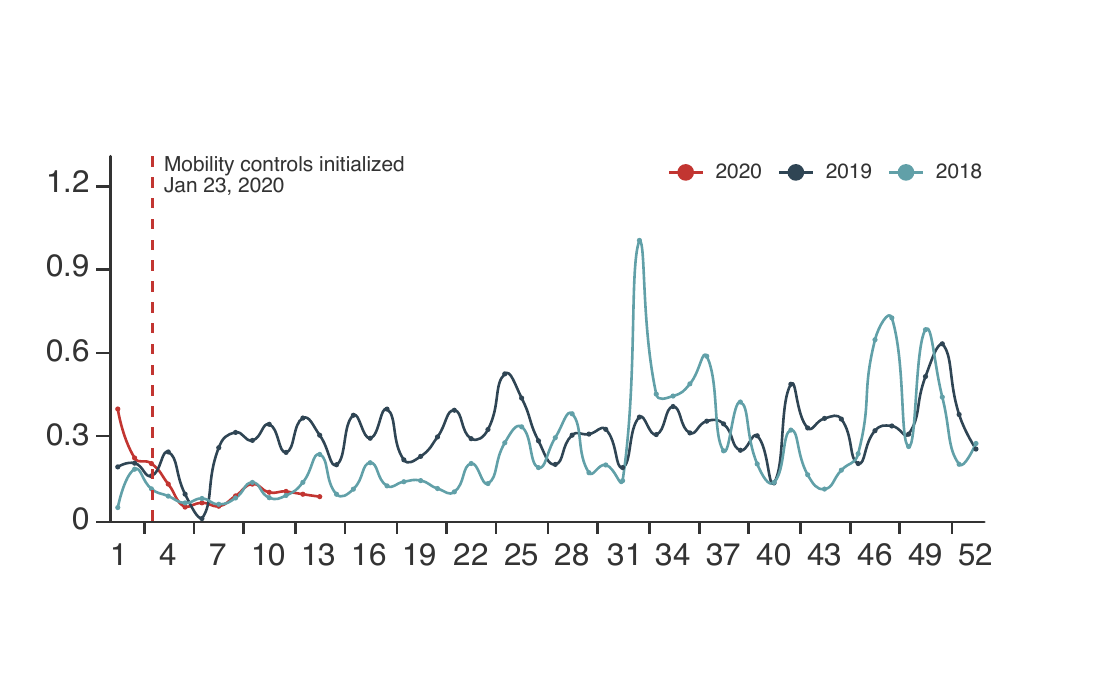}}\
    \subfloat[The \vvv in Hotels]{\label{fig:l-hotel-v3}\includegraphics[width=0.49\textwidth,trim={0.48cm 1.08cm 0.58cm 1.08cm},clip]{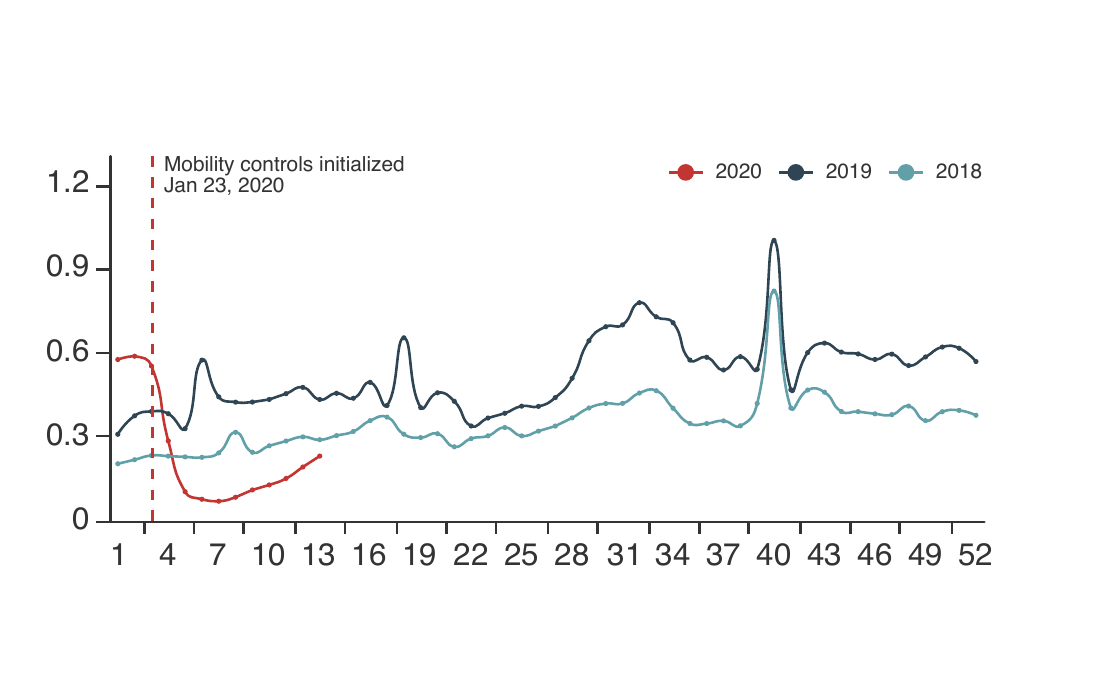}}\
    \subfloat[The \nvc for Hotels]{\label{fig:l-hotel-nvc}\includegraphics[width=0.49\textwidth,trim={0.48cm 1.08cm 0.58cm 1.08cm},clip]{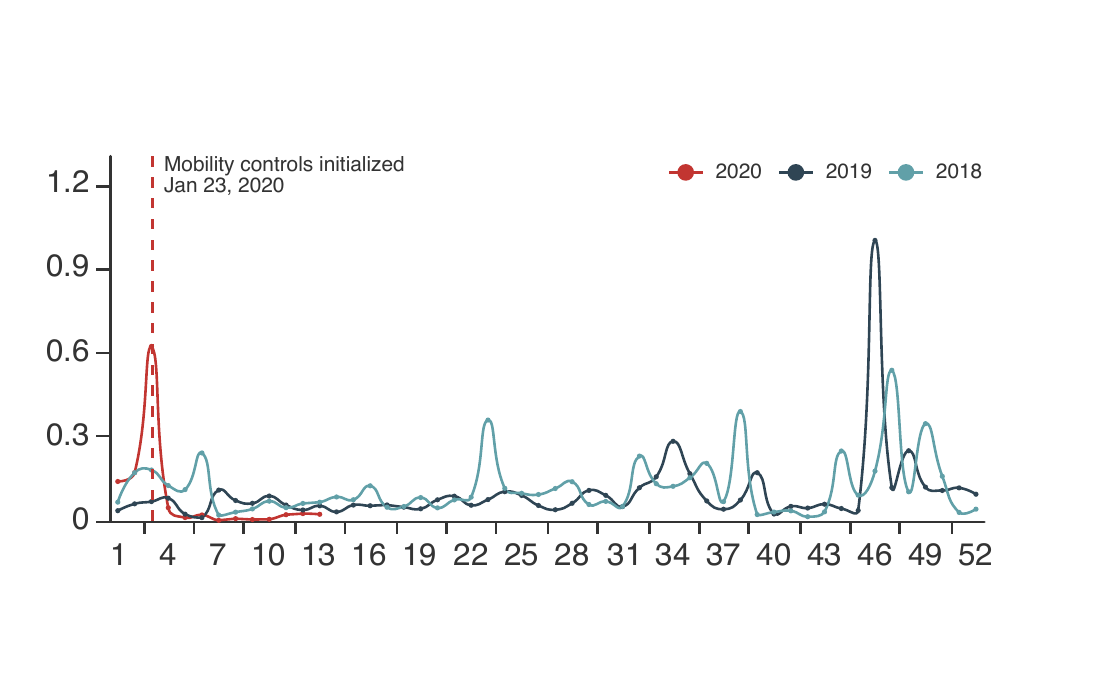}}\
    \caption{The L-shaped recession of the \vvv in industrial sectors of Mainland China.}
    \label{fig:l-shaped-cats}
\end{figure}

\begin{figure}[th!]
    \centering
    \subfloat[The \vvv in Restaurants \& Beverages]{\label{fig:ck-restaurants-v3}\includegraphics[width=0.49\textwidth,trim={0.48cm 1.08cm 0.58cm 1.08cm},clip]{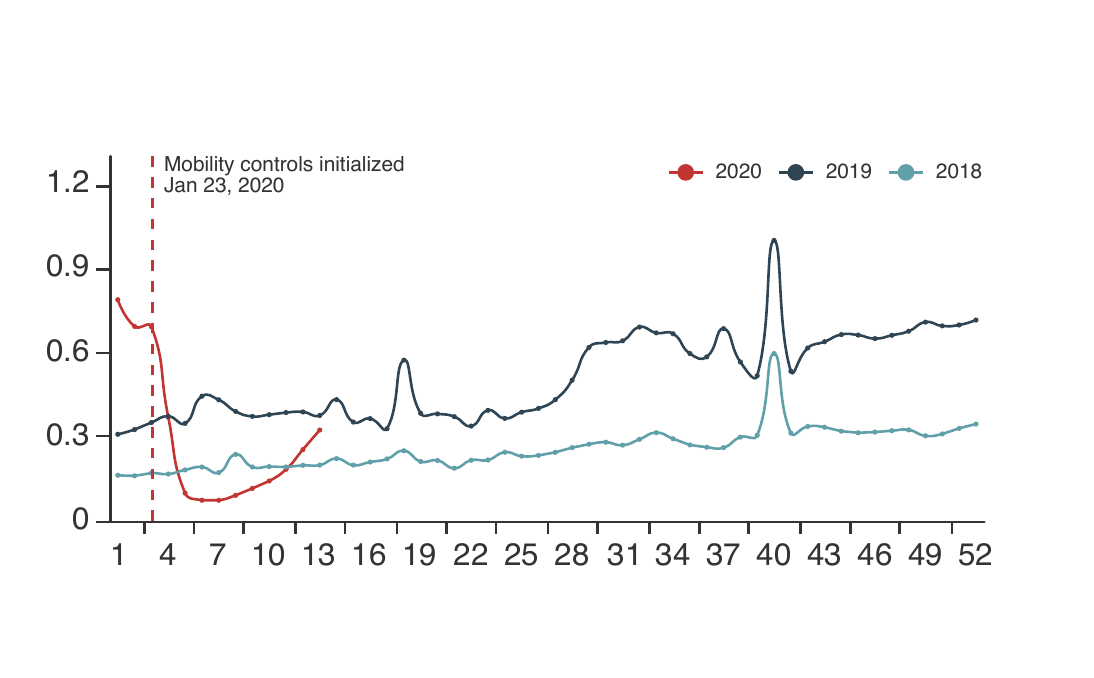}}\
    \subfloat[The \nvc for Restaurants \& Beverages]{\label{fig:ck-restaurants-nvc}\includegraphics[width=0.49\textwidth,trim={0.48cm 1.08cm 0.58cm 1.08cm},clip]{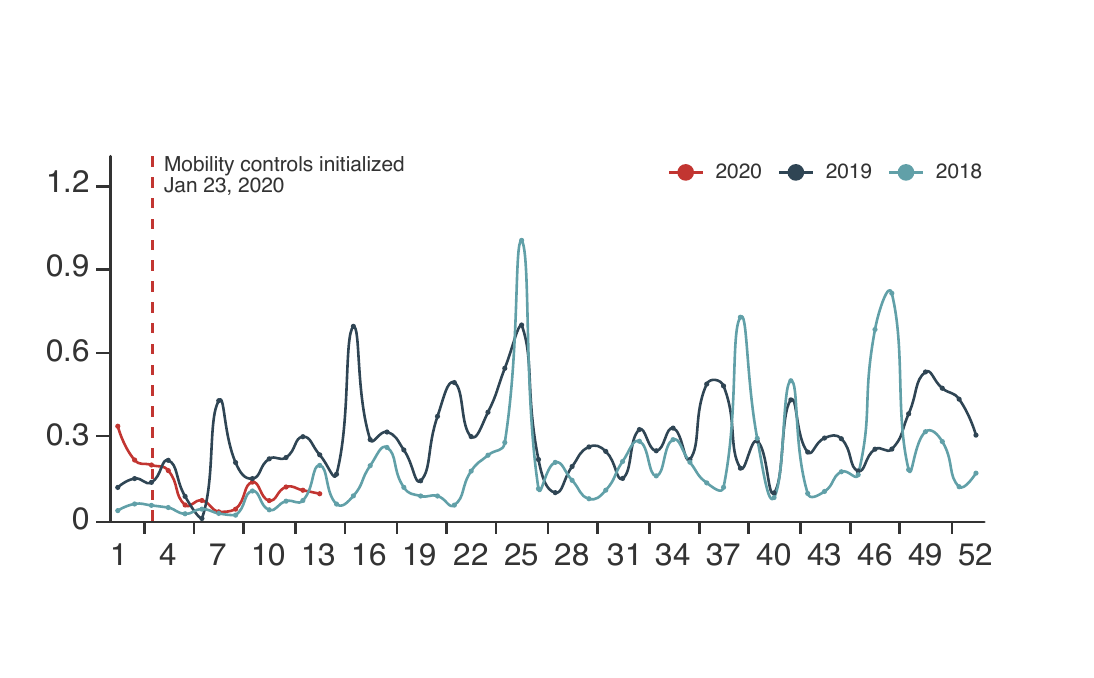}}\
    \subfloat[The \vvv in Recreation Sites]{\label{fig:ck-recreation-v3}\includegraphics[width=0.49\textwidth,trim={0.48cm 1.08cm 0.58cm 1.08cm},clip]{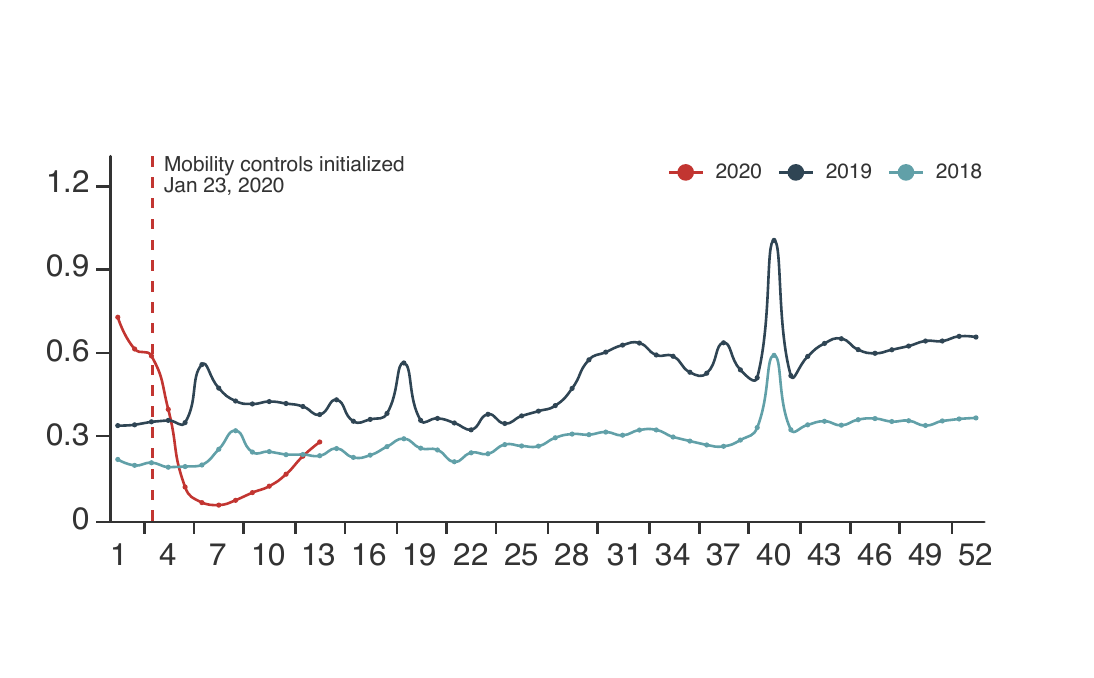}}\
    \subfloat[The \nvc in Recreation Sites]{\label{fig:ck-recreation-nvc}\includegraphics[width=0.49\textwidth,trim={0.48cm 1.08cm 0.58cm 1.08cm},clip]{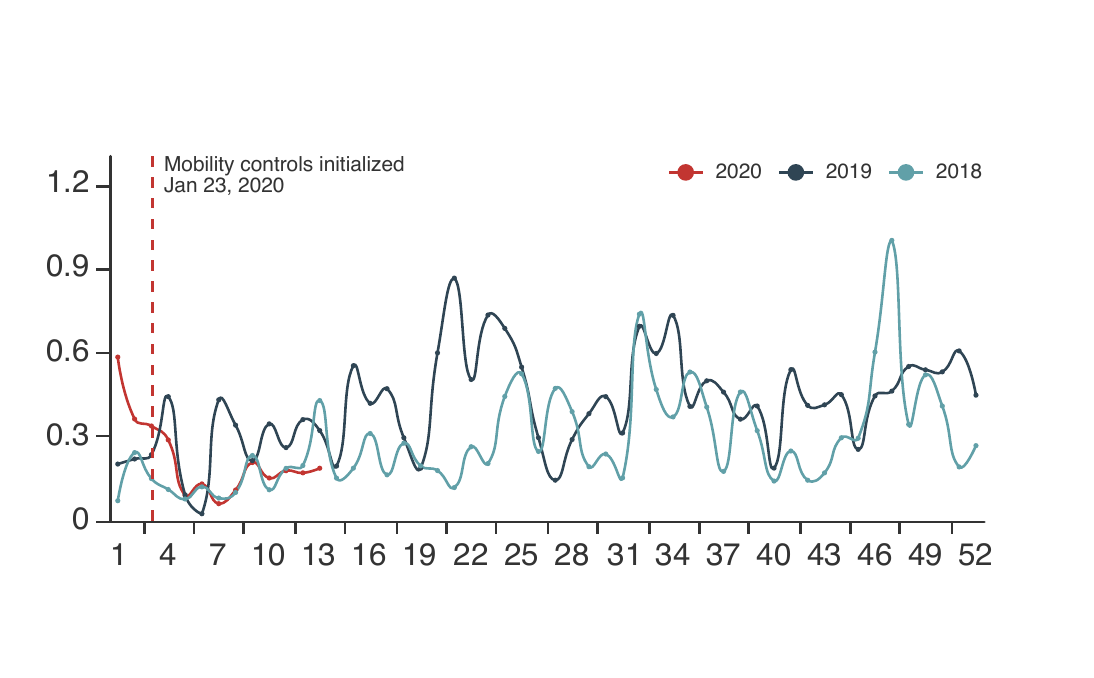}}\
    \subfloat[The \vvv in Parks \& Scenic Spots]{\label{fig:ck-parks-v3}\includegraphics[width=0.49\textwidth,trim={0.48cm 1.08cm 0.58cm 1.08cm},clip]{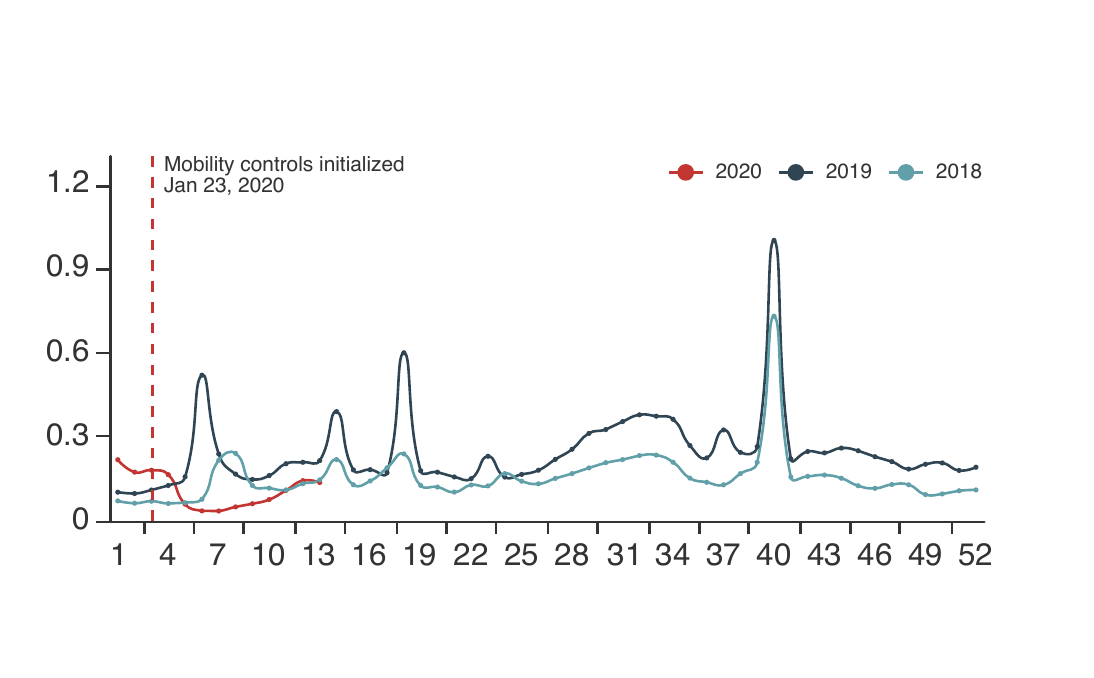}}\
    \subfloat[The \nvc in Parks \& Scenic Spots]{\label{fig:ck-parks-nvc}\includegraphics[width=0.49\textwidth,trim={0.48cm 1.08cm 0.58cm 1.08cm},clip]{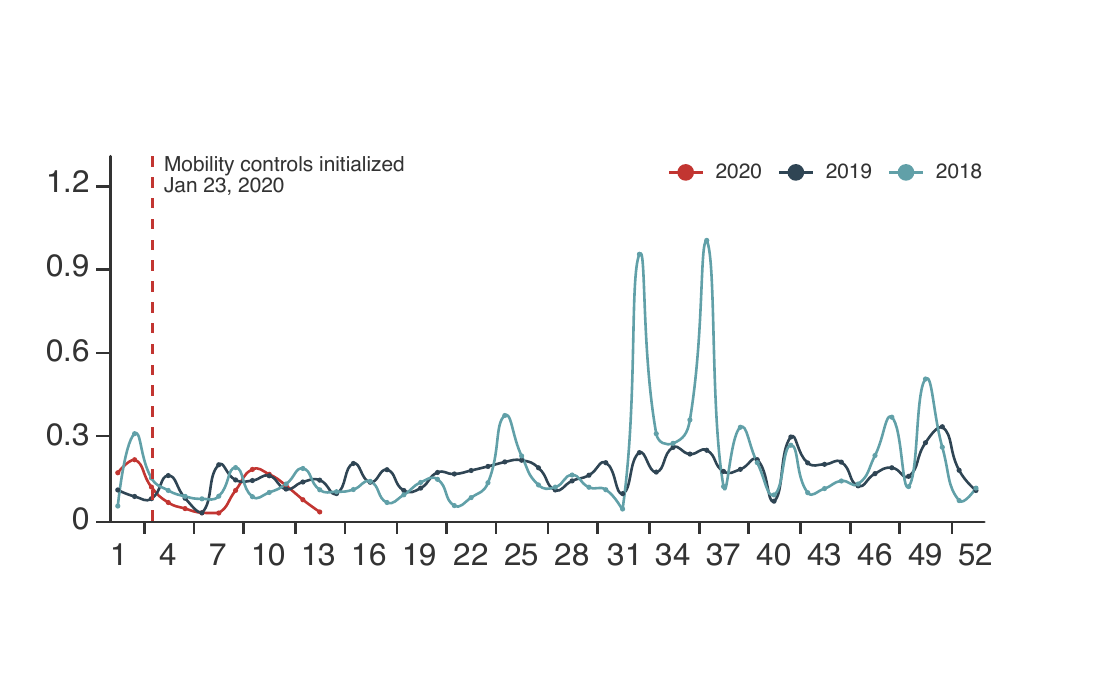}}\
    \caption{The vibrant, Check-mark-shaped recovery of the \vvv in industrial sectors of Mainland China.}
    \label{fig:ck-shaped-cats}
\end{figure}

\begin{figure}[th!]
    \centering
    \subfloat[The \vvv in Workplaces]{\label{fig:v-workplaces-v3}\includegraphics[width=0.49\textwidth,trim={0.48cm 1.08cm 0.58cm 1.08cm},clip]{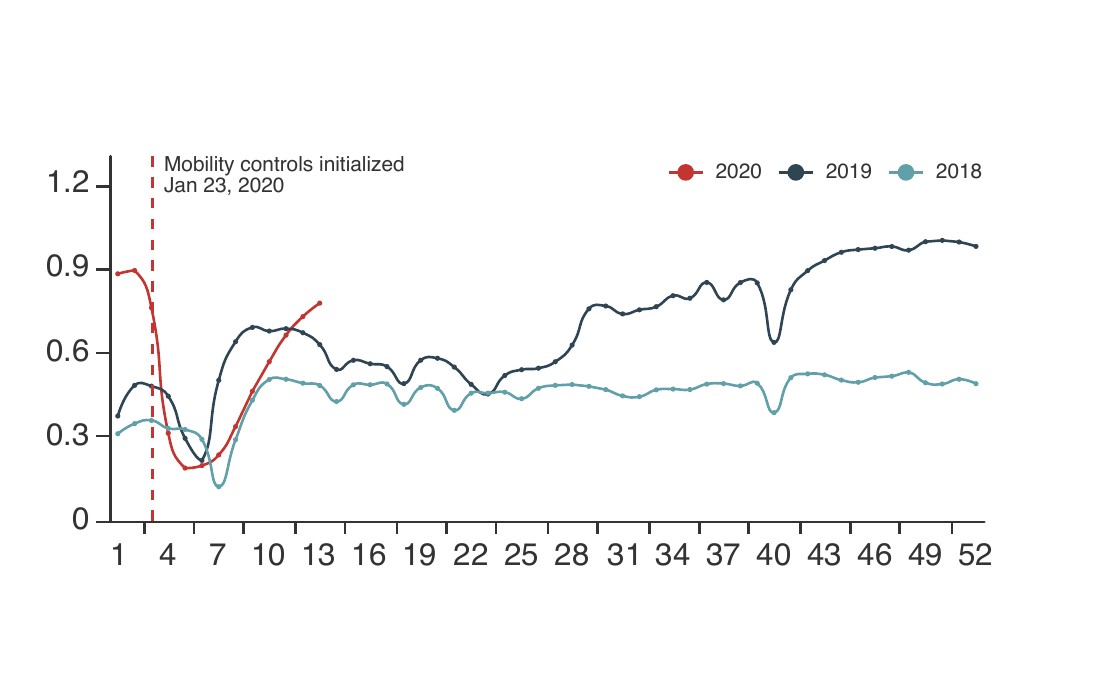}}
    \subfloat[The \nvc for Workplaces]{\label{fig:v-workplaces-nvc}\includegraphics[width=0.49\textwidth,trim={0.48cm 1.08cm 0.58cm 1.08cm},clip]{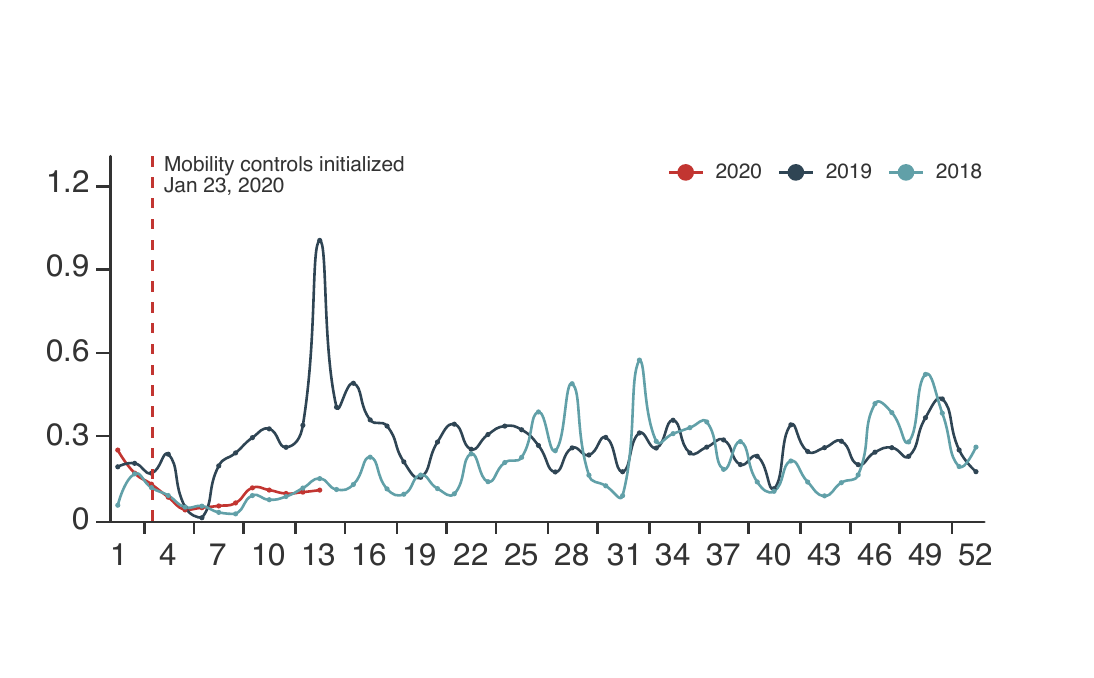}}\
    \subfloat[The \vvv in Residential Areas]{\label{fig:v-residential-v3}\includegraphics[width=0.49\textwidth,trim={0.48cm 1.08cm 0.58cm 1.08cm},clip]{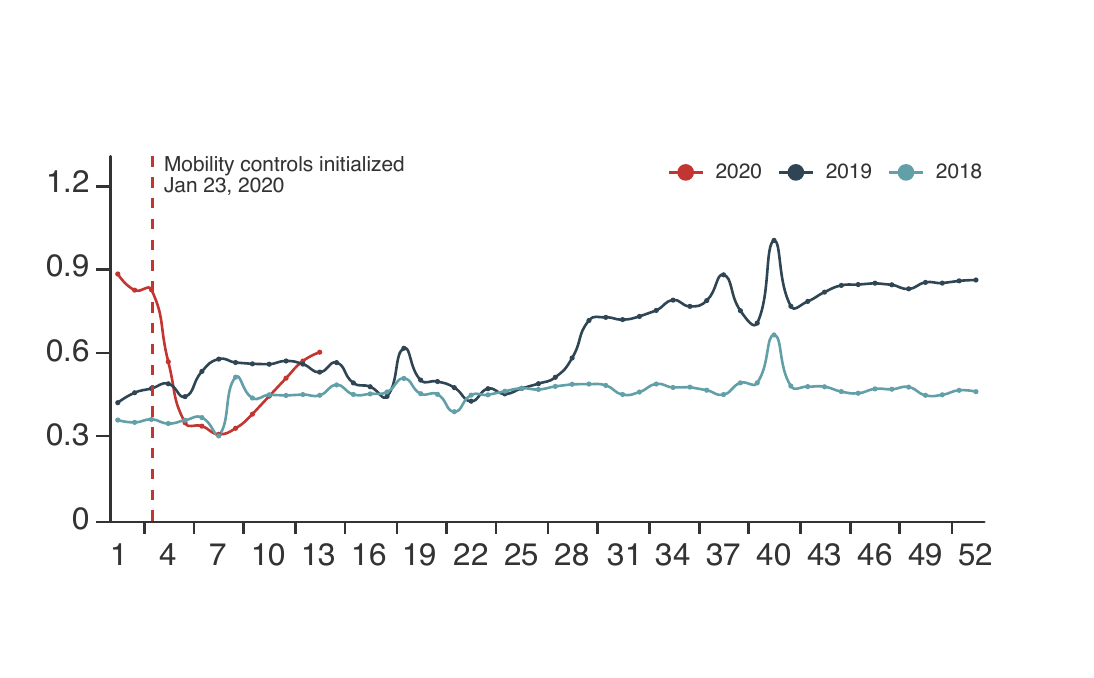}}
    \subfloat[The \nvc for Residential Areas]{\label{fig:v-residential-nvc}\includegraphics[width=0.49\textwidth,trim={0.48cm 1.08cm 0.58cm 1.08cm},clip]{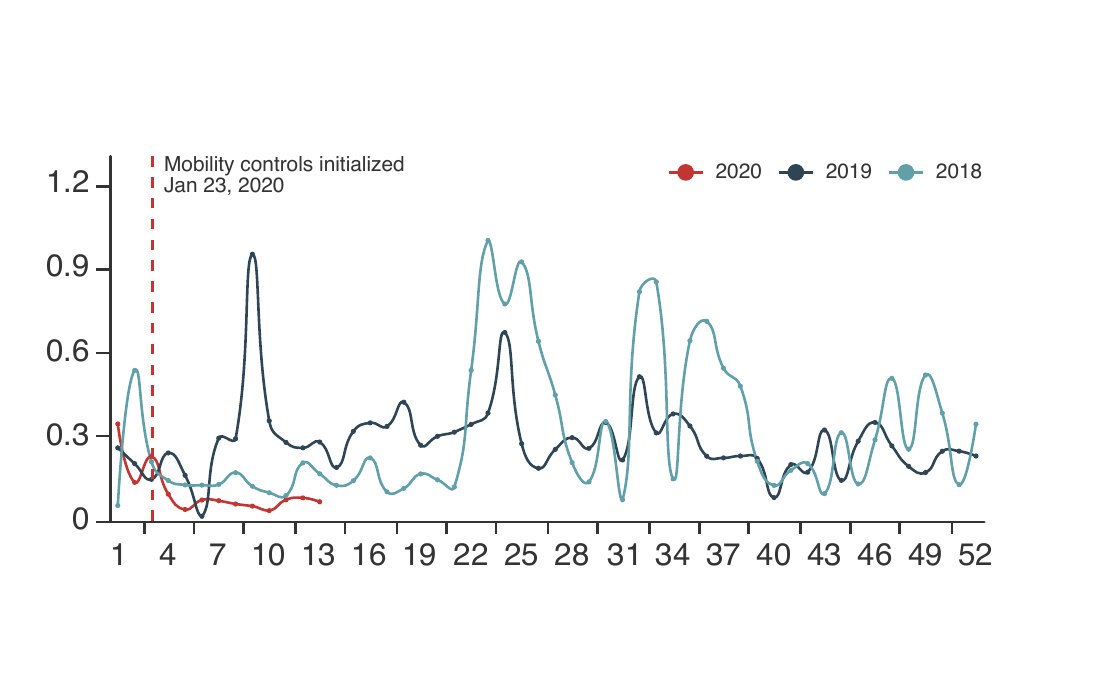}}\
    \subfloat[The \vvv in Shopping \& Markets]{\label{fig:v-shopping-v3}\includegraphics[width=0.49\textwidth,trim={0.48cm 1.08cm 0.58cm 1.08cm},clip]{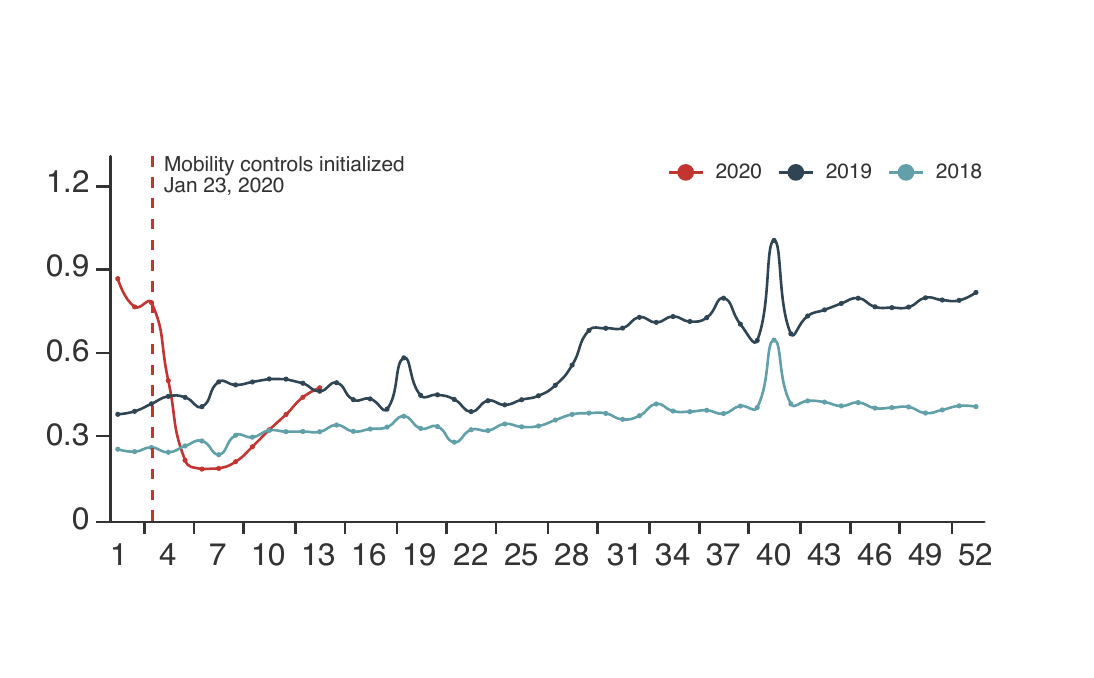}}\
    \subfloat[The \nvc for Shopping \& Markets]{\label{fig:v-shopping-nvc}\includegraphics[width=0.49\textwidth,trim={0.48cm 1.08cm 0.58cm 1.08cm},clip]{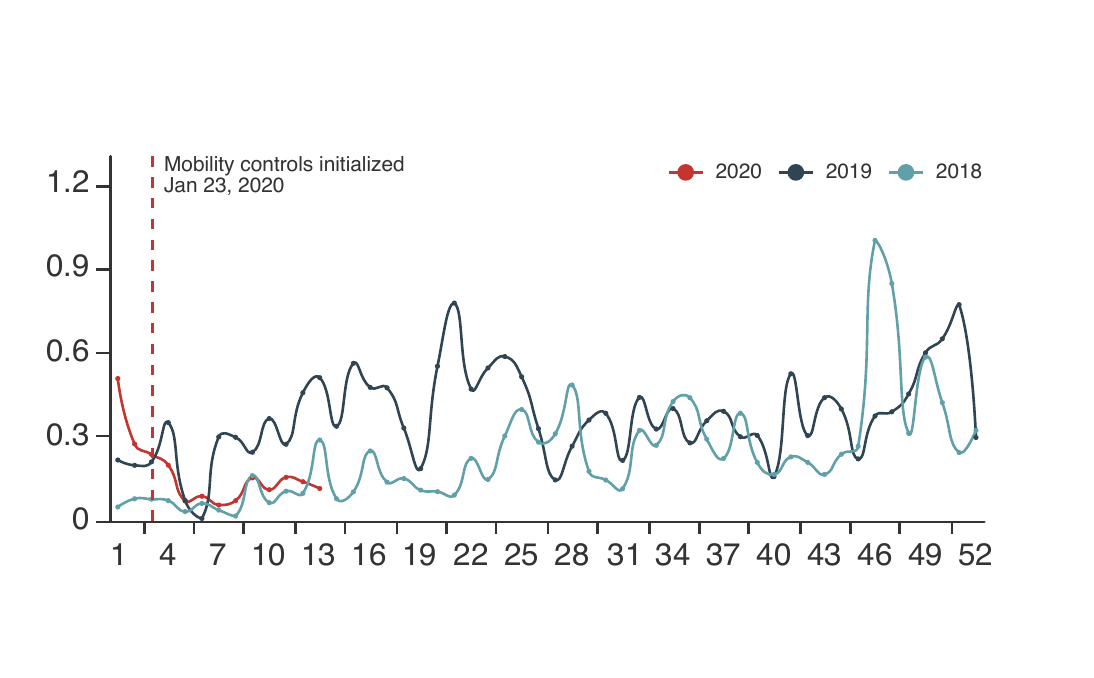}}\
    \subfloat[The \vvv in Hospitals \& Pharmacies]{\label{fig:v-hospital-v3}\includegraphics[width=0.49\textwidth,trim={0.48cm 1.08cm 0.58cm 1.08cm},clip]{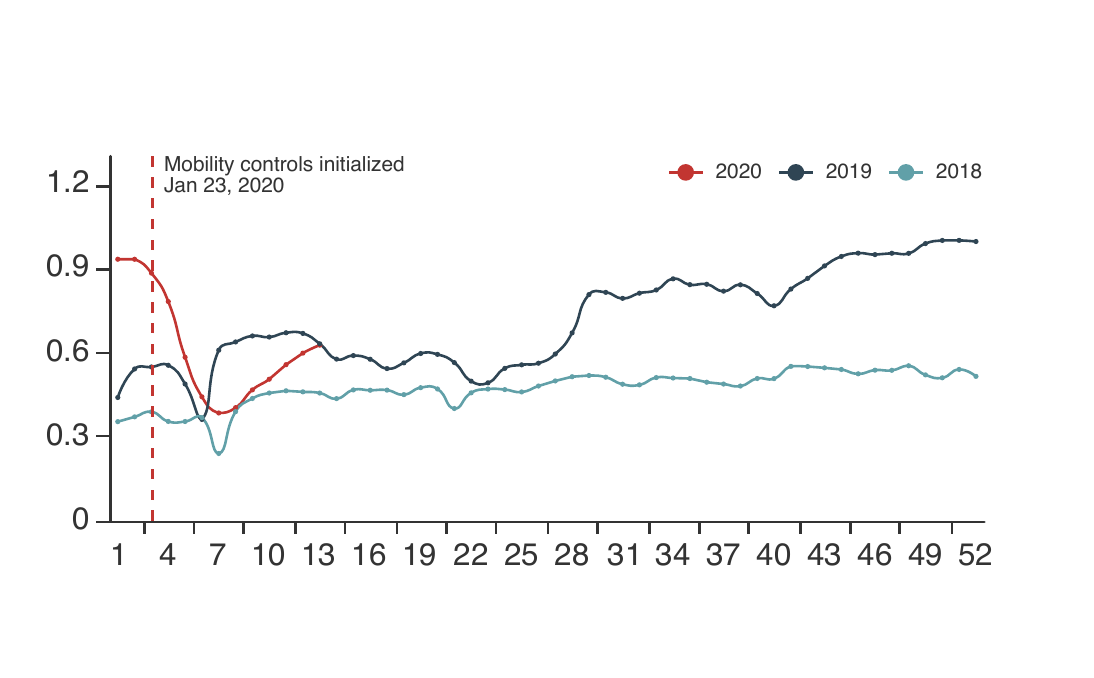}}\
    \subfloat[The \nvc for Hospitals \& Pharmacies]{\label{fig:v-hospital-nvc}\includegraphics[width=0.49\textwidth,trim={0.48cm 1.08cm 0.58cm 1.08cm},clip]{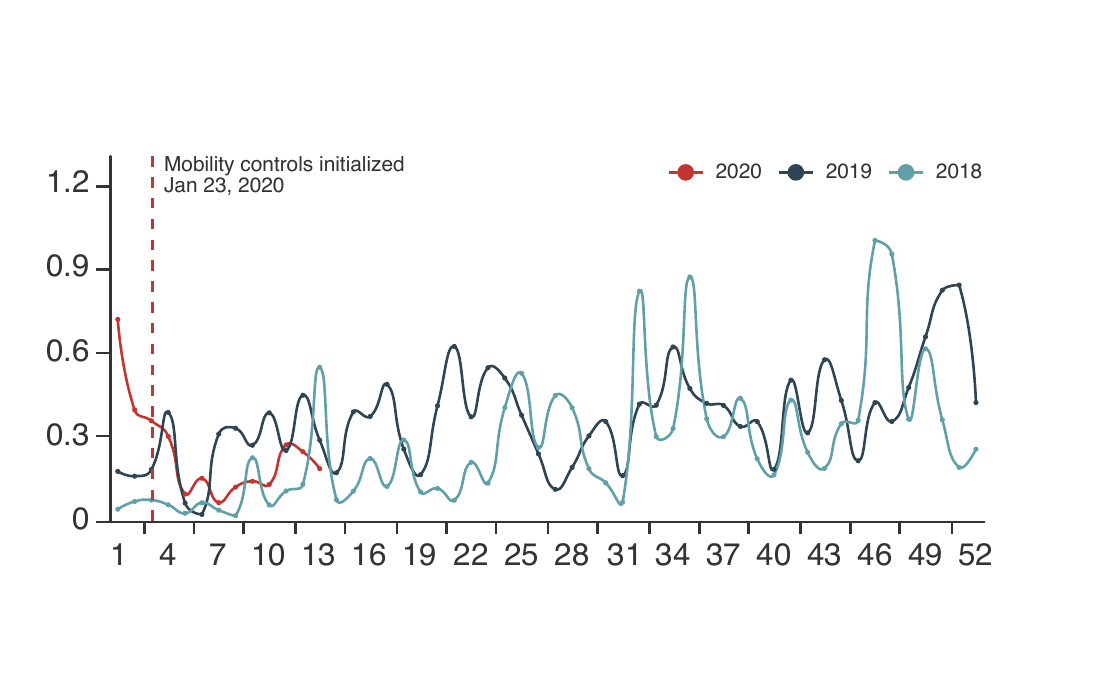}}\
    \caption{The emerging, V-shaped bounces of the \vvv in the industrial sectors of Mainland China.}
    \label{fig:v-shaped-cats}
\end{figure}

We mapped recent trends of \vvv and \nvc of 2020 and the historical traces onto the industrial sectors of 11 major categories, ranging from shopping centers, to hotels, to restaurants, to transportation, to workplaces, to residential areas, and so on. Figures~\ref{fig:l-shaped-cats},~\ref{fig:ck-shaped-cats}, and~\ref{fig:v-shaped-cats} illustrate the trends and records of \vvv and \nvc in those 11 major categories.
For all these categories, we can observe the significant declines in the recent trends of \vvv and \nvc since the 3$^{rd}$ week of 2020, while significant bounces could be found in \vvv for all categories since the 7$^{th}$ week. More specifically, we categorized the economic performance of these sectors, according to the shapes\cite{lshaperec} of their recent \vvv trends, as follows.

(\RNum{1})\emph{~RECESSION: We document a clear, L-shaped recession in travel-dependent public sectors, which are likely to request long-term support, toward returning to the status quo of 2018.} 

Figures~\ref{fig:l-shaped-cats} presents the recent and historical trends of \vvv and \nvc for four L-shaped categories, where the recent trends of both \vvv and \nvc are still lower than the same period of 2018. The trends of \vvv and \nvc for the categories of airports (Figures~\ref{fig:l-airports-v3} and \ref{fig:l-airports-nvc}) and train stations (Figures~\ref{fig:l-trainst-v3} and \ref{fig:l-trainst-nvc}) reveal the decline in demand for transportation with people amid COVID-19 lockdown, which would inevitably lead to the decline in demand for gas and oil. The recent drop of crude oil prices\cite{covid19onoil} helps to validate this observation from the consumption perspectives, such that the crude oil price turned negative for the first time in history and settled at -\$37.63/bbl on April 20, 2020\cite{worldoildropneg}. A recent study demonstrated that the aircraft location data can be used as real-time estimates of GDP in both the UK and the U.S.\cite{Miller2020}, which might also validate our observation from another side.
We observe  similar patterns for the categories of educational institutes and hotels in Figures~\ref{fig:l-eduins-v3},~\ref{fig:l-eduins-nvc},~\ref{fig:l-hotel-v3}, and~\ref{fig:l-hotel-nvc}, respectively. This might be because these two industries might rely on the availability of public transportation, and because they also involve the aggregation of large crowds. The performance of hotels looks better than the performance of the transportation sectors, partially due to the occupation of hotels by Chinese authorities to quarantine the travelers immediately after they entered any Chinese cities. The evidence suggests a clear, but relatively weak, bounce, in terms of transportation and hospitality activities. There would be a long way to go for the full recovery in both consumption and investments.

(\RNum{2})\emph{~RECOVERY: We document a vibrant, Check-mark-shaped ({\large\rotatebox[origin=c]{-120}{\sffamily 7}}) recovery in the sectors of restaurants \& beverages, recreation, and parks \& scenic, which have returned to or even surpassed the status quo of 2018,  but which remain relatively below 2019.} 

In contrast to airports, train stations, educational institutions, and hotels, we can still observe vibrant recovery, in terms of on-site consumption (represented by \vvvtail), in the sectors that are more closely integrated into daily life, such as parks \& scenic venues, restaurants \& beverages, and recreation. In Figures~\ref{fig:ck-restaurants-v3}~and~\ref{fig:ck-restaurants-nvc}, the recent trends of \nvc of restaurants \& beverages have recovered to the level of the same period in 2018, while the \vvv has returned to the status quo of 2019. 
In Figures~\ref{fig:ck-recreation-v3},~\ref{fig:ck-recreation-nvc},~\ref{fig:ck-parks-v3},~and~\ref{fig:ck-parks-nvc}, we can clearly observe that, while the performance of the investments (represented by \nvctail) has been marginally close to the trends in 2018 with oscillations, the visiting frequencies (represented by \vvvtail) to the recreation sites and the parks \& scenic spots have returned to the status quo of 2018. Compared to the trends of public transportation sectors, we could conclude that there has been a better bounce, indicative of greater recovery, in local recreation businesses.
These observations suggested that there has already been a solid recovery of on-site consumption in the sectors of restaurants \& beverages, recreation, and parks \& scenic spots, while the recovery of investment therein would still be on its way. 

(\RNum{3})\emph{~REBOUND / BOUNCES: We document the emerging, V-shaped bounces in the sectors of workplaces, residential, shopping \& markets, and hospitals \& pharmacies, which have surpassed the status quo of both 2018 and 2019.} 

For the workplace category, it has been observed that the recent \vvv has been growing rapidly, and it has even surpassed the volumes in the same period of 2019, while the \nvc was still lower than the performance of 2019 (but comparable to the same period of 2018). We observed the similar \emph{fall-and-rebound} patterns in the recent \vvv and \nvc trends of residential areas. The evidence suggests a strong recovery, and even a year-to-year growth (compared to the same period of 2019) in the consumption activities, within the categories of workplaces and residential areas, while there was no data to support the recovery of investments.
Compared to the trends of the above two sectors, the recent trends within the sectors of (a) shopping \& markets and (b) hospitals \& pharmacies show relatively lower levels of recovery, but their \vvv reach slightly beyond the levels of the same period of 2019. 
In Figures~\ref{fig:v-shopping-v3},~\ref{fig:v-shopping-nvc},~\ref{fig:v-hospital-v3},~and~\ref{fig:v-hospital-nvc}, the recent trends of \nvc of shopping \& markets and of hospitals \& pharmacies have recovered to the level of the same period in 2018, while \vvv has recovered and even earned a year-to-year growth, compared to the same period of 2019. Note that, as the venues from the V-shaped and Check-mark-shaped sectors constitute the largest proportions of all venue types, the trends of \vvv and \nvc in these sectors majorly shaped the nationwide economic recovery demonstrated in Figure~\ref{fig:two-factors-china}.

All above observations suggested four basic conclusions. (1) Investment activities represented by \nvctail, such as the establishment of new hotels, new markets, and new recreation sites, in all categories of industries, have been crushed by the pandemic, and these have not yet recovered to the status quo of 2019. (2) On-site consumption activities, such as check-ins into hotels, restaurants, and scenic visits, in all categories, have recovered, but to varying degrees. (3) For the sectors that are mandatory components of daily life, such as shopping, housing, and working, the bounce back toward recovery of consumption activities has been strong. (4) For the public transportation and travel-dependent sectors, more patience, or even financial liquidity and/or financial support, should be provided to these sectors, to strengthen their recovery.

\begin{figure}[ht!]
    \centering
    \subfloat[The \vvv in the Southern Coast region]{\label{fig:ck-f3-sc-v3}\includegraphics[width=0.49\textwidth,trim={0.48cm 1.08cm 0.58cm 1.08cm},clip]{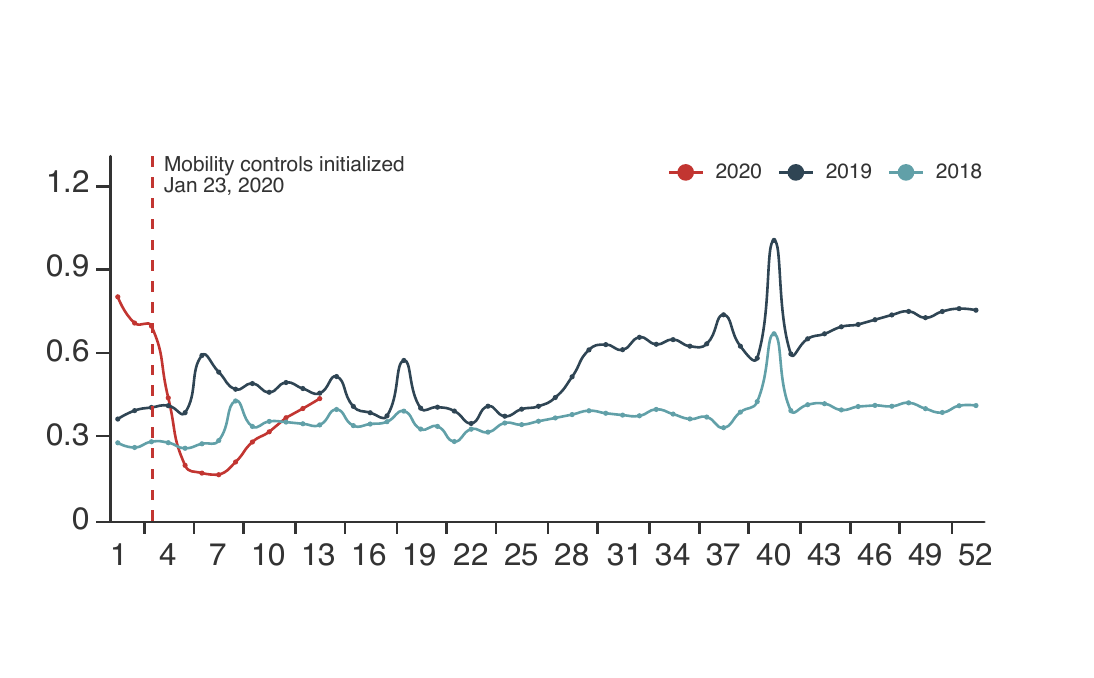}}\
    \subfloat[The \nvc for the Southern Coast region]{\label{fig:ck-f3-sc-nvc}\includegraphics[width=0.49\textwidth,trim={0.48cm 1.08cm 0.58cm 1.08cm},clip]{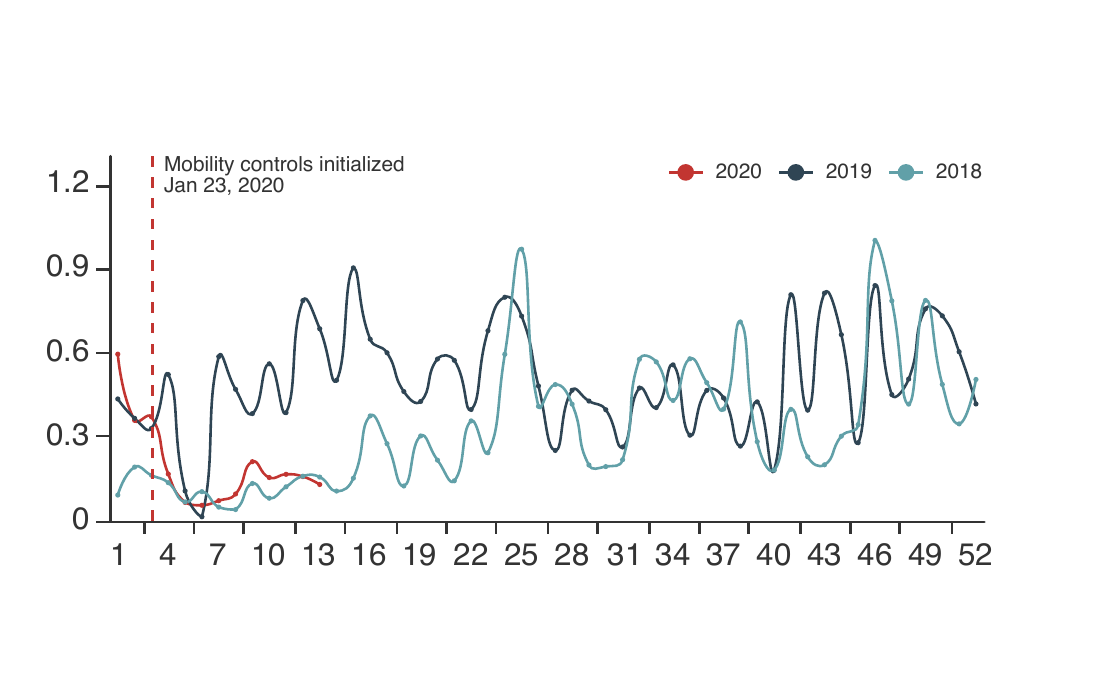}}\
    \subfloat[The \vvv in the Middle Yellow River region]{\label{fig:ck-f3-hh-v3}\includegraphics[width=0.49\textwidth,trim={0.48cm 1.08cm 0.58cm 1.08cm},clip]{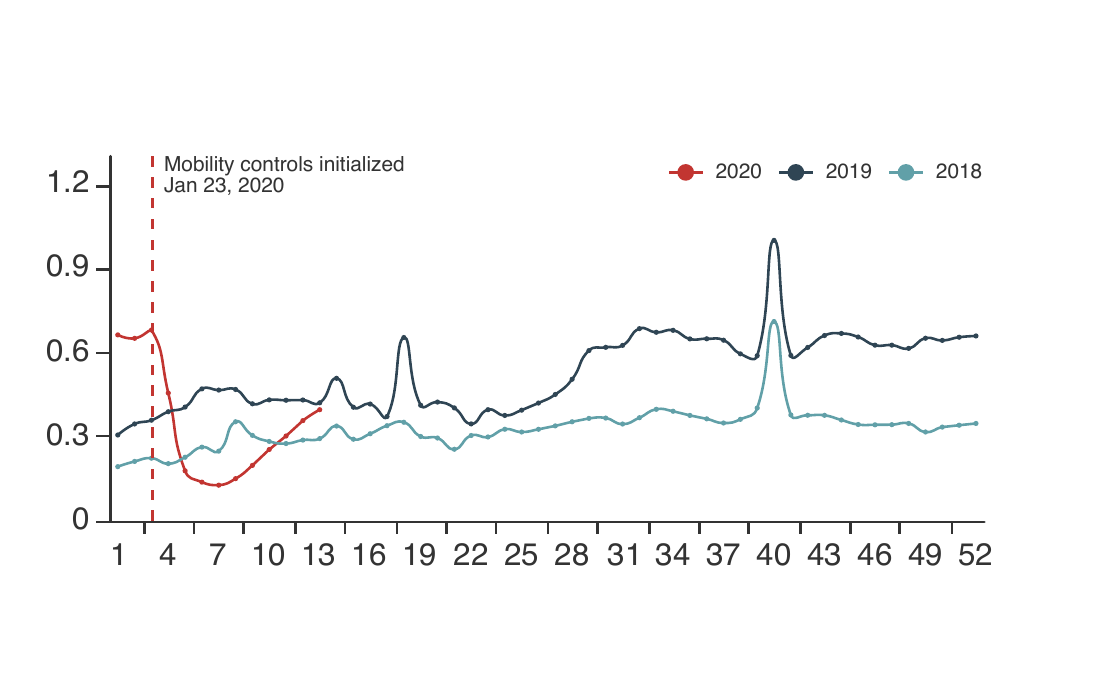}} \
    \subfloat[The \nvc for the Middle Yellow River region]{\label{fig:ck-f3-hh-nvc}\includegraphics[width=0.49\textwidth,trim={0.48cm 1.08cm 0.58cm 1.08cm},clip]{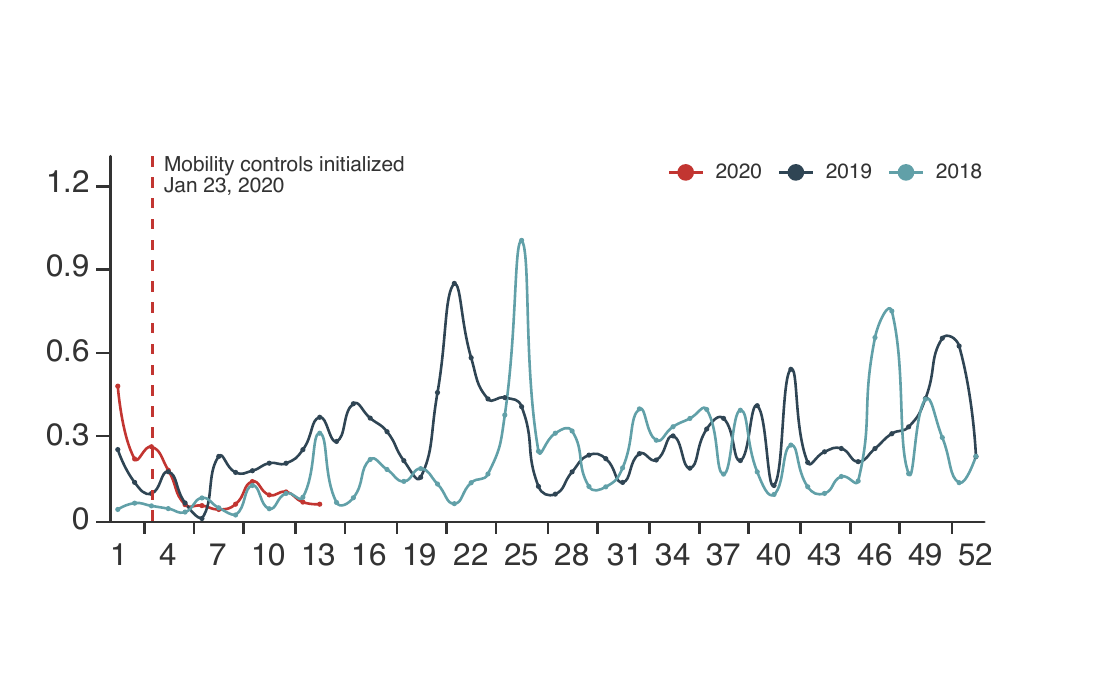}}\
    \subfloat[The \vvv in the Eastern Coast region]{\label{fig:ck-f3-ec-v3}\includegraphics[width=0.49\textwidth,trim={0.48cm 1.08cm 0.58cm 1.08cm},clip]{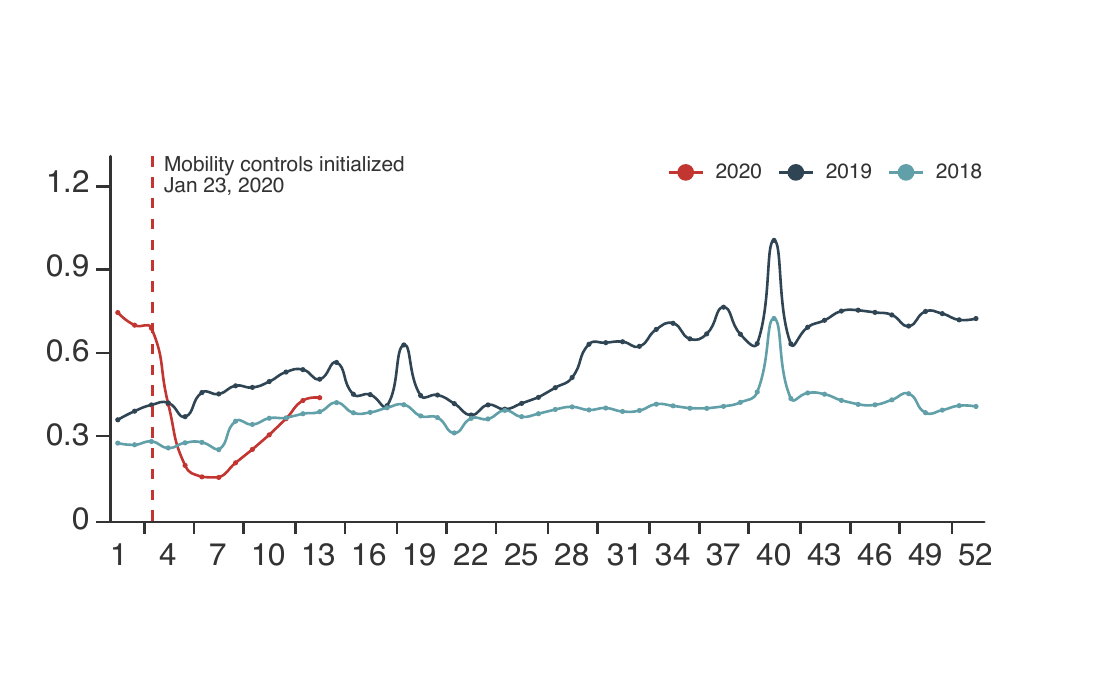}}\
    \subfloat[The \nvc for the Eastern Coast region]{\label{fig:ck-f3-ec-nvc}\includegraphics[width=0.49\textwidth,trim={0.48cm 1.08cm 0.58cm 1.08cm},clip]{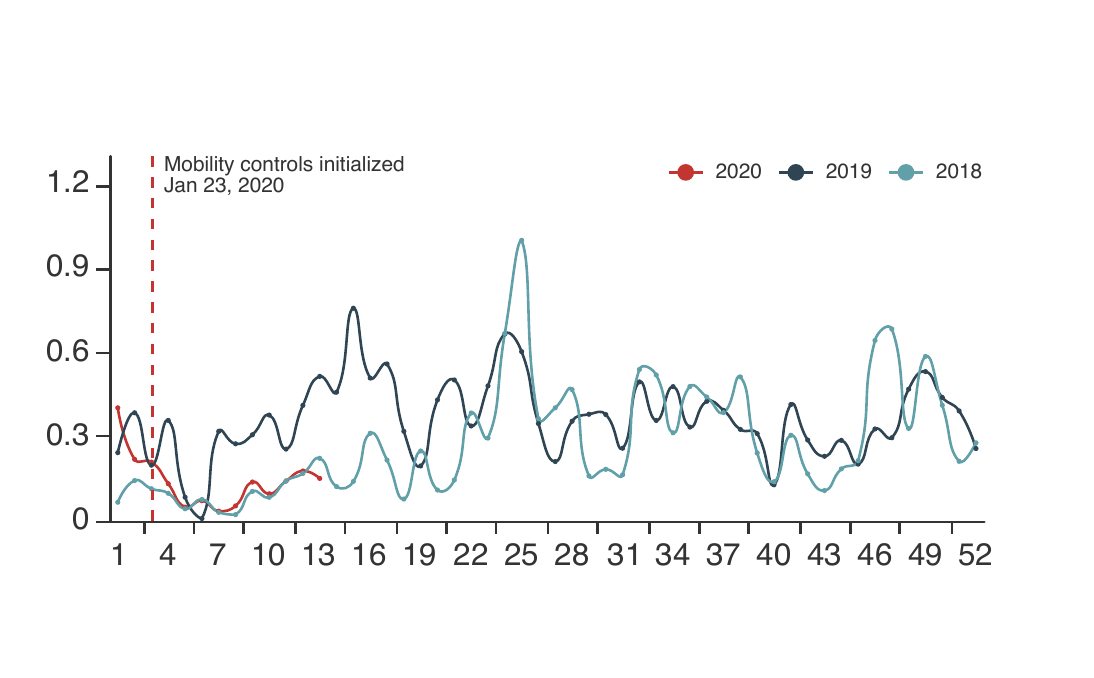}}\
    \subfloat[The \vvv in the Middle Yangtze River region]{\label{fig:ck-f3-cj-v3}\includegraphics[width=0.49\textwidth,trim={0.48cm 1.08cm 0.58cm 1.08cm},clip]{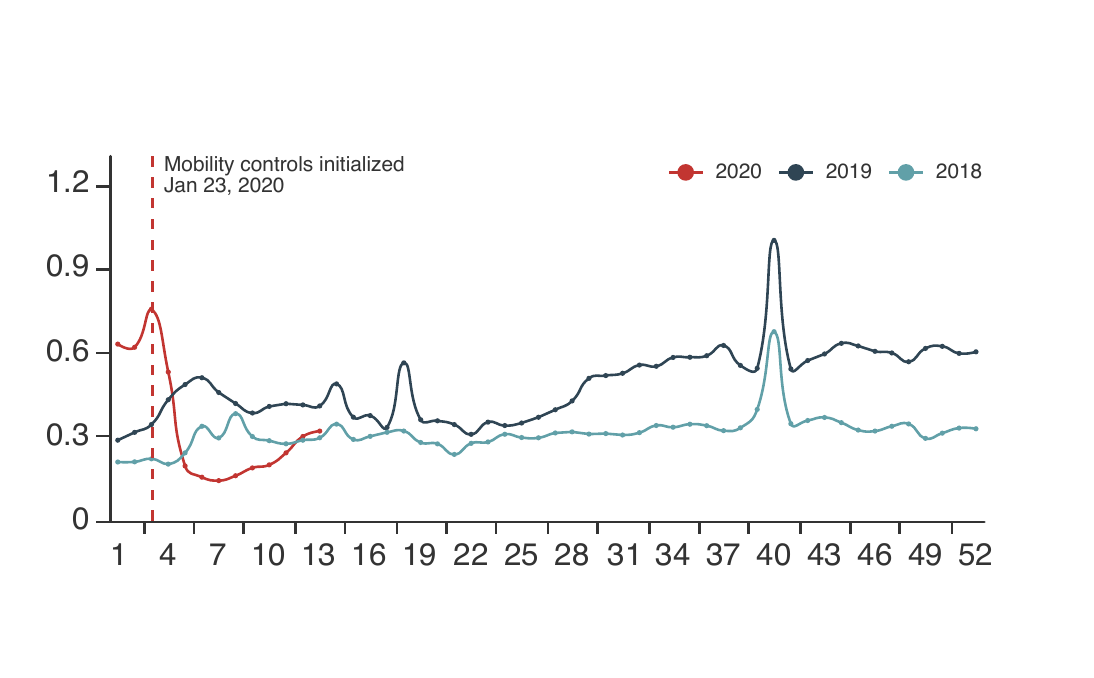}} \
    \subfloat[The \nvc for the Middle Yangtze River region]{\label{fig:ck-f3-cj-nvc}\includegraphics[width=0.49\textwidth,trim={0.48cm 1.08cm 0.58cm 1.08cm},clip]{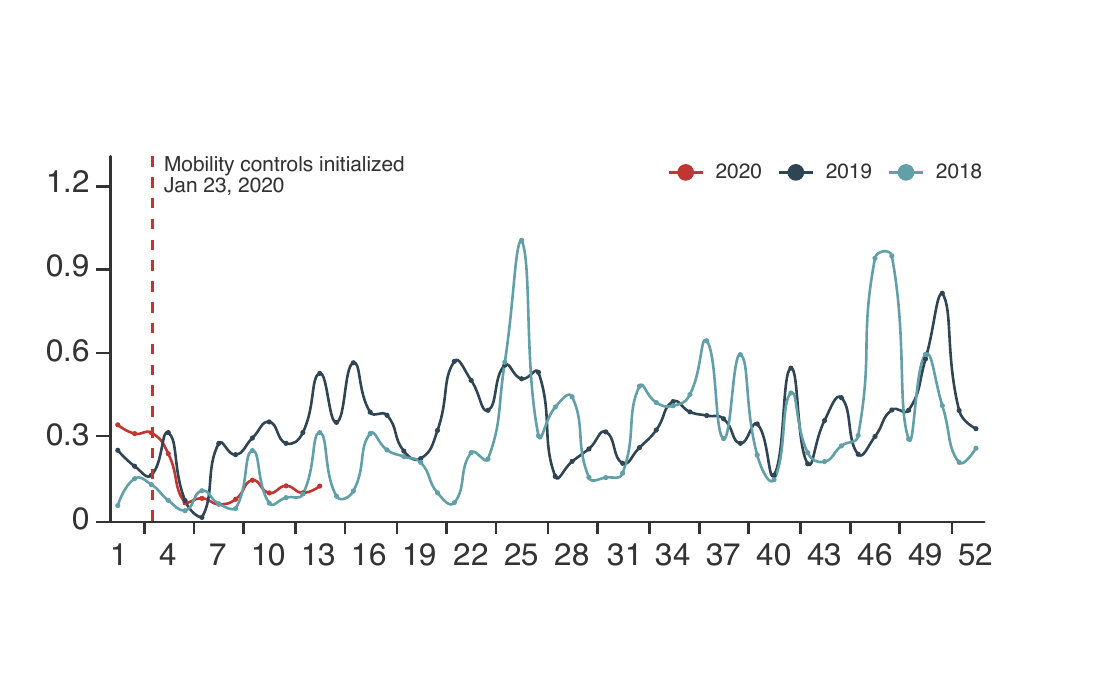}}\
    \caption{\label{fig:ck-shaped-areas}The vibrant, Check-mark-shaped recovery of the \vvv in five economic regions of Mainland China. (1 of 2)}
\end{figure}

\begin{figure}[ht!]
    \centering
    \ContinuedFloat
    \subfloat[The \vvv in the Northern Coast region]{\label{fig:ck-f3-nc-v3}\includegraphics[width=0.49\textwidth,trim={0.48cm 1.08cm 0.58cm 1.08cm},clip]{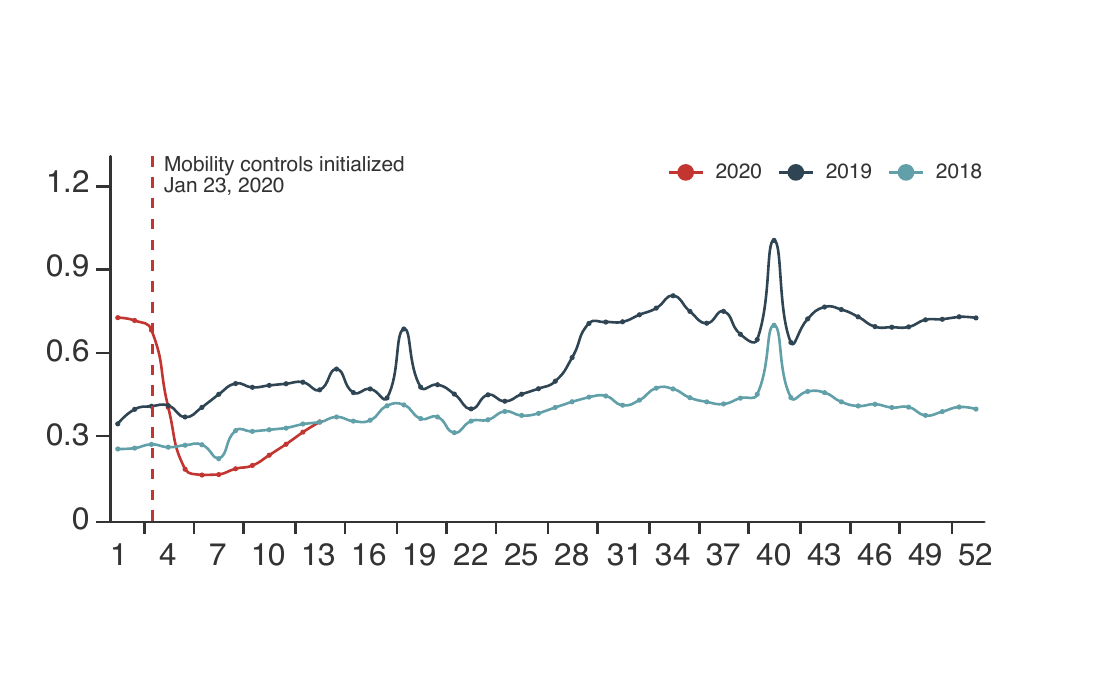}}\
    \subfloat[The \nvc for the Northern Coast region]{\label{fig:ck-f3-nc-nvc}\includegraphics[width=0.49\textwidth,trim={0.48cm 1.08cm 0.58cm 1.08cm},clip]{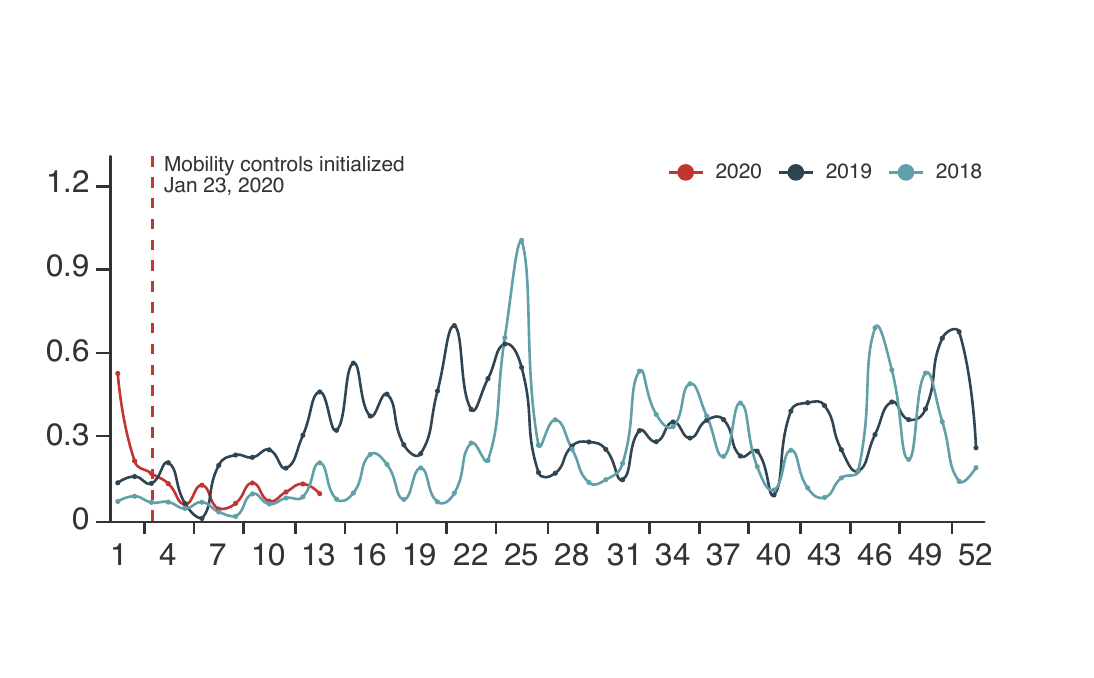}}\
    \caption{The vibrant, Check-mark-shaped recovery of the \vvv in five economic regions of Mainland China. (2 of 2)}
\end{figure}

\subsection*{Economic impact of COVID-19 on the provinces}

While the pandemic clearly crushed the economy of each province, all provinces have recovered from the darkest hours, as viewed from the \vvv and \nvc perspectives (see Supplementary Materials). Given the diverse nature of Chinese economies, across the different provinces, the degrees of recession caused by the COVID-19 pandemic, as well as the rates of progress of recovery, are quite different. In general, we found that the current levels of progress in some underdeveloped provinces, such as those in the Big Northwestern, Northeastern, and Southwestern areas, are much better than those in the other areas, in terms of comparisons with their past records.  On the other hand, the provinces along the Eastern and Southern coasts of China, which used to be the most powerful areas with the strongest economic growth, have not recovered yet, compared to the status quo of 2019. More specifically, we categorize the provinces of Mainland China using the shape of recession and recovery, as follows.

(\RNum{1})\emph{~RECESSION: We document a painful, L-shaped recession, found in the data of Hubei, Beijing, and Tianjin, probably due to the continuous impact of ongoing mobility restrictions.} 

As shown in Figure~\ref{fig:a-tale-of-three-cities}, we can clearly observe an L-shaped recession of \vvv in the data of Hubei, Beijing, and Tianjin (\vvv in these provinces/provincial cities have not yet recovered to the status quo of 2018), while in Hubei, the rebound of \nvc has not even started yet. As the epicenter of COVID-19 in Mainland China, Wuhan --- the capital of Hubei province --- has become the city with the most infection cases and the highest death toll in the country. The province-wide mobility restrictions have been removed in late March, and Wuhan finally ended its quarantine on the 8$^{th}$ of April, 2020. We can observe the continuous shrinkage of \vvv from the 3$^{rd}$ week to the 10$^{th}$ week of 2020, while nationwide, such an  indicator has bounced back up significantly, since the 8$^{th}$ week, as shown in Figure~\ref{fig:two-factors-china}. The \vvv trends in Hubei could be divided into three periods: (a) the cliff fall (the 3$^{rd}$--5$^{th}$ week); (b) the slow sliding (the 6$^{th}$--10$^{th}$ week); and (c) the hill climbing (11$^{th}$--present). We found  that the trends in Hubei, which incorporate a slow sliding period of significant length, are unique, when compared to the rest of the provinces or areas in Mainland China. In terms of \nvctail, Hubei has been demonstrated to be at a low level, compared to the past records. 

On the other hand, Beijing, which is thousands of kilometers away from Wuhan, and which has a much lower infection rate, also demonstrated a weak economic recovery, with a consistent but slow bounce, since the 7$^{th}$ week of 2020. This is likely because Beijing carried out a set of strict mobility control policies, which were comparable to those of Wuhan\cite{wikihubeilockd}. Furthermore, the slow economic recovery of Beijing also influenced the pace of recover of its satellite provincial city, Tianjin. In general, Hubei, Beijing, and Tianjin have been badly hit by the outbreaks of COVID-19. Their consumption activities (represented by \vvvtail) have recovered from their lowest points, but they have not yet returned to the status quo of 2018. Meanwhile, the investment activities (represented by \nvctail) of Beijing and Tianjin have rebounded. 

(\RNum{2})\emph{~RECOVERY: We document a clear, Check-mark-shaped ({\large\rotatebox[origin=c]{-120}{\sffamily 7}}) recovery in the data of provinces of the Southern Coast, the Middle Yellow River, the Eastern Coast, the Middle Yangtze River, and the Northern Coast economic regions of Mainland China.} 

We mapped the trends of \vvv and \nvc onto the eight economic regions of Mainland China, including the Eastern Coast, the Southern Coast, the Northern Coast, the Big Northwest, the Northeast, the Southwest, the Middle Yellow River, and the Middle Yangtze River. Each region consists of multiple provinces/provincial cities. These regions were defined by the National Bureau of Statistics of China\cite{chinagdpbyregion}.
We observed a clear, Check-mark-shaped recovery in five of them. More specifically, the recent trends of \vvv of these five regions have bounced back up to the status quo of 2018, albeit that they are still below the same period of 2019. These five regions include the richest areas of Mainland China, such as the provinces in the Southern Coast and Eastern Coast regions, as well as the traditional agricultural and industrial backbones, such as the provinces in the Middle Yellow River region, the Middle Yangtze River region and the Northern Coast region. The economic fundamentals\cite{Lueaau9413} have helped these provinces recover rapidly. 
Note that Hubei province is part of the Middle Yangtze River region, while Beijing and Tianjing are part of the Northern Coast region. We did not perform the statistics separately. 

(\RNum{3})\emph{~REBOUNDING: We discovered emerging, V-shaped bounces in the data of provinces in the Big Northwest, Northeast, and Southwest economic regions of Mainland China.} 

We evidenced that the Big Northwest, Northeast, and Southwest economic regions have rapidly rebounded from the bottom and have returned to levels close to or even beyond the status quo of 2019. 
These three regions had already been under the macroeconomic recession before the pandemic\cite{ftreport}, or were even considered to be the underdeveloped areas\cite{feng2019combating,wei2017regional} in Mainland China. Their rapid recovery with strong rebounds might suggest that the COVID-19 pandemic has not made any huge impact on the economics in these regions, partially due to the long distances from these regions to epicenters\cite{Chinazzieaba9757}.

The spatial dissection of the Chinese economy indicates that, compared to the Southern Coast and Eastern Coast regions of Mainland China, provinces in the Big Northwest, Northeast, and Southwest regions majorly rely on internal production and consumption, with sufficient resources (e.g., energy) and sufficient supplies\cite{wu2019decoupling} (e.g., foods). On the other hand, the industries in the coastal regions of Mainland China heavily rely on the export business and the global supply chain\cite{wang2017chinese,chen2017coastal}, while these business opportunities have been crushed by the global pandemic\cite{globalsc}. In this way, we are able to conclude that \emph{the self-sufficient and self-sustainable economic regions\cite{mayhew1995aristotle}, with internal supplies, internal production, and internal consumption, could recover from the pandemic even faster.}

\section*{Conclusion}
In this study, we attempt to quantify the economic impact of COVID-19 with the nationwide mobility data, including the volumes of visits to venues (\vvvtail) of all industrial sectors and the number of newly created venues (\nvctail),  from Baidu Maps. We consider \vvv and \nvc as important economic indicators,  related to consumption and investment activities, respectively. We then discover the strongly  positive correlations between the indicators and the national and provincial GDPs of Mainland China. We further use the drifts of these two indicators during the pandemic, together with the year-to-year changes compared to the past two years (2018 and 2019), to understand the impact of COVID-19 upon Chinese economies. The results show that industrial sectors that are mandatory to human life have recovered faster than the rest of the sectors, while the self-sufficiency of a region might help to avoid the hurt caused by the outbreak, due to being more insulated from, and less impacted by, the decline of global supply chains.

\section*{Data availability}
All experiments in this paper were carried out using anonymous data and secure data analytics provided by Baidu Data Federation Platform (Baidu FedCube). For data accesses and usages, contact us through \url{http://shubang.baidu.com/page/page_en.html}.

\section*{Author contributions statement}
Jizhou Huang formulated the research problems, detailed the research design, originated the framework of the article, and drafted parts of the manuscript. Haifeng Wang proposed the research, coordinated the research efforts, and oversaw the whole research process. Haoyi Xiong and Miao Fan contributed to writing the article and performed data analysis. An Zhuo collected the experimental data from Baidu Maps and carried out the data visualization. Ying Li managed the user privacy protocol and was in charge of data anonymization. Dejing Dou contributed to revising the article.

\bibliography{main}

\appendix
\section*{Supplementary Materials}
Here we first introduce the details on data normalization, and then we elaborate upon our analysis of the \vvvtail-vs-GDP and \nvctail-vs-GDP correlations, as well as upon the trends of \vvv and \nvc in 31 provinces of Mainland China.

\subsection*{Data normalization}
For the sake of preserving the data privacy of the users in Baidu Maps, we used normalized data to plot all the figures in this article. The data that need to be normalized include GDP, \vvvtail, and \nvctail, and we adopt the following formula to normalize them:

\begin{equation}
    normalized(\textbf{x}) = \frac{\textbf{x}}{\text{max}(\textbf{x})}
\end{equation}
where $\textbf{x}$ represents the time series data on GDP, \vvvtail, and \nvctail.

\subsection*{Correlation analysis of the GDP with \vvv and \nvc}
We conducted correlation analysis of the GDP of 31 provinces with \vvv and \nvctail. Figure~\ref{fig:f1-GDP-corr} illustrates the normalized values of GDP, \vvvtail, and \nvc from the first quarter of the year 2018 to the fourth quarter of the year 2019. To be specific, the sub-figures in Figure~\ref{fig:f1-GDP-corr} are ranked by the \vvvtail-vs-GDP correlations in descending order. In addition, it should be noted that the quarterly GDP data of 31 provinces for 2020 Q1 have not yet been released by the National Bureau of Statistics of China as of April 20, 2020. The results show that both \vvv and \nvc have significantly (i.e., p-values $<$ 0.01) positive correlations with GDP in all $31$ provinces of Mainland China.

\begin{figure}
    \centering
    \subfloat[Correlations between the GDP and  \nvctail/\vvv of Tianjin]{\label{fig:f1-GDP-tianjin}\includegraphics[width=0.685\textwidth,trim={0.8cm 1.1cm 1.0cm 1.35cm},clip]{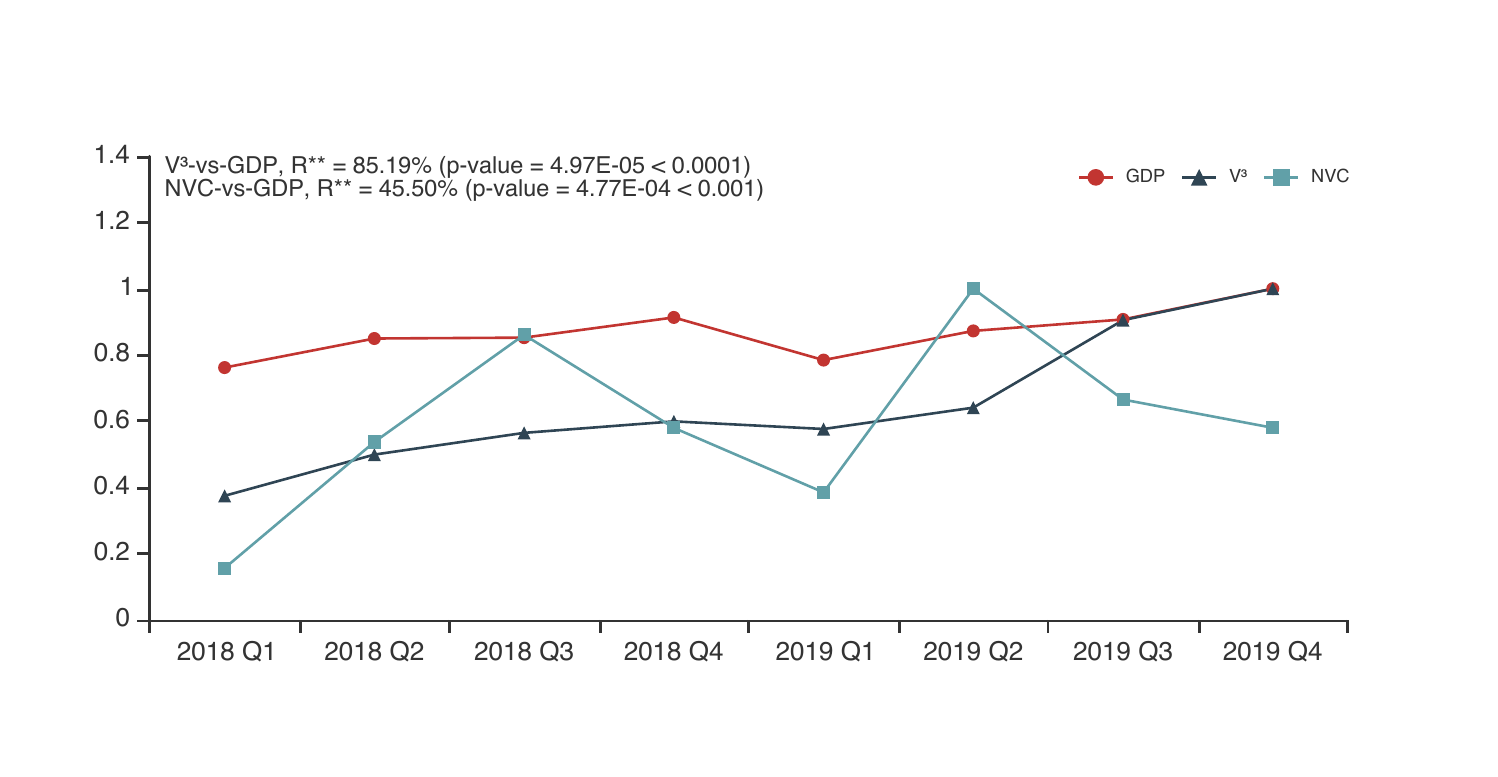} } \
    \subfloat[Correlations between the GDP and  \nvctail/\vvv of Hebei]{\label{fig:f1-GDP-hebei}\includegraphics[width=0.685\textwidth,trim={0.8cm 1.1cm 1.0cm 1.35cm},clip]{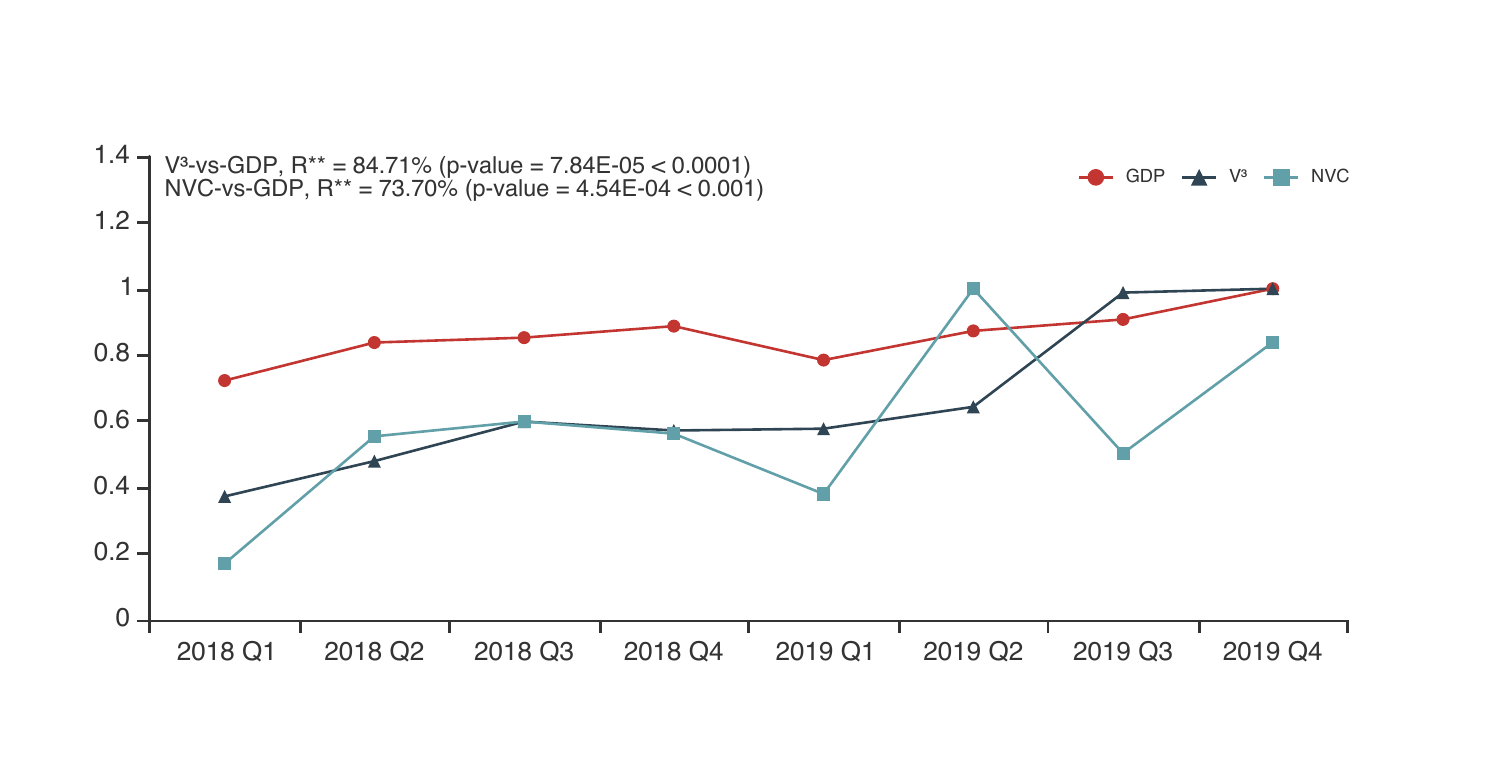} } \
    \subfloat[Correlations between the GDP and  \nvctail/\vvv of Shandong]{\label{fig:f1-GDP-shandong}\includegraphics[width=0.685\textwidth,trim={0.8cm 1.1cm 1.0cm 1.35cm},clip]{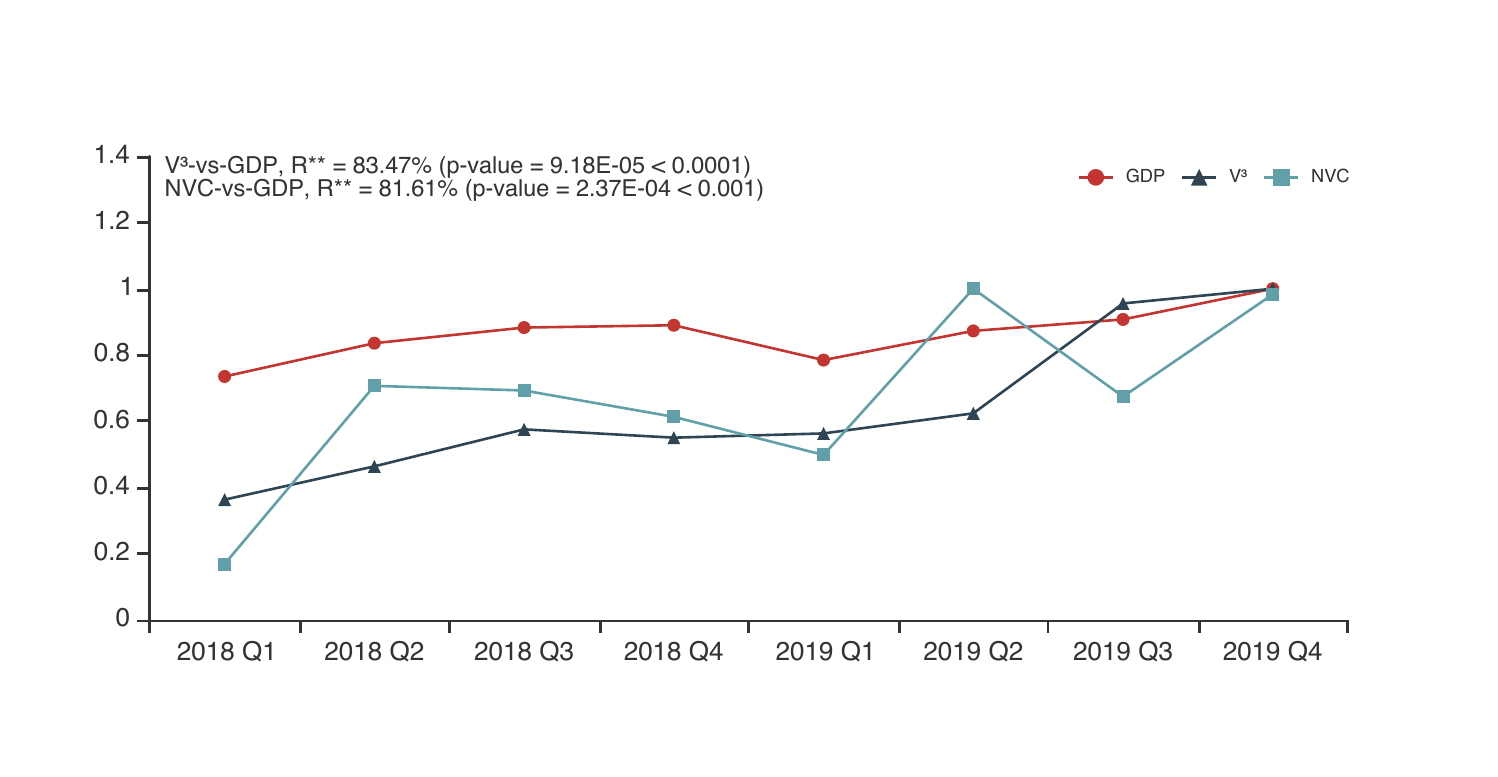} } \
    \subfloat[Correlations between the GDP and  \nvctail/\vvv of Fujian]{\label{fig:f1-GDP-fujian}\includegraphics[width=0.685\textwidth,trim={0.8cm 1.1cm 1.0cm 1.35cm},clip]{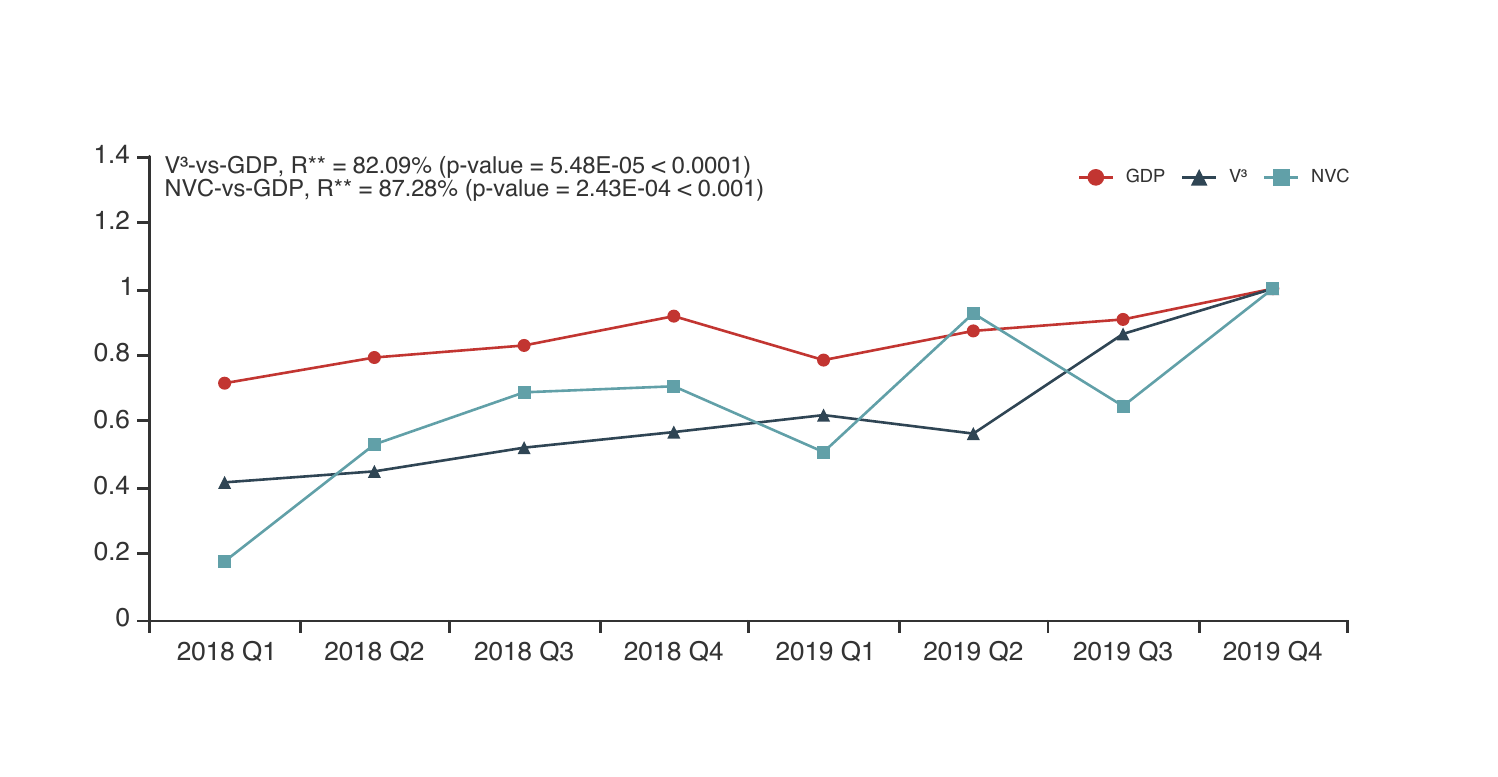} } \
    \caption{\label{fig:f1-GDP-corr}Correlations between the GDP and  \nvctail/\vvv of 31 provinces. (1 of 8)}
\end{figure}

\begin{figure}
    \centering    
    \ContinuedFloat
    \subfloat[Correlations between the GDP and  \nvctail/\vvv of Beijing]{\label{fig:f1-GDP-beijing}\includegraphics[width=0.685\textwidth,trim={0.8cm 1.1cm 1.0cm 1.35cm},clip]{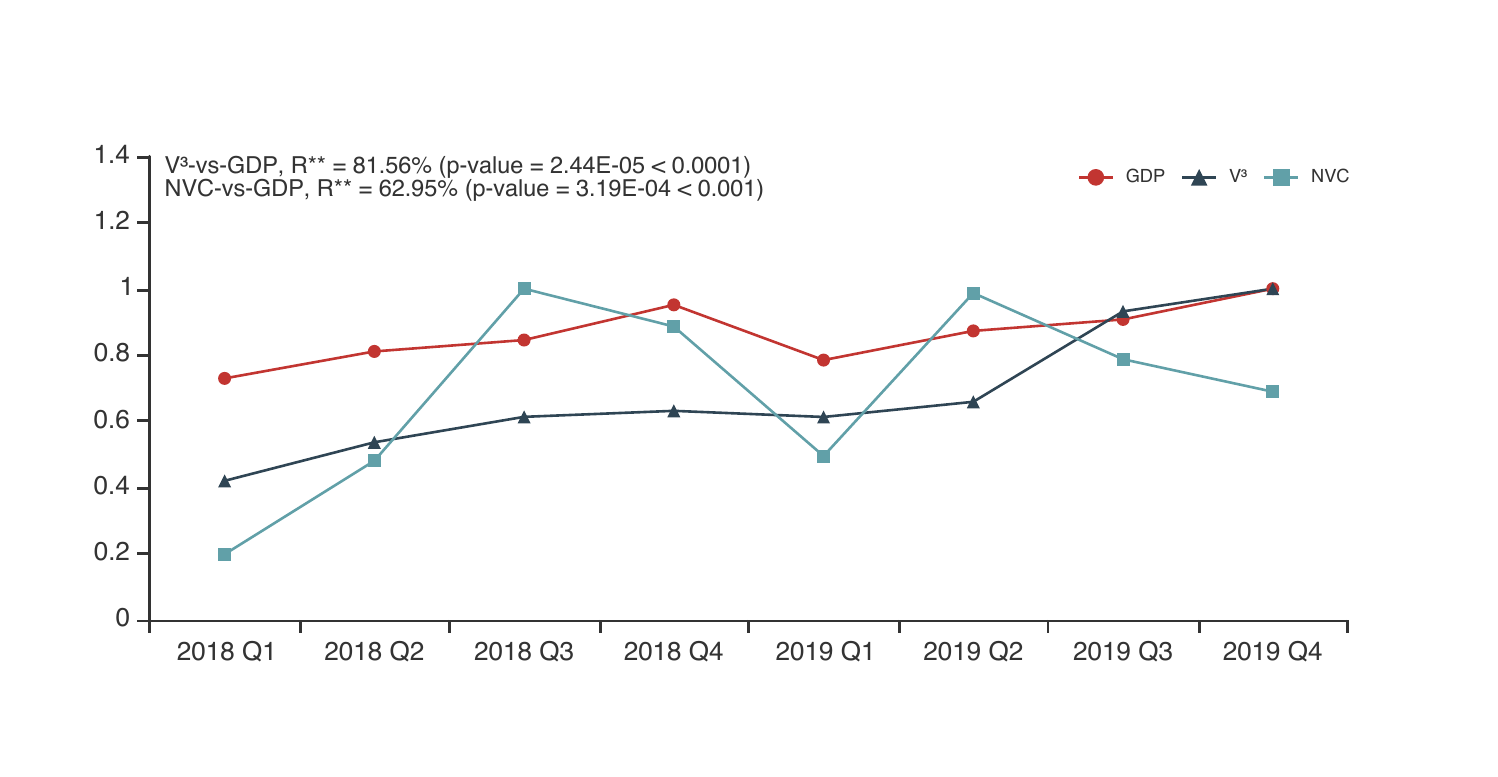} } \
    \subfloat[Correlations between the GDP and  \nvctail/\vvv of Yunnan]{\label{fig:f1-GDP-yunnan}\includegraphics[width=0.685\textwidth,trim={0.8cm 1.1cm 1.0cm 1.35cm},clip]{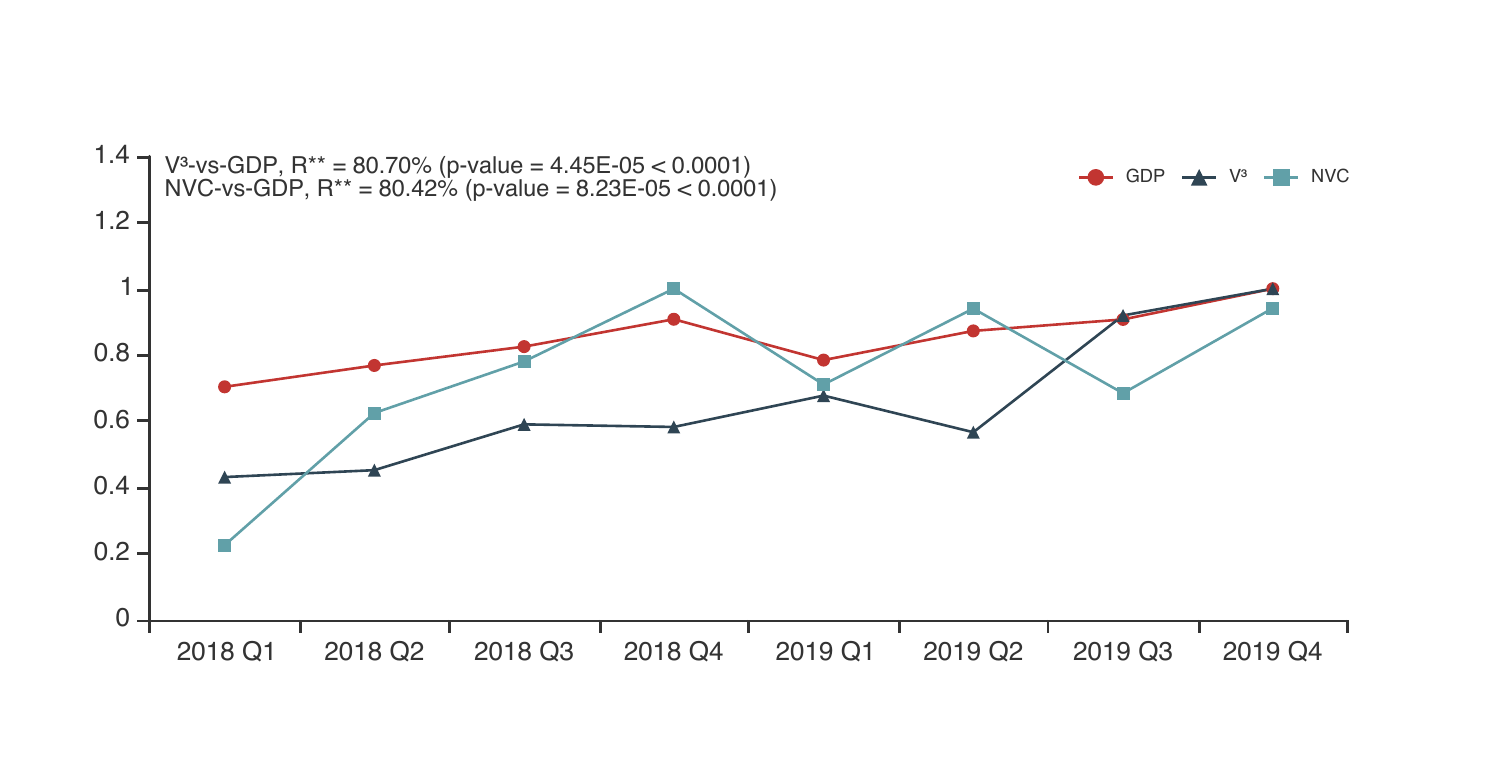} } \
    \subfloat[Correlations between the GDP and  \nvctail/\vvv of Jilin]{\label{fig:f1-GDP-jilin}\includegraphics[width=0.685\textwidth,trim={0.8cm 1.1cm 1.0cm 1.35cm},clip]{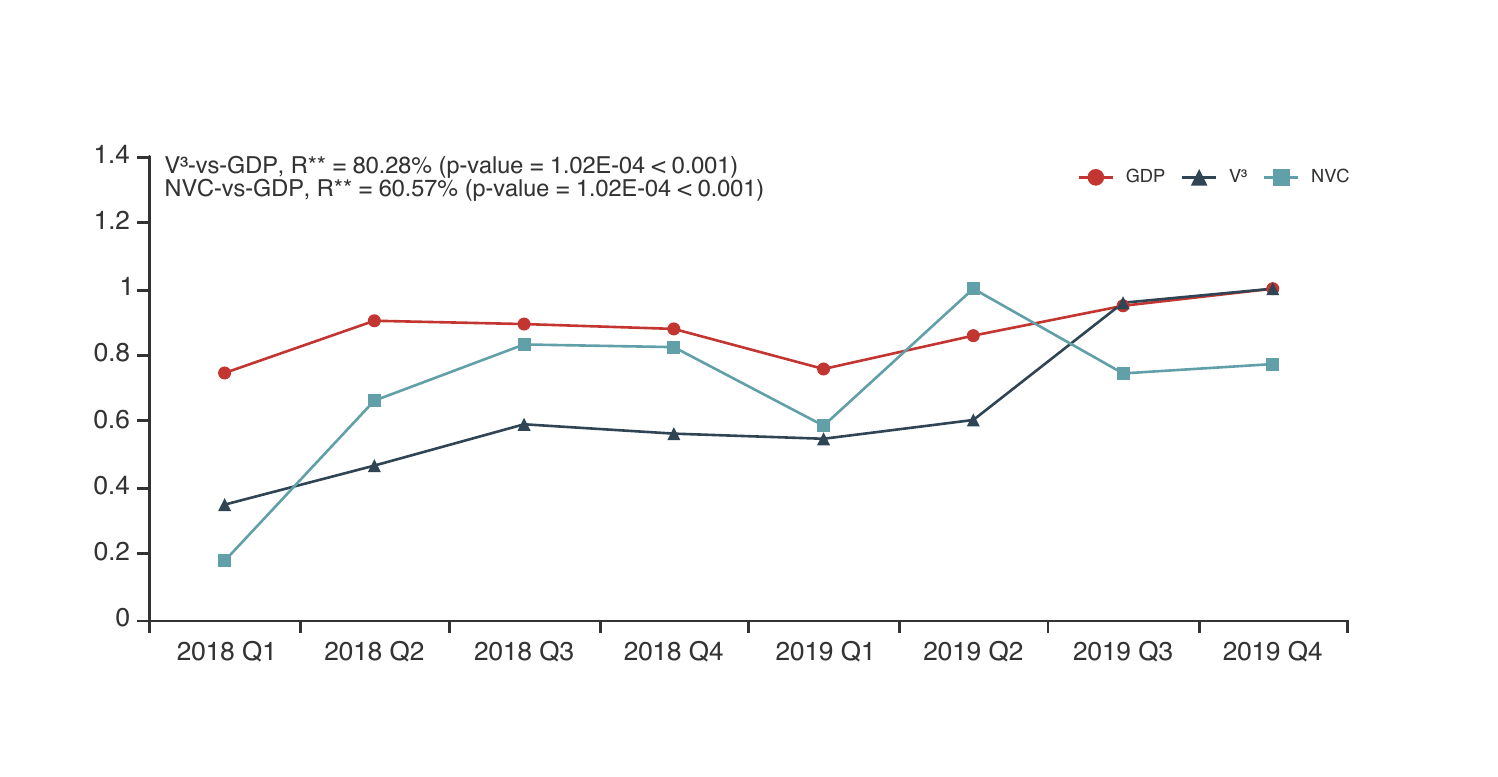} } \
    \subfloat[Correlations between the GDP and  \nvctail/\vvv of Hunan]{\label{fig:f1-GDP-hunan}\includegraphics[width=0.685\textwidth,trim={0.8cm 1.1cm 1.0cm 1.35cm},clip]{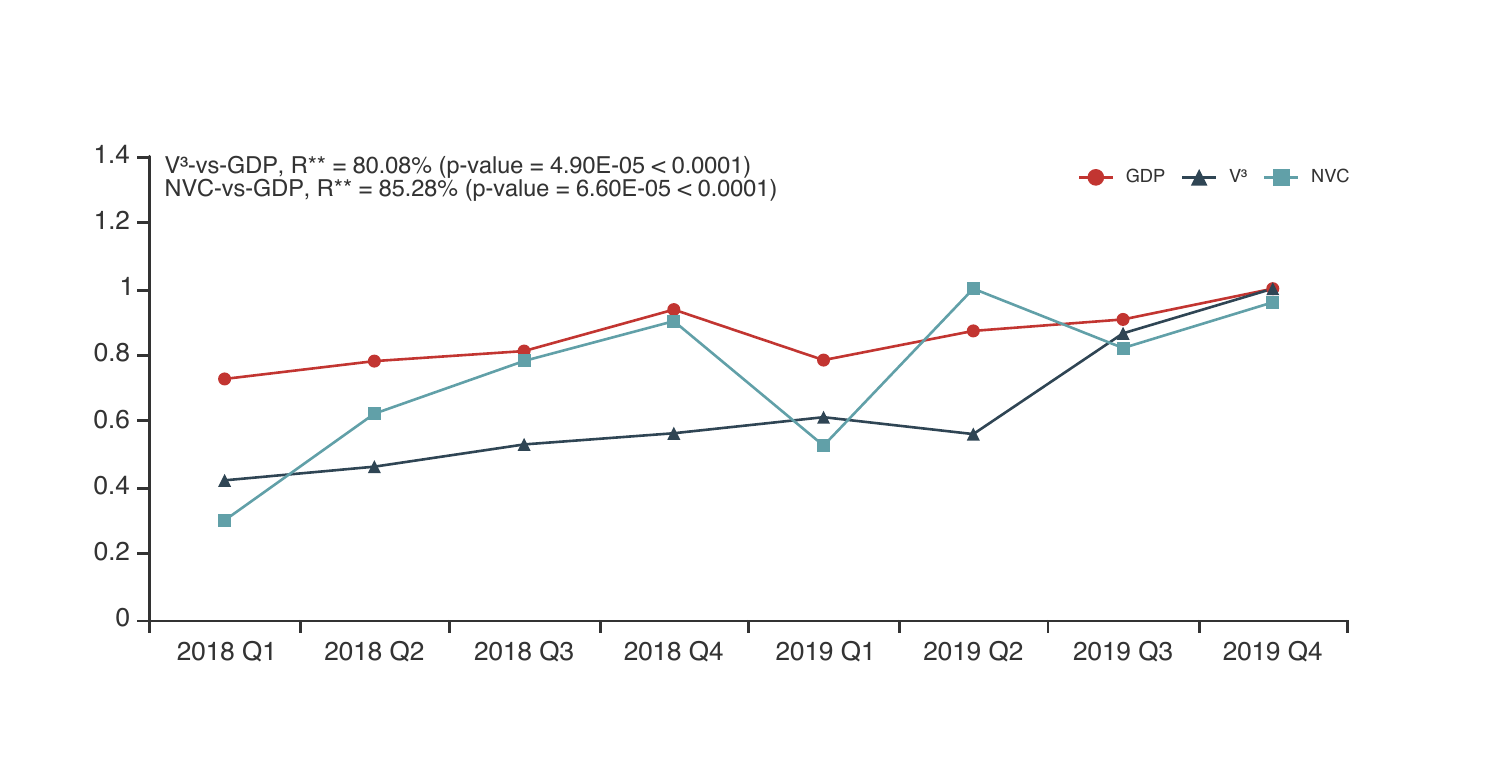} } \
    \caption{Correlations between the GDP and  \nvctail/\vvv of 31 provinces. (2 of 8)}
\end{figure}

\begin{figure}
    \centering
    \ContinuedFloat
    \subfloat[Correlations between the GDP and  \nvctail/\vvv of Jiangsu]{\label{fig:f1-GDP-jiangsu}\includegraphics[width=0.685\textwidth,trim={0.8cm 1.1cm 1.0cm 1.35cm},clip]{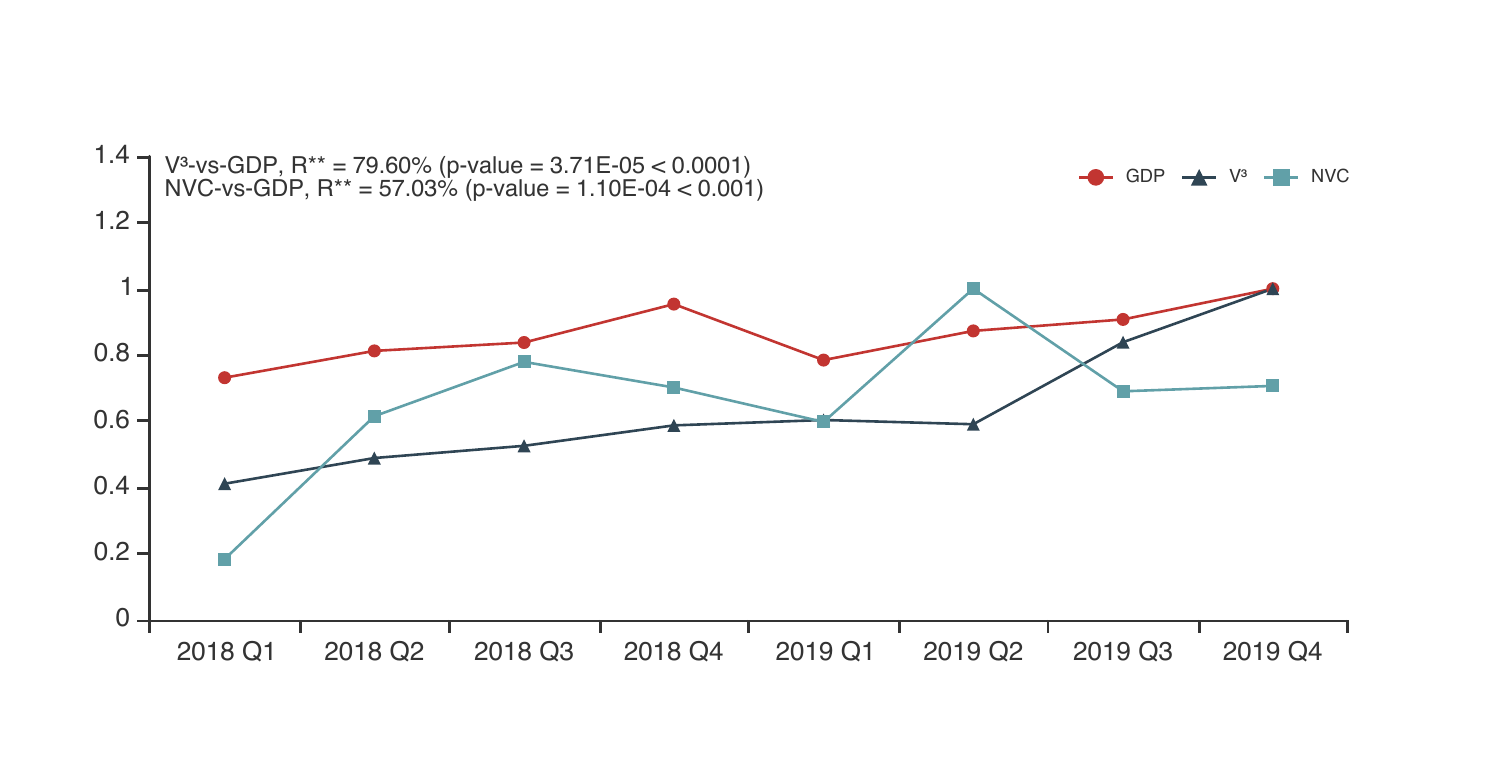} } \
    \subfloat[Correlations between the GDP and  \nvctail/\vvv of Hubei]{\label{fig:f1-GDP-hubei}\includegraphics[width=0.685\textwidth,trim={0.8cm 1.1cm 1.0cm 1.35cm},clip]{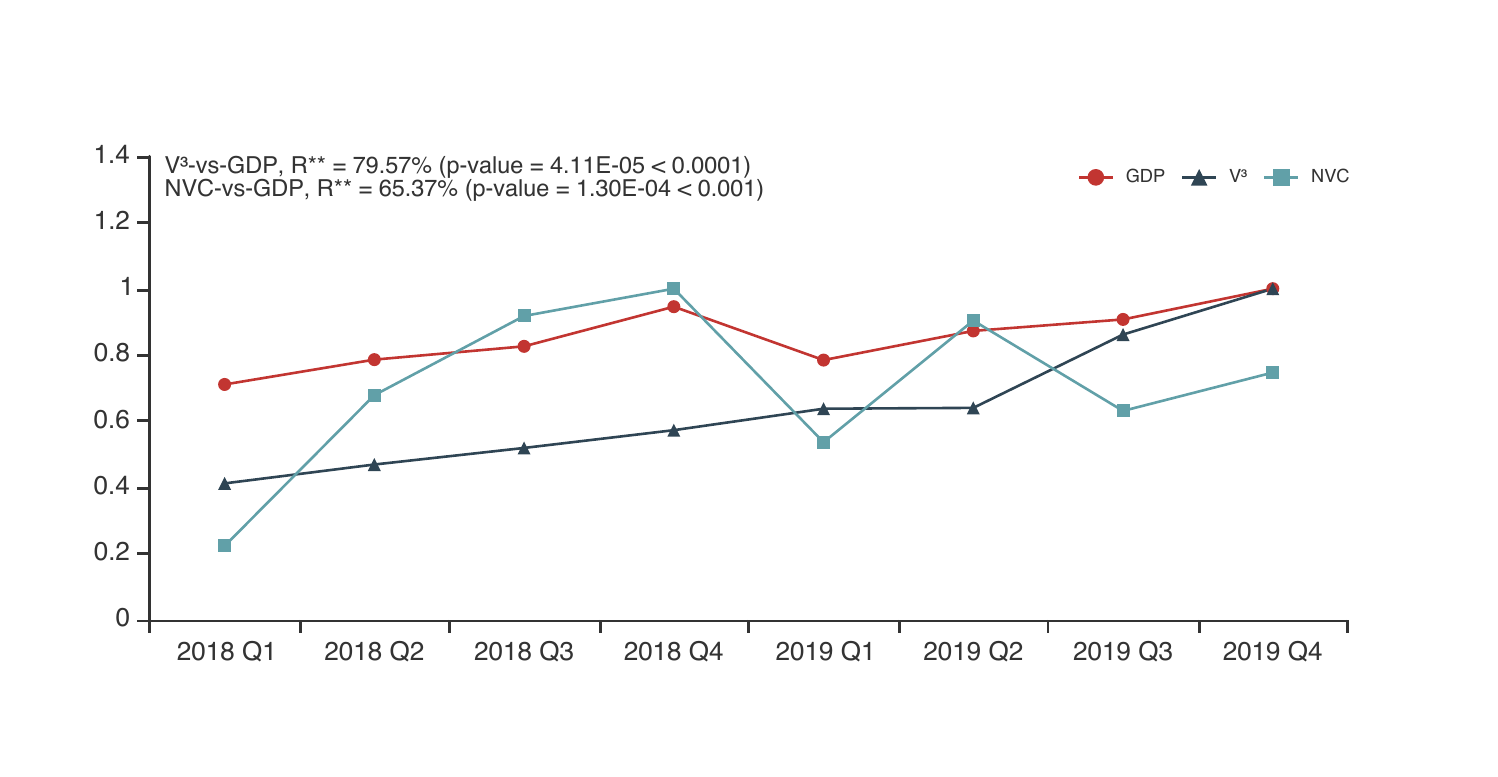} } \
    \subfloat[Correlations between the GDP and  \nvctail/\vvv of Zhejiang]{\label{fig:f1-GDP-zhejiang}\includegraphics[width=0.685\textwidth,trim={0.8cm 1.1cm 1.0cm 1.35cm},clip]{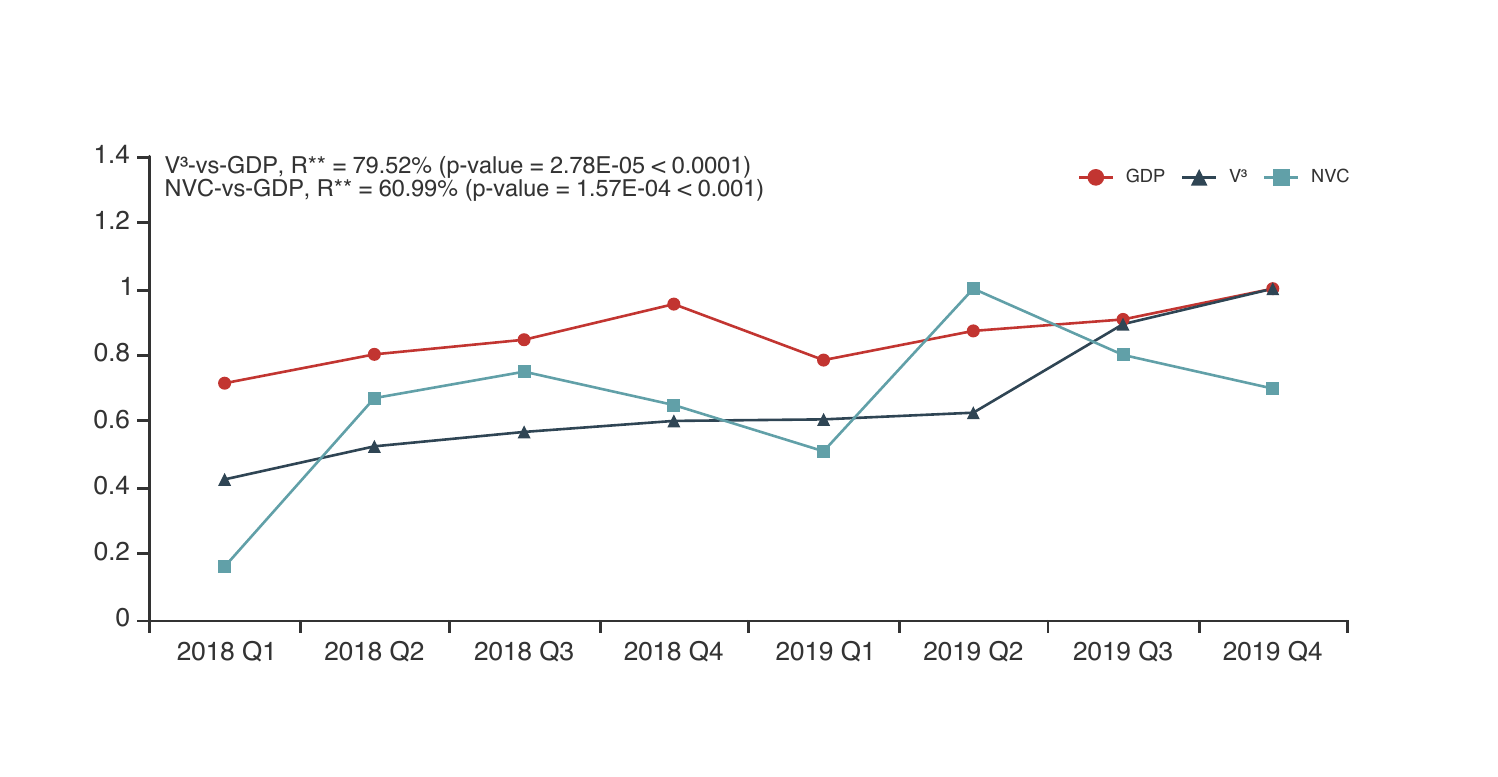} } \
    \subfloat[Correlations between the GDP and  \nvctail/\vvv of Shaanxi]{\label{fig:f1-GDP-shaanxi}\includegraphics[width=0.685\textwidth,trim={0.8cm 1.1cm 1.0cm 1.35cm},clip]{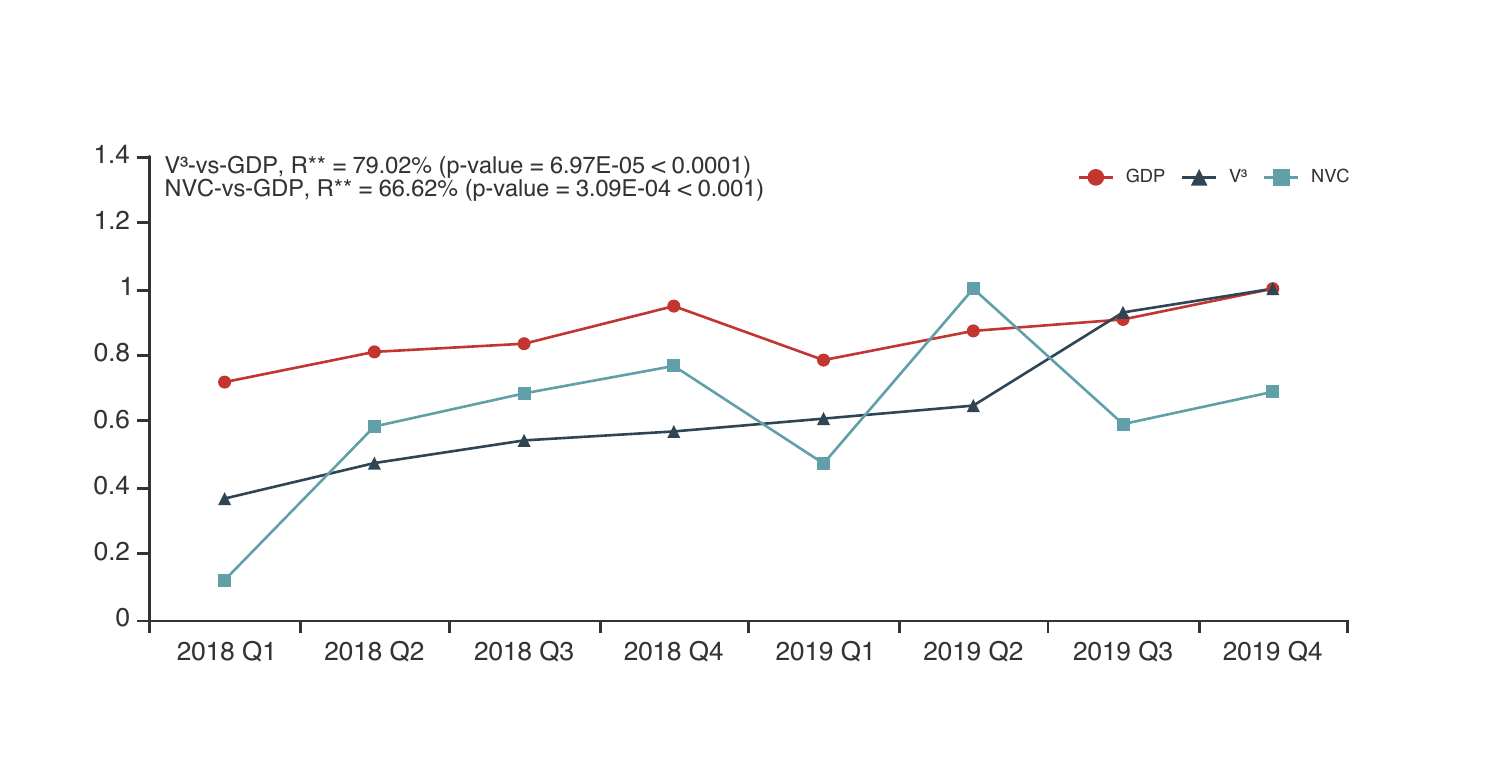} } \
    \caption{Correlations between the GDP and  \nvctail/\vvv of 31 provinces. (3 of 8)}
\end{figure}

\begin{figure}
    \centering
    \ContinuedFloat
    \subfloat[Correlations between the GDP and  \nvctail/\vvv of Henan]{\label{fig:f1-GDP-henan}\includegraphics[width=0.685\textwidth,trim={0.8cm 1.1cm 1.0cm 1.35cm},clip]{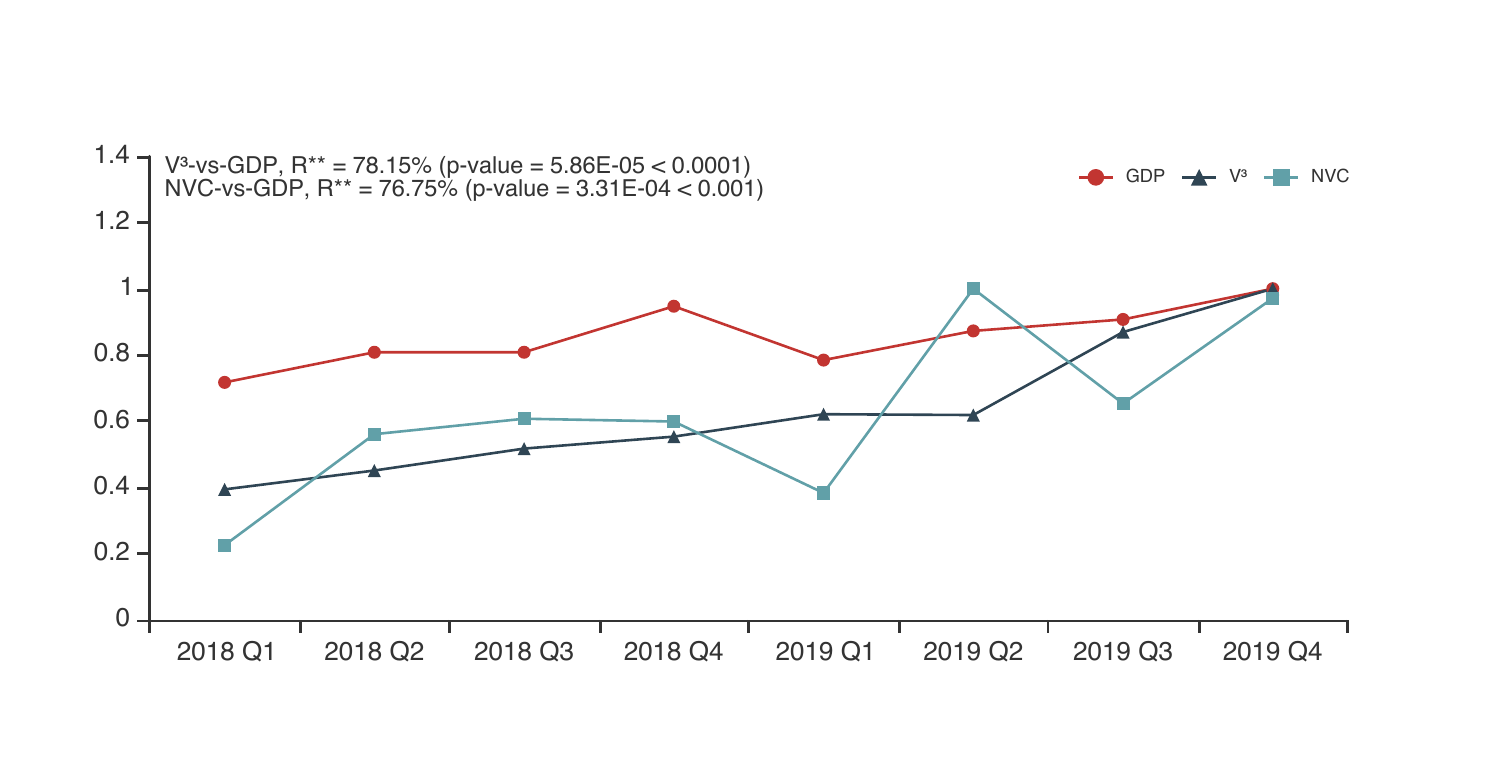} } \
    \subfloat[Correlations between the GDP and  \nvctail/\vvv of Anhui]{\label{fig:f1-GDP-anhui}\includegraphics[width=0.685\textwidth,trim={0.8cm 1.1cm 1.0cm 1.35cm},clip]{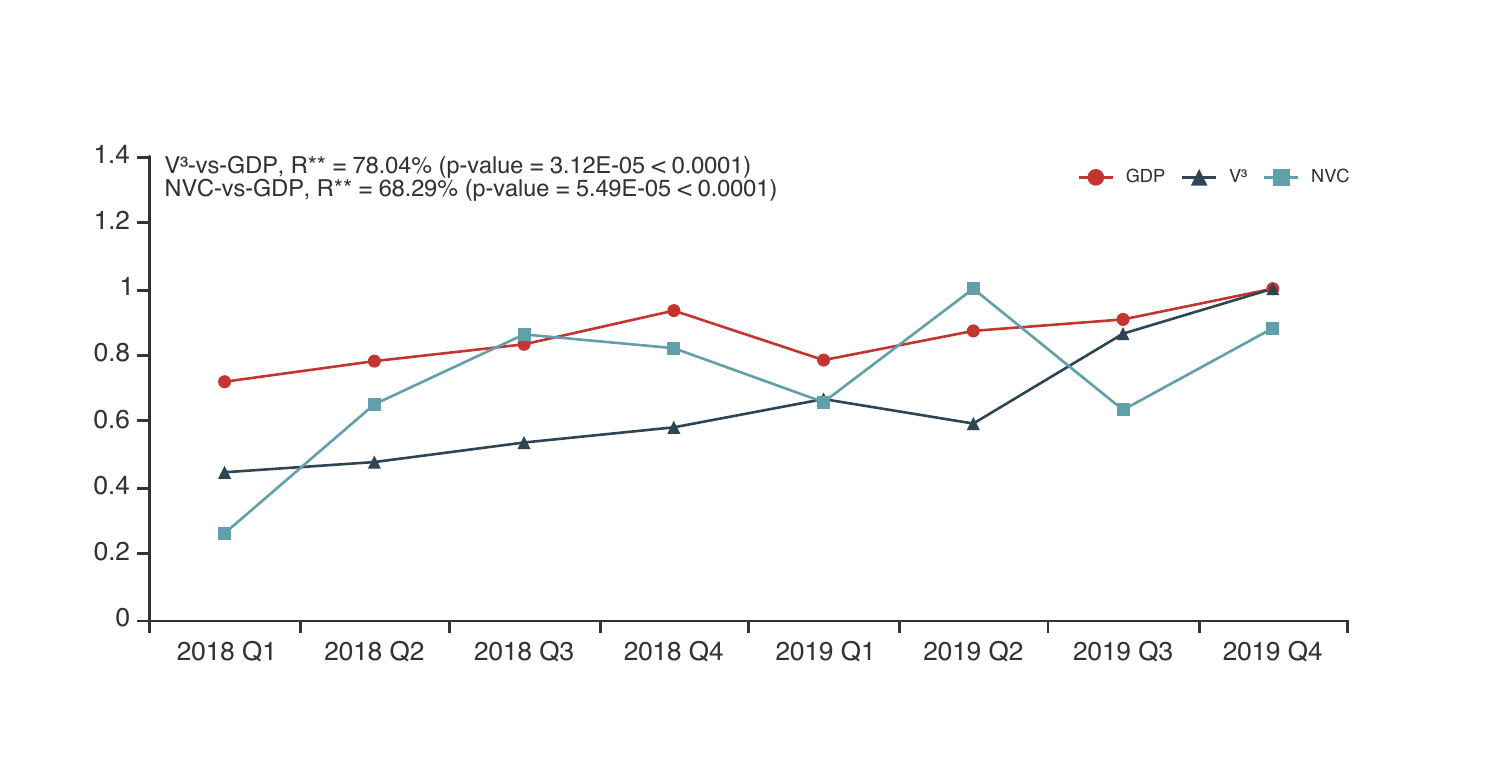} } \
    \subfloat[Correlations between the GDP and  \nvctail/\vvv of Guangdong]{\label{fig:f1-GDP-guangdong}\includegraphics[width=0.685\textwidth,trim={0.8cm 1.1cm 1.0cm 1.35cm},clip]{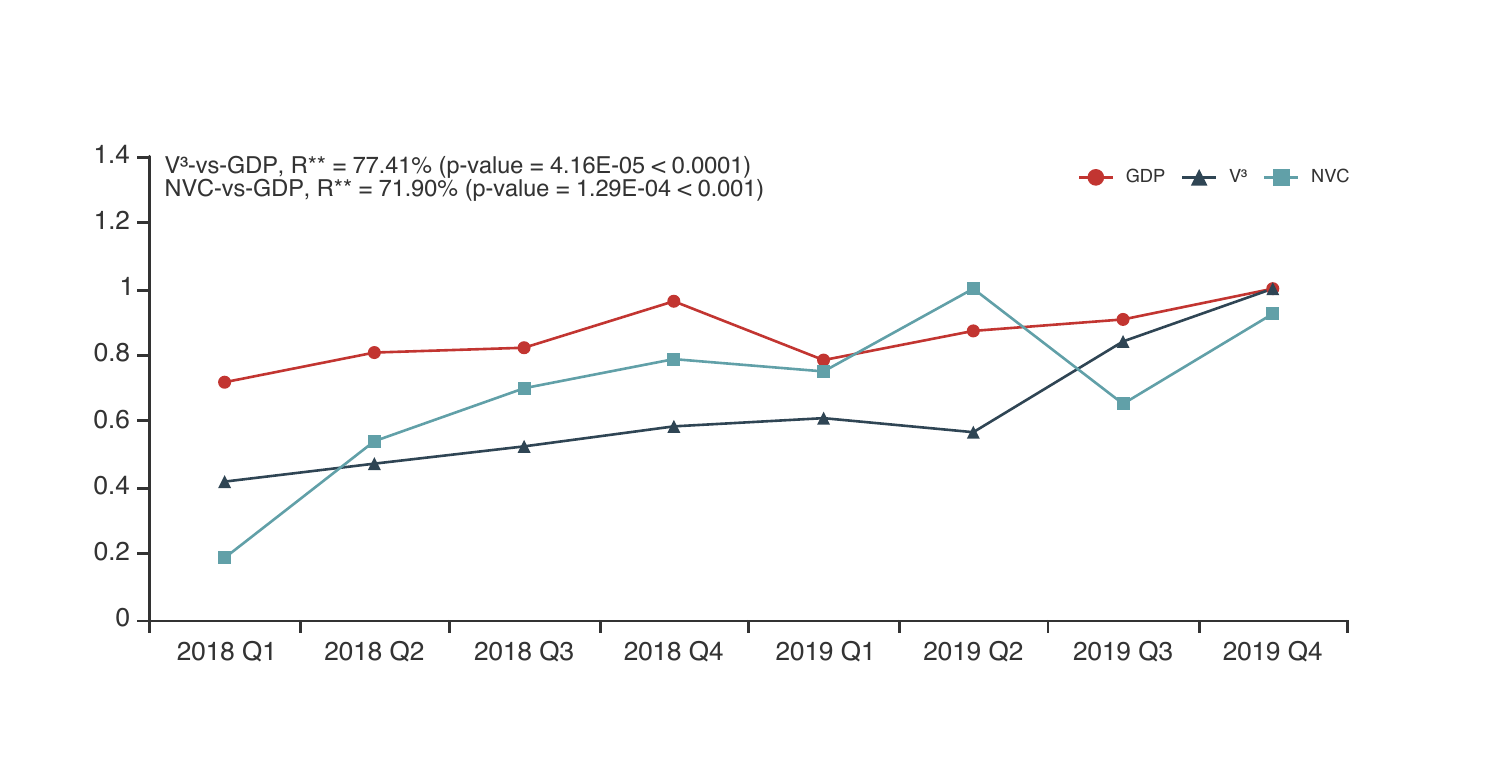} } \
    \subfloat[Correlations between the GDP and  \nvctail/\vvv of Chongqing]{\label{fig:f1-GDP-chongqing}\includegraphics[width=0.685\textwidth,trim={0.8cm 1.1cm 1.0cm 1.35cm},clip]{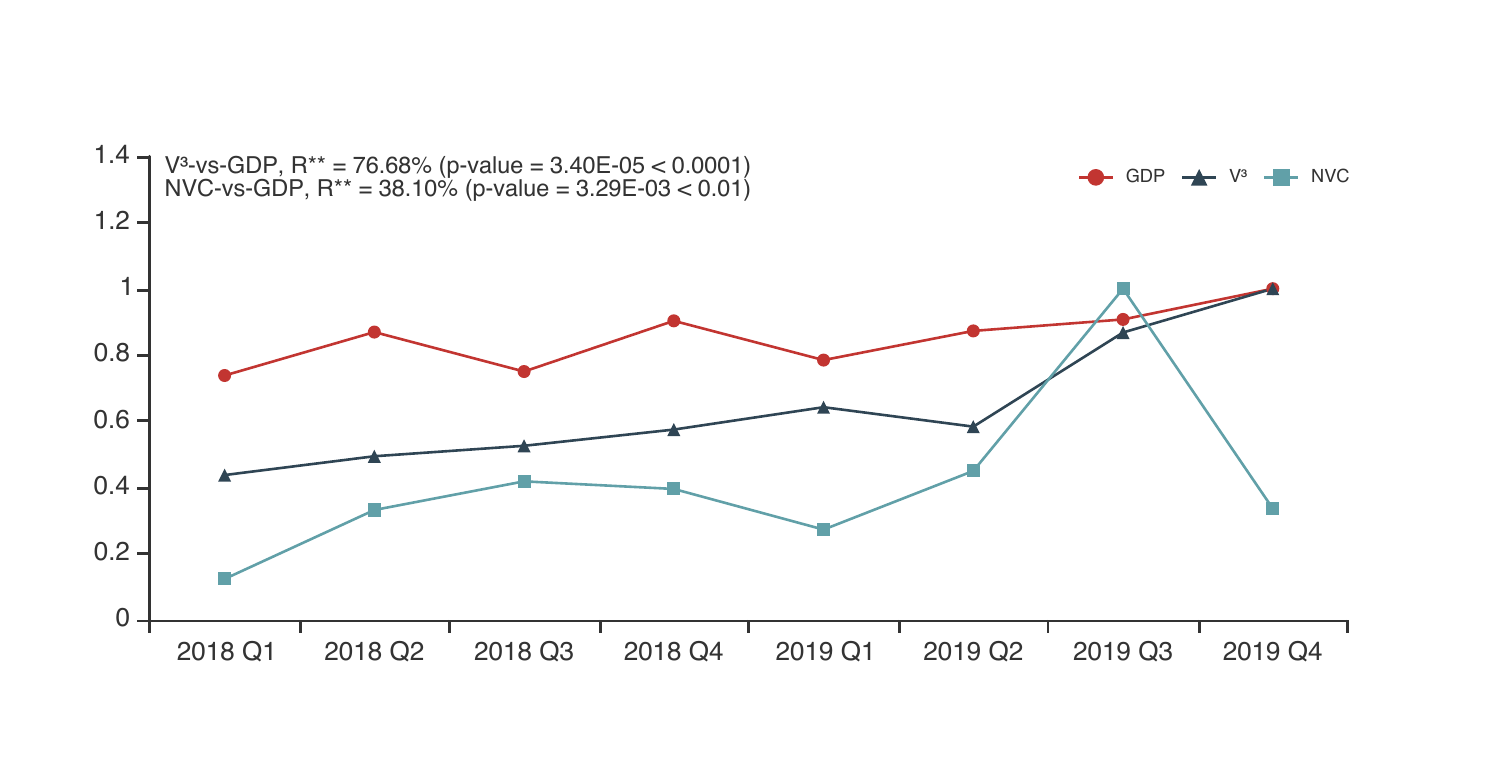} } \
    \caption{Correlations between the GDP and  \nvctail/\vvv of 31 provinces. (4 of 8)}
\end{figure}

\begin{figure}
    \centering    
    \ContinuedFloat
    \subfloat[Correlations between the GDP and  \nvctail/\vvv of Jiangxi]{\label{fig:f1-GDP-jiangxi}\includegraphics[width=0.685\textwidth,trim={0.8cm 1.1cm 1.0cm 1.35cm},clip]{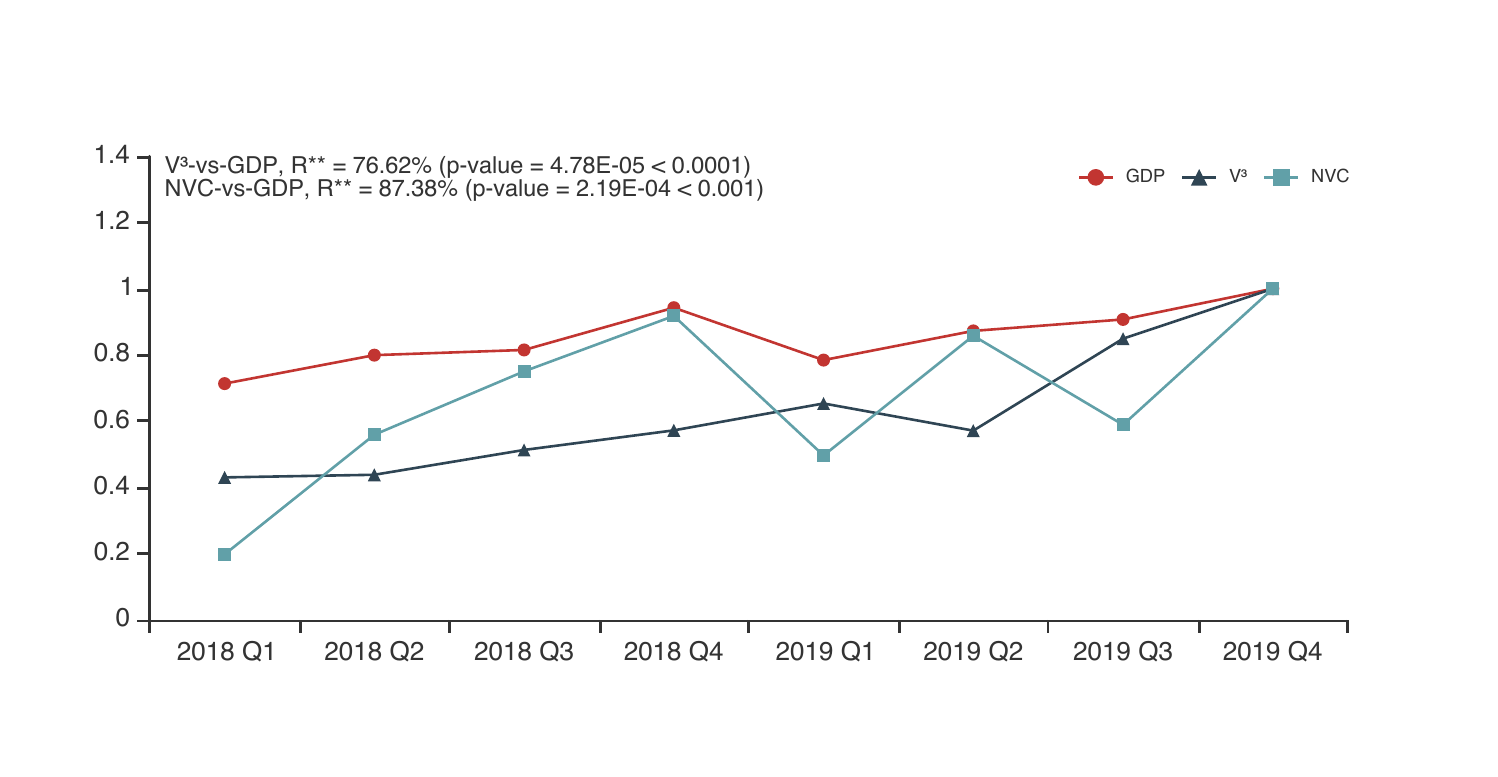} } \
    \subfloat[Correlations between the GDP and  \nvctail/\vvv of Liaoning]{\label{fig:f1-GDP-liaoning}\includegraphics[width=0.685\textwidth,trim={0.8cm 1.1cm 1.0cm 1.35cm},clip]{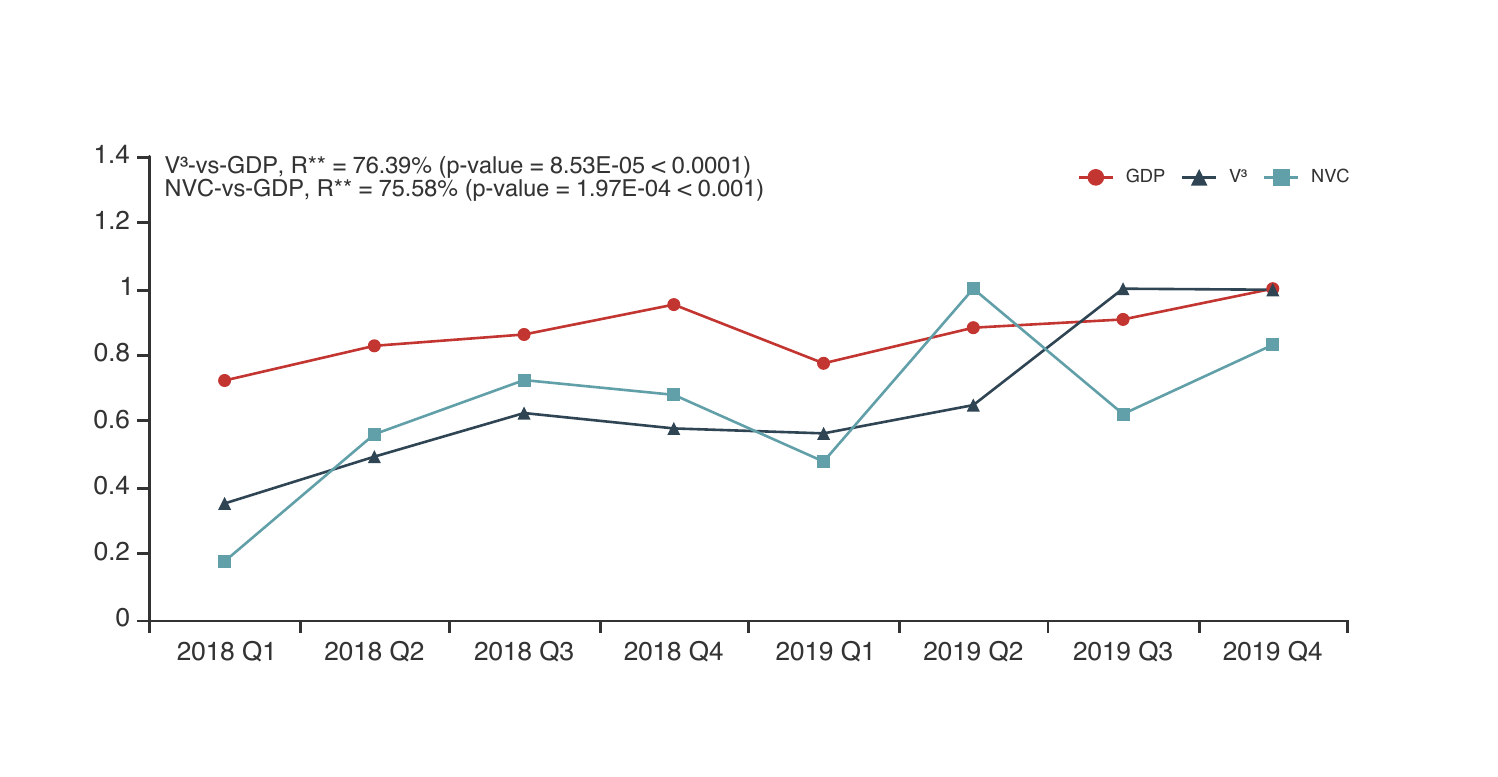} } \
    \subfloat[Correlations between the GDP and  \nvctail/\vvv of Shanghai]{\label{fig:f1-GDP-shanghai}\includegraphics[width=0.685\textwidth,trim={0.8cm 1.1cm 1.0cm 1.35cm},clip]{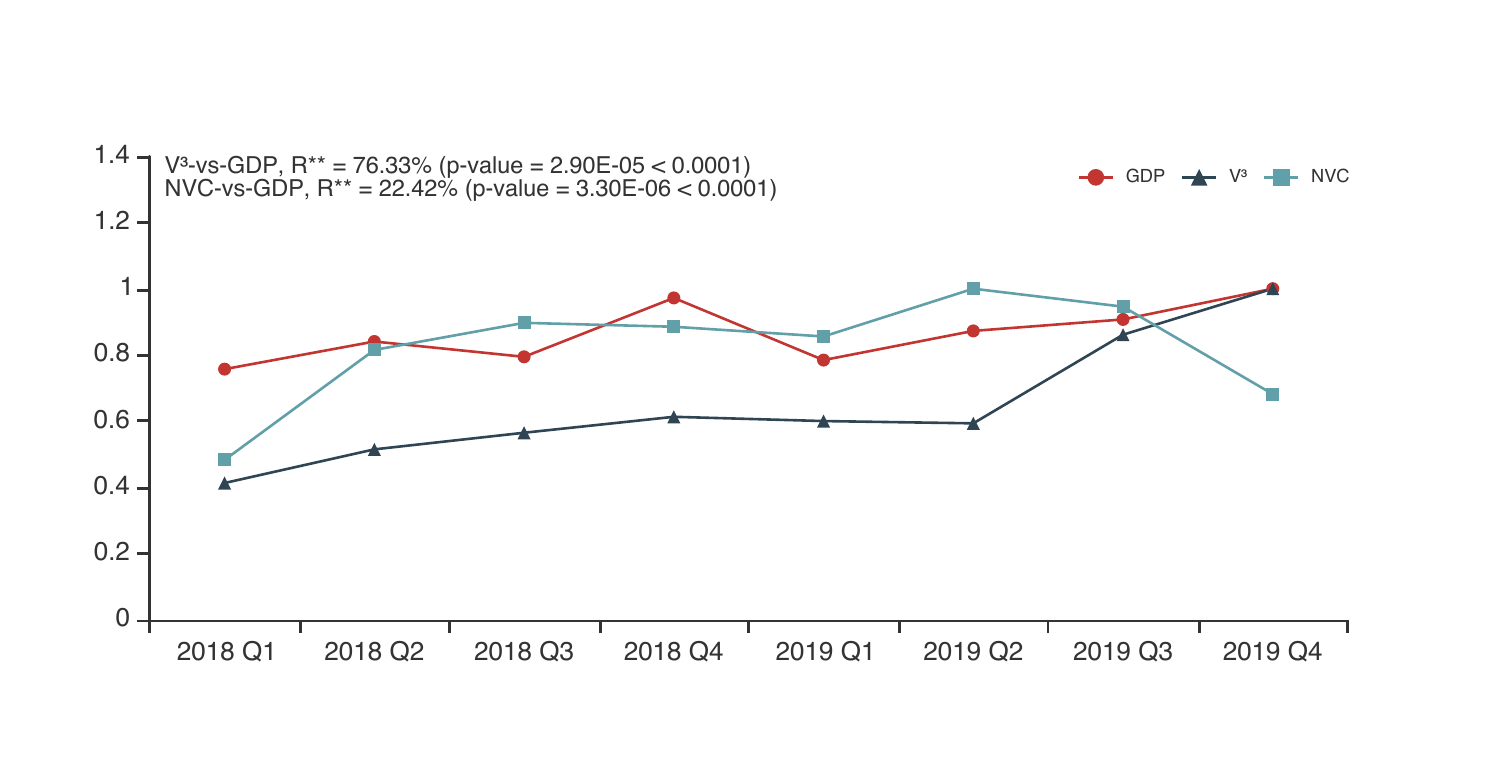} } \
    \subfloat[Correlations between the GDP and  \nvctail/\vvv of Sichuan]{\label{fig:f1-GDP-sichuan}\includegraphics[width=0.685\textwidth,trim={0.8cm 1.1cm 1.0cm 1.35cm},clip]{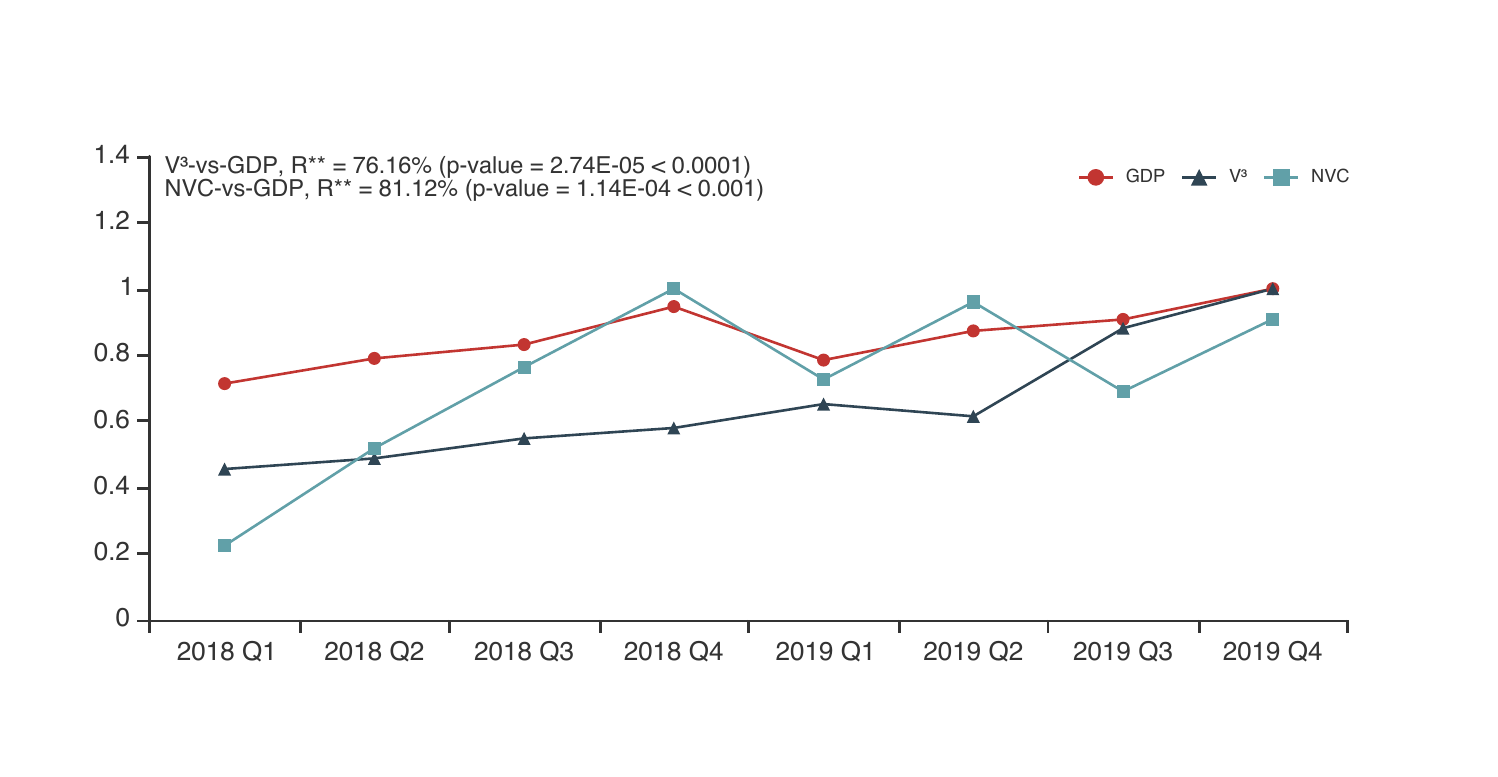} } \
    \caption{Correlations between the GDP and  \nvctail/\vvv of 31 provinces. (5 of 8)}
\end{figure}

\begin{figure}
    \centering
    \ContinuedFloat
    \subfloat[Correlations between the GDP and  \nvctail/\vvv of Gansu]{\label{fig:f1-GDP-gansu}\includegraphics[width=0.685\textwidth,trim={0.8cm 1.1cm 1.0cm 1.35cm},clip]{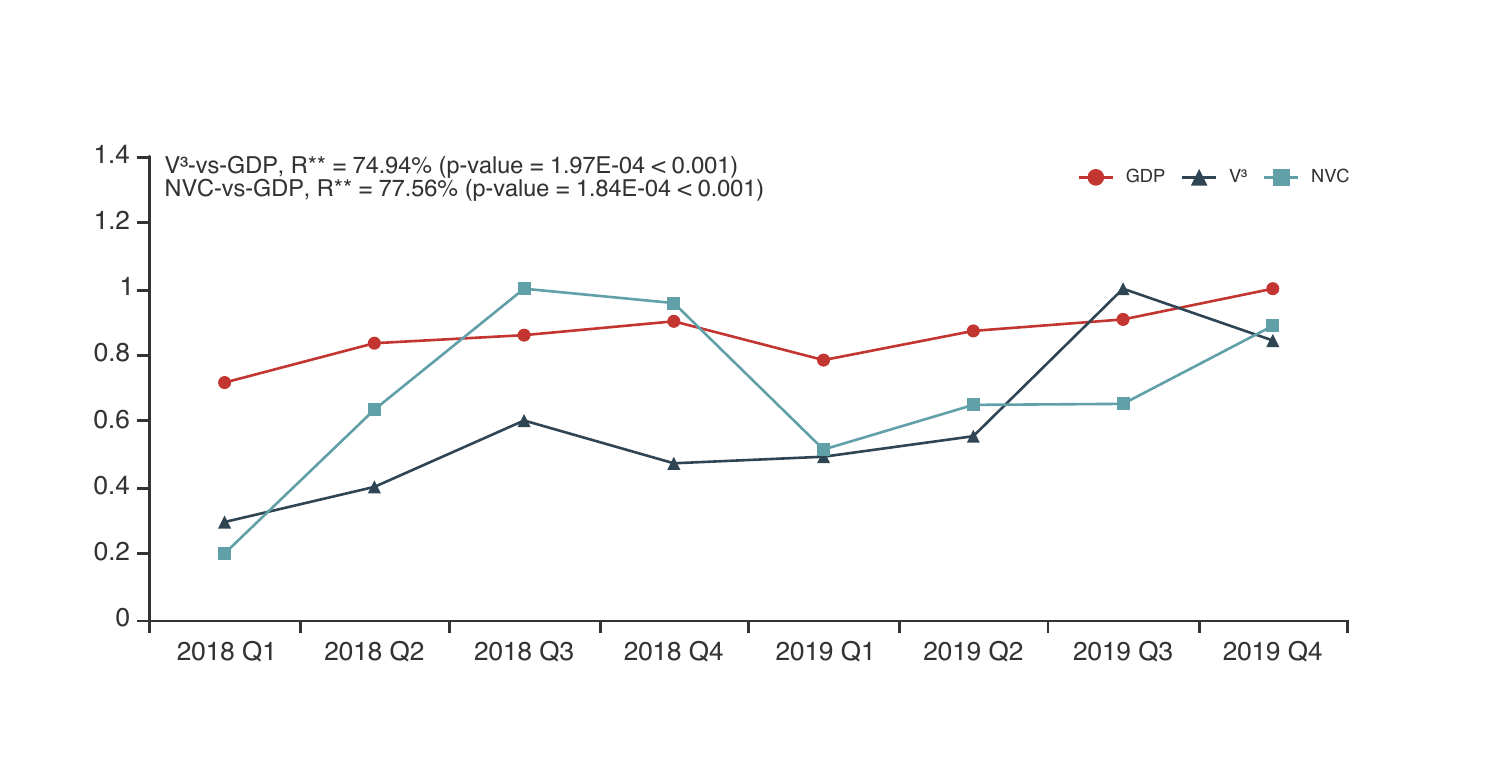} } \
    \subfloat[Correlations between the GDP and  \nvctail/\vvv of Ningxia]{\label{fig:f1-GDP-ningxia}\includegraphics[width=0.685\textwidth,trim={0.8cm 1.1cm 1.0cm 1.35cm},clip]{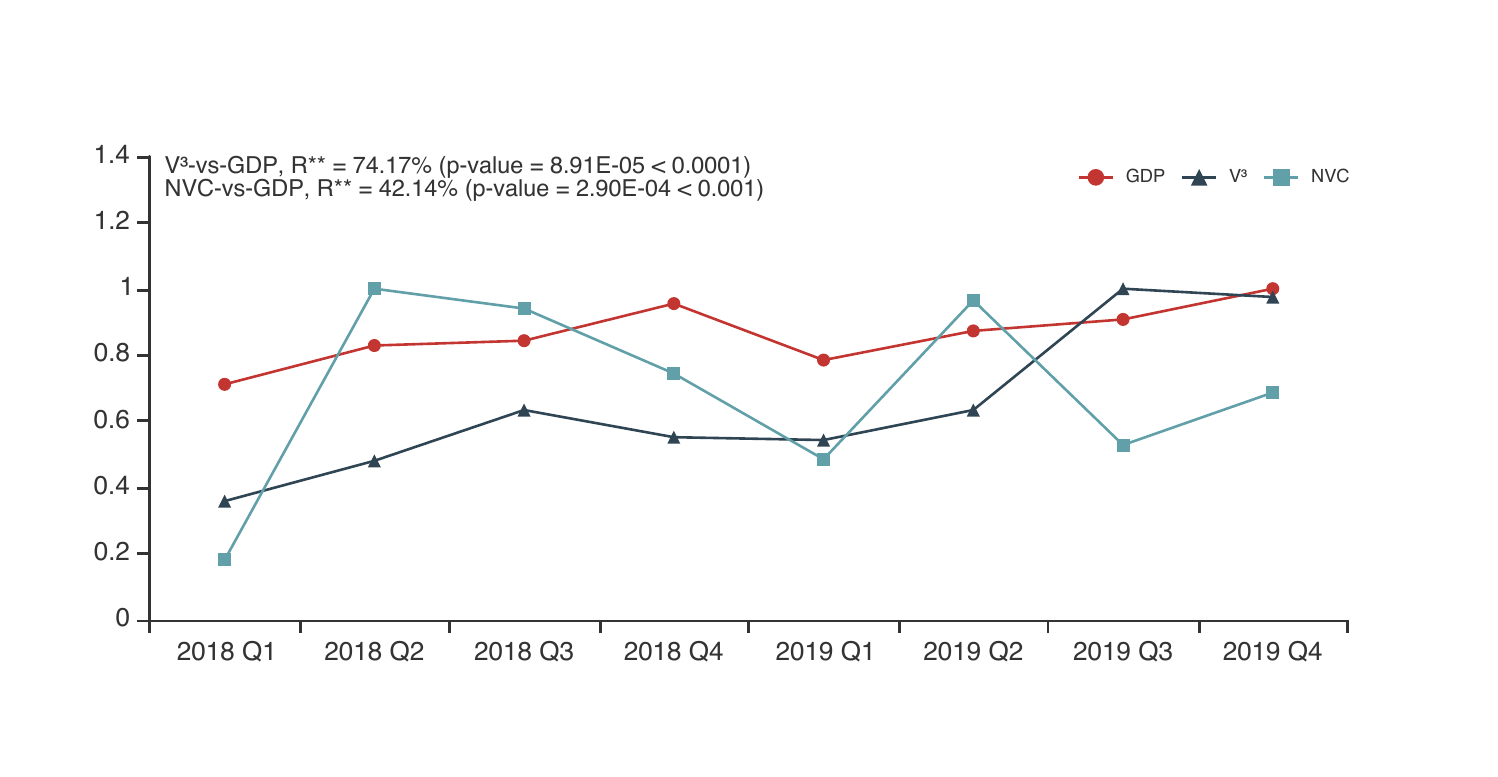} } \
    \subfloat[Correlations between the GDP and  \nvctail/\vvv of Inner Mongolia]{\label{fig:f1-GDP-inner-mongolia}\includegraphics[width=0.685\textwidth,trim={0.8cm 1.1cm 1.0cm 1.35cm},clip]{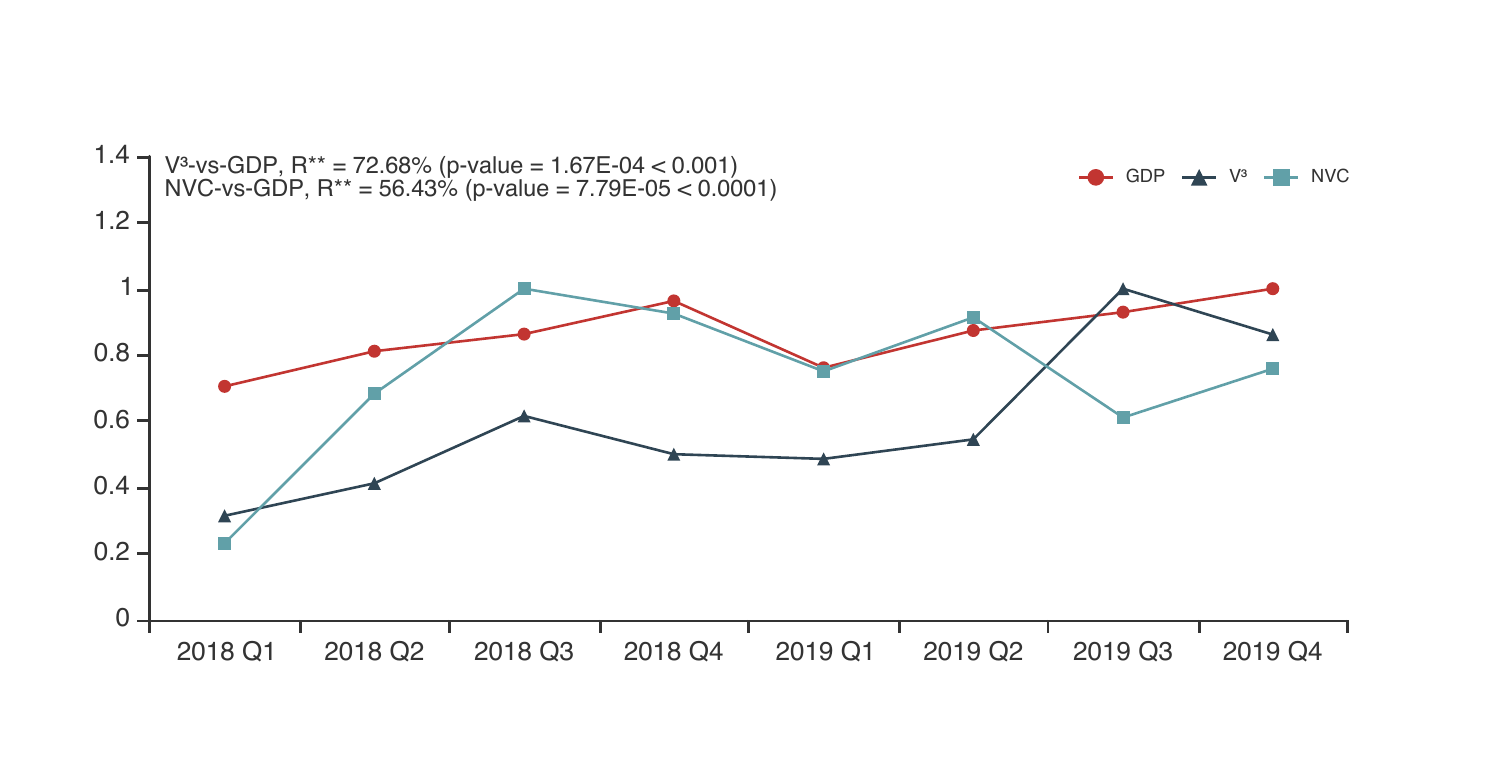}} \
    \subfloat[Correlations between the GDP and  \nvctail/\vvv of Shanxi]{\label{fig:f1-GDP-shanxi}\includegraphics[width=0.685\textwidth,trim={0.8cm 1.1cm 1.0cm 1.35cm},clip]{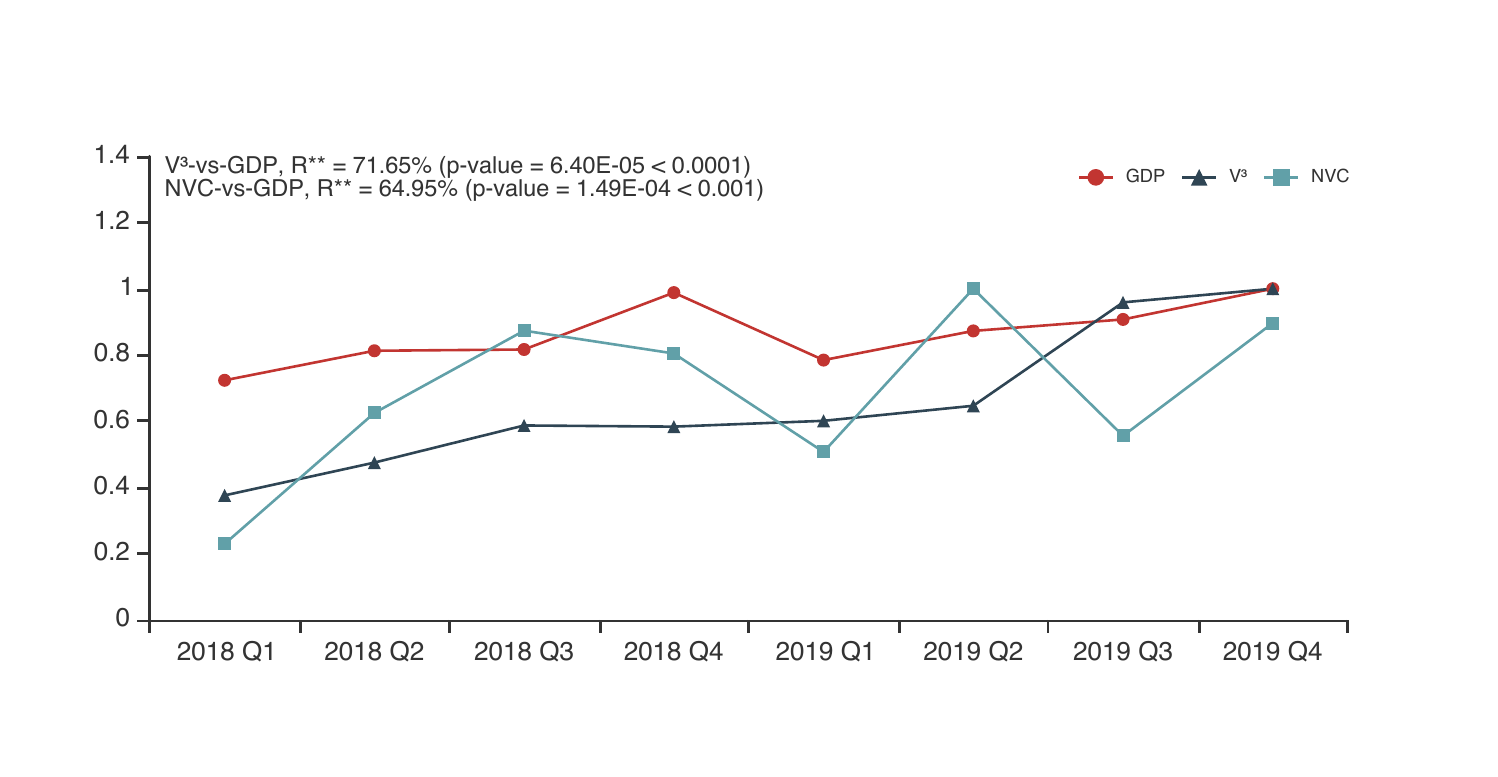} } \
    \caption{Correlations between the GDP and  \nvctail/\vvv of 31 provinces. (6 of 8)}
\end{figure}

\begin{figure}
    \centering
    \ContinuedFloat
    \subfloat[Correlations between the GDP and  \nvctail/\vvv of Guizhou]{\label{fig:f1-GDP-guizhou}\includegraphics[width=0.685\textwidth,trim={0.8cm 1.1cm 1.0cm 1.35cm},clip]{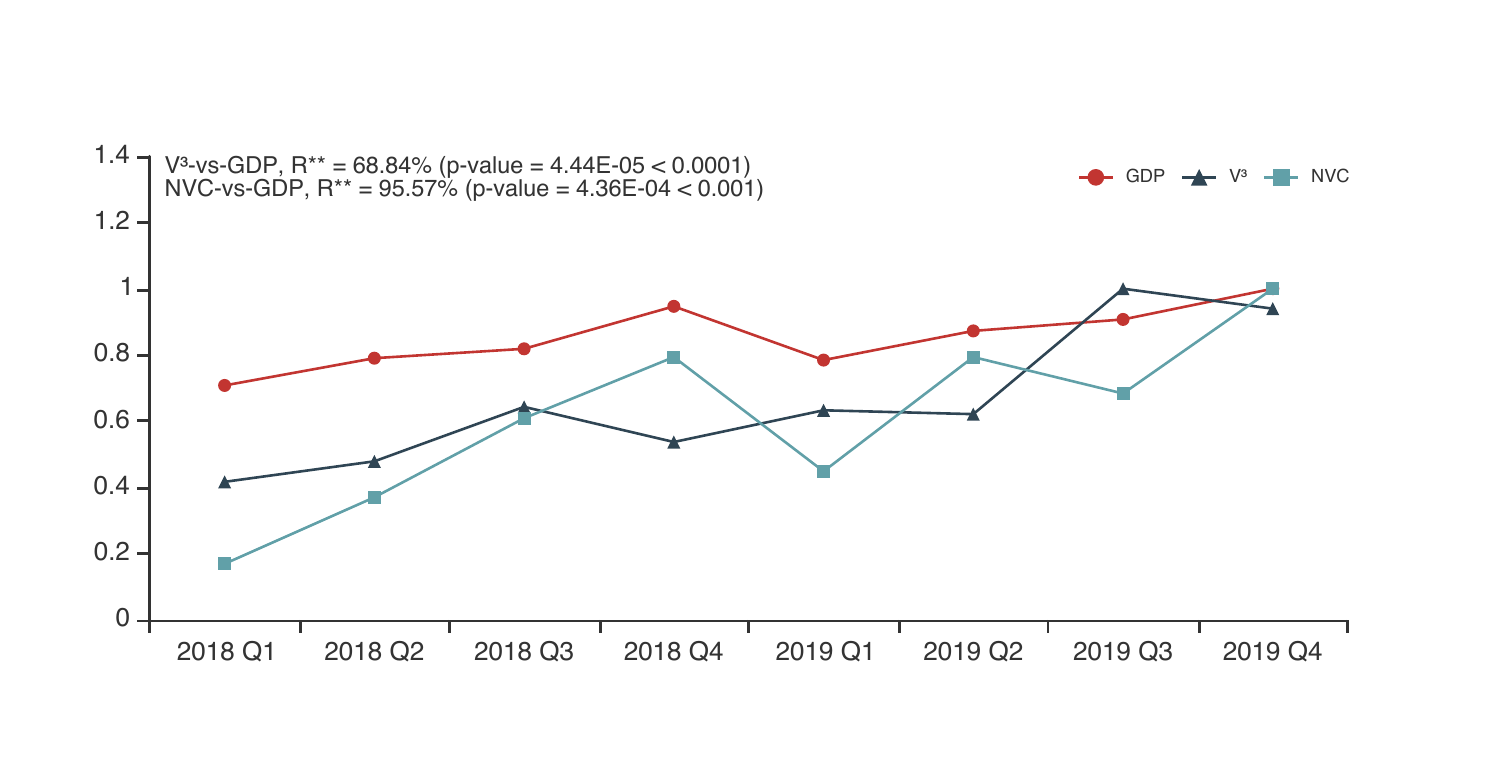} } \
    \subfloat[Correlations between the GDP and  \nvctail/\vvv of Xinjiang]{\label{fig:f1-GDP-xinjiang}\includegraphics[width=0.685\textwidth,trim={0.8cm 1.1cm 1.0cm 1.35cm},clip]{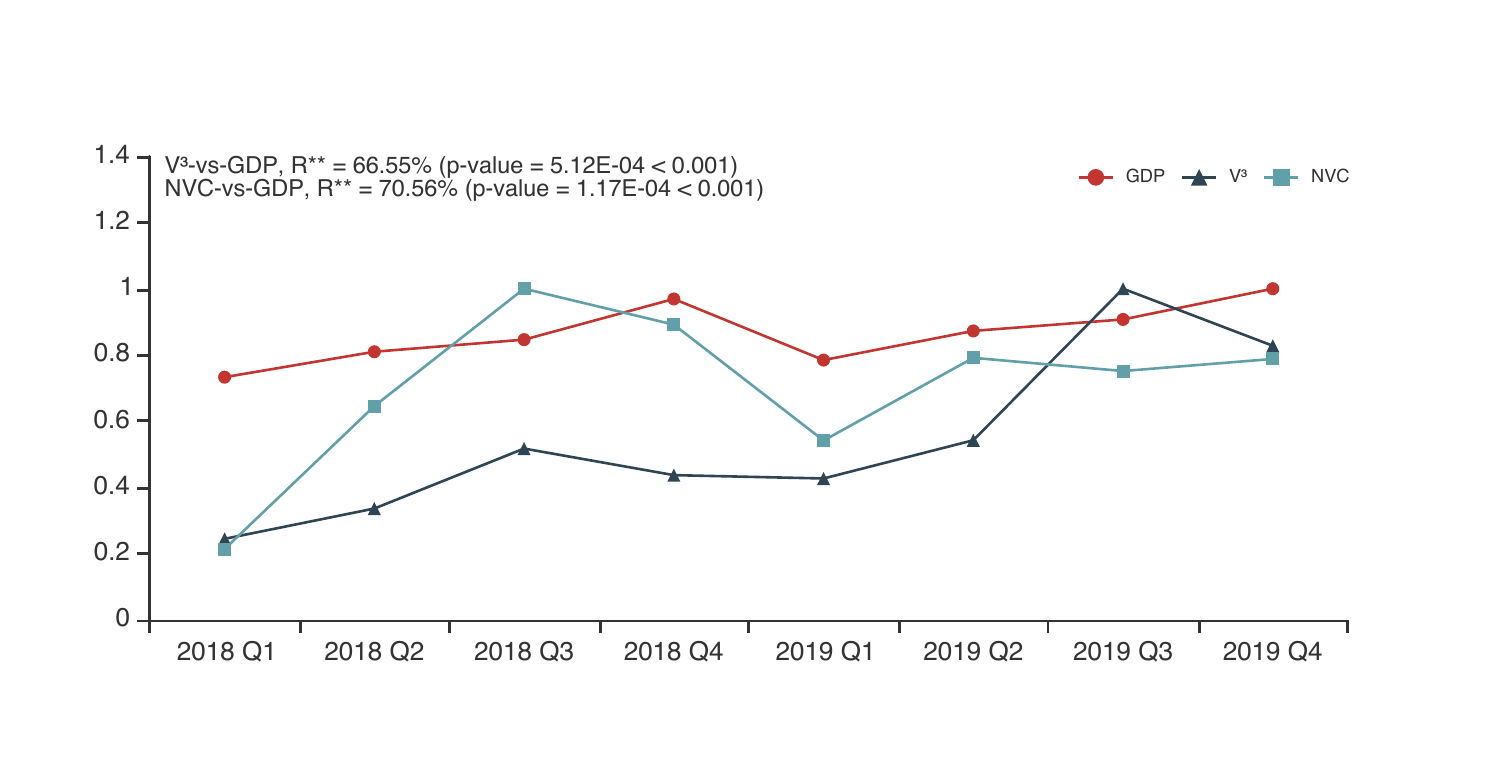} } \
    \subfloat[Correlations between the GDP and  \nvctail/\vvv of Guangxi]{\label{fig:f1-GDP-guangxi}\includegraphics[width=0.685\textwidth,trim={0.8cm 1.1cm 1.0cm 1.35cm},clip]{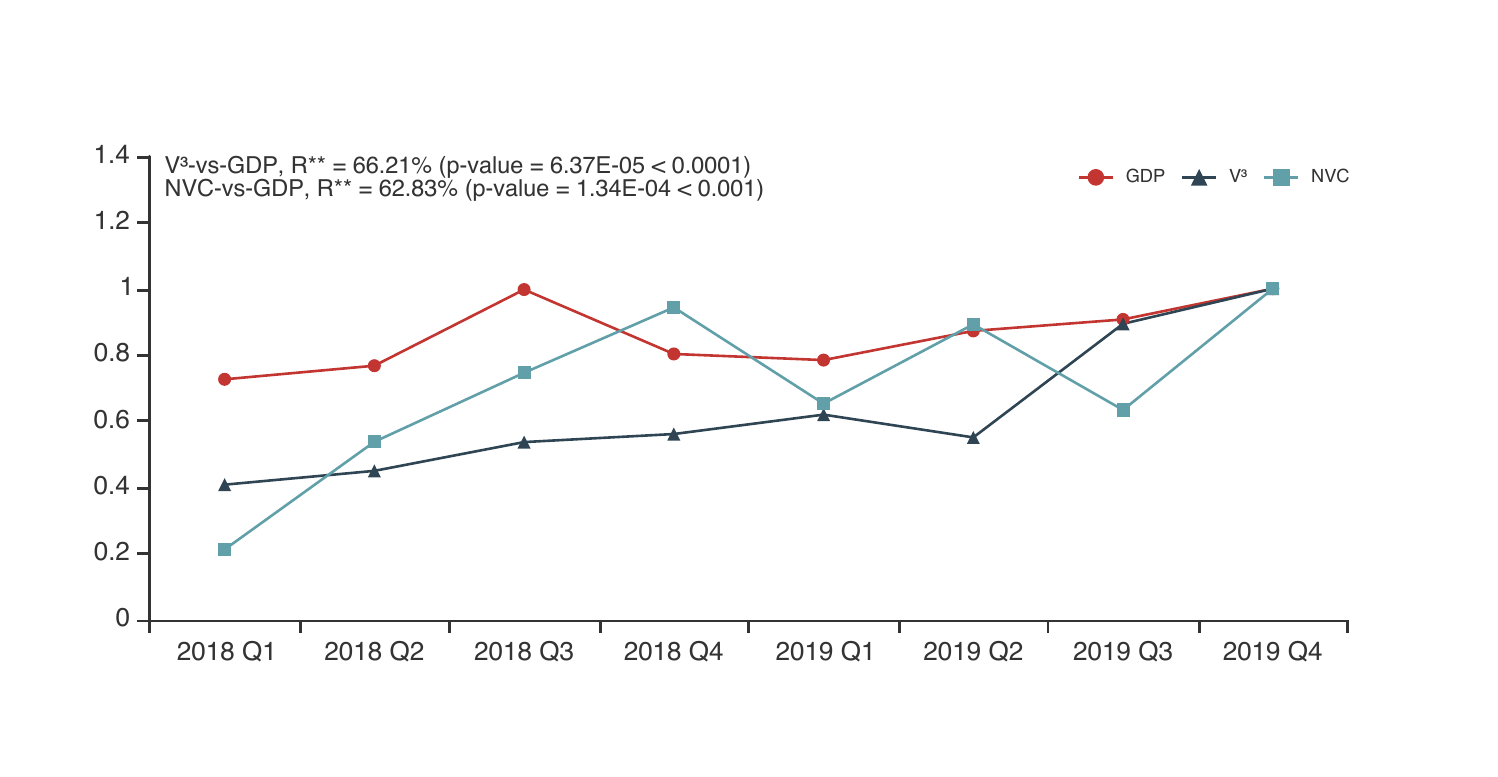} } \
    \subfloat[Correlations between the GDP and  \nvctail/\vvv of Xizang]{\label{fig:f1-GDP-xizang}\includegraphics[width=0.685\textwidth,trim={0.8cm 1.1cm 1.0cm 1.35cm},clip]{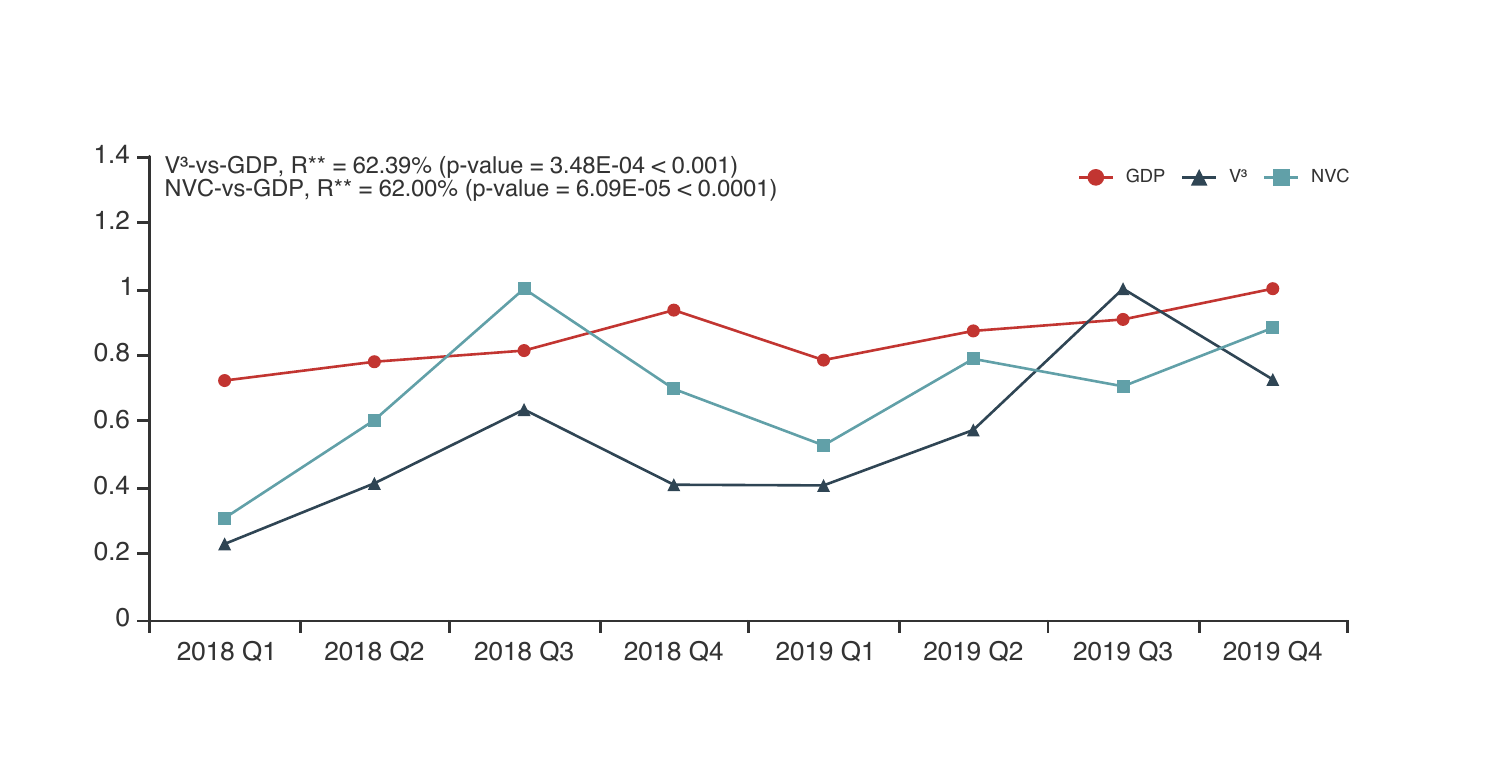} } \
    \caption{Correlations between the GDP and  \nvctail/\vvv of 31 provinces. (7 of 8)}
\end{figure}

\begin{figure}
    \centering
    \ContinuedFloat
    \subfloat[Correlations between the GDP and  \nvctail/\vvv of Hainan]{\label{fig:f1-GDP-hainan}\includegraphics[width=0.685\textwidth,trim={0.8cm 1.1cm 1.0cm 1.35cm},clip]{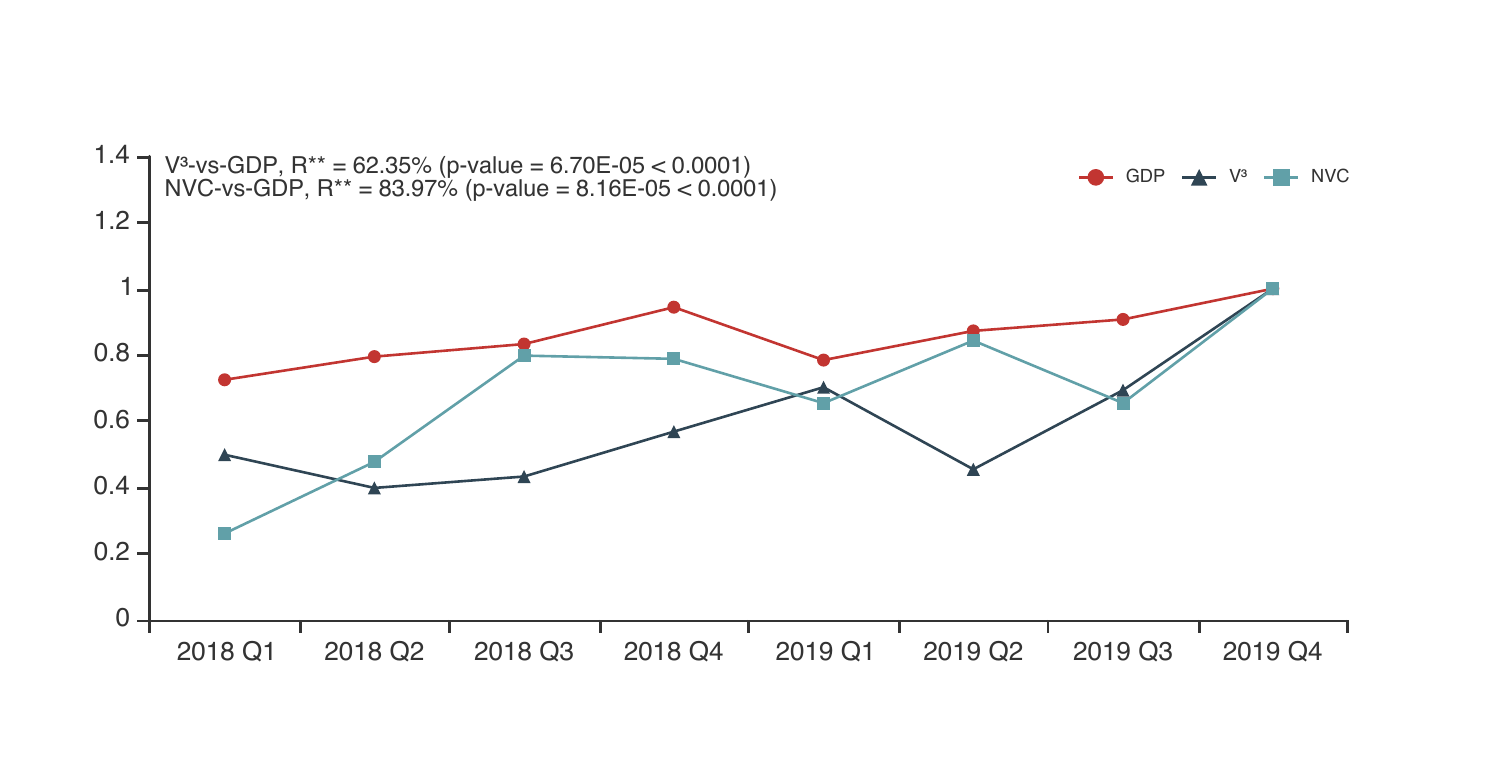} } \
    \subfloat[Correlations between the GDP and  \nvctail/\vvv of Qinghai]{\label{fig:f1-GDP-qinghai}\includegraphics[width=0.685\textwidth,trim={0.8cm 1.1cm 1.0cm 1.35cm},clip]{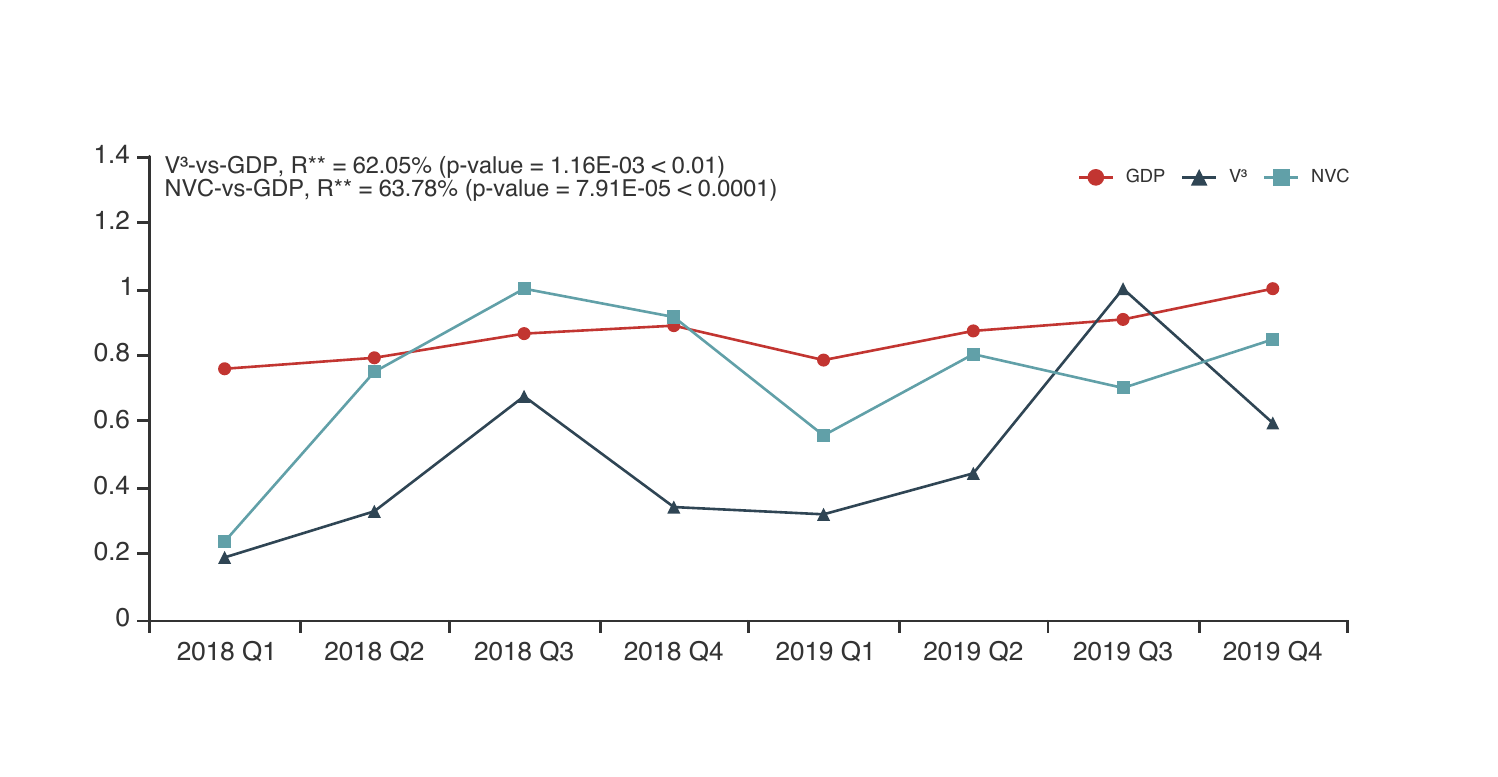} } \
    \subfloat[Correlations between the GDP and  \nvctail/\vvv of Heilongjiang]{\label{fig:f1-GDP-heilongjiang}\includegraphics[width=0.685\textwidth,trim={0.8cm 1.1cm 1.0cm 1.35cm},clip]{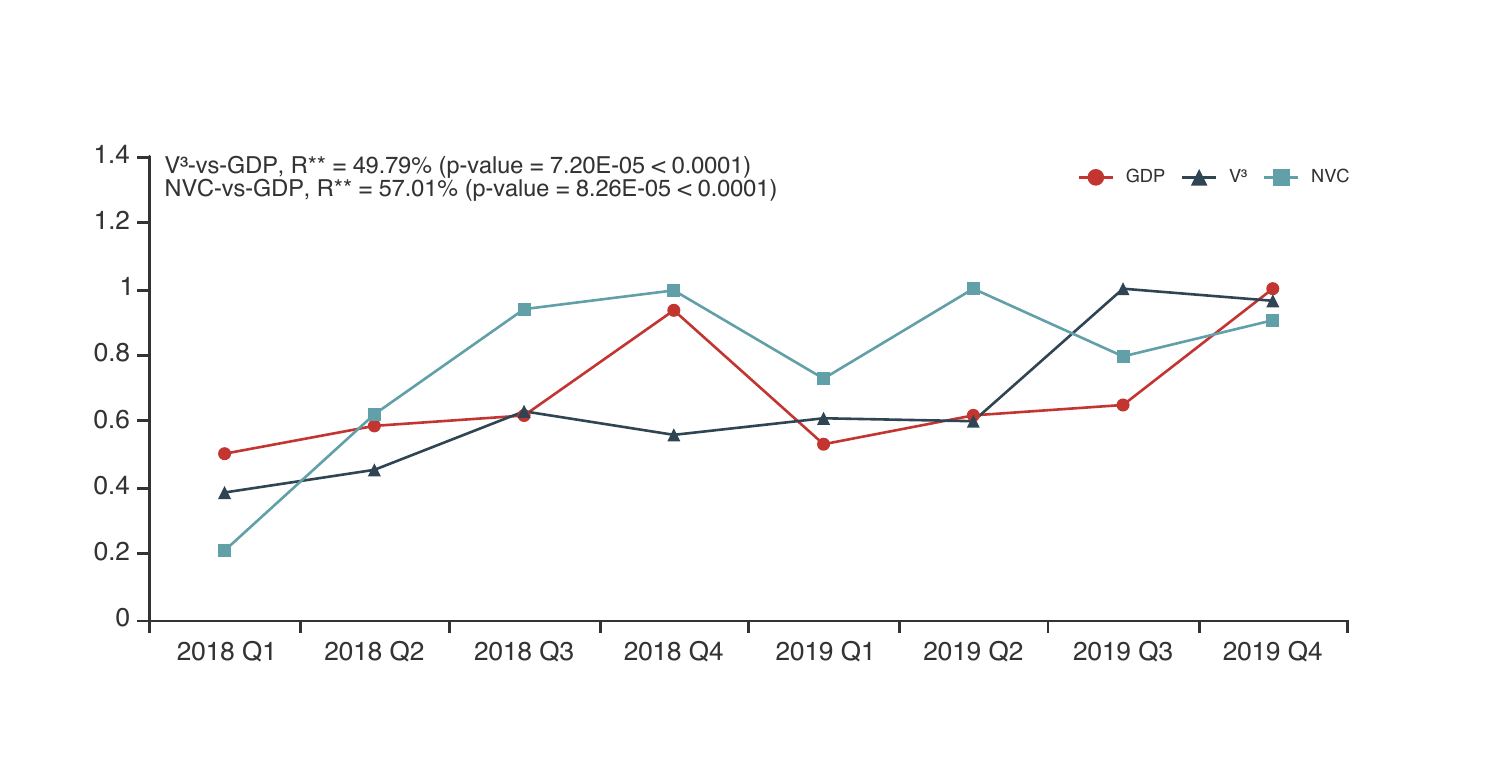} } \
    \caption{Correlations between the GDP and  \nvctail/\vvv of 31 provinces. (8 of 8)}
\end{figure}

\subsection*{Trends of different provinces with \vvv and \nvc}
Figure~\ref{fig:v3-nvc-31-provinces} illustrates the trends and weekly records of the two indicators, i.e., \vvv and \nvctail, for 31 provinces excluding Hong Kong, Macao, and Taiwan of China, from the 1$^{st}$ week of 2018 to the 13$^{th}$ week of 2020. The sub-figures in Figure~\ref{fig:v3-nvc-31-provinces} are ranked by the difference between the value of \vvv in the 13$^{th}$ week of the year 2019 and the year 2020 in ascending order. From the figure, we can observe the three shapes (i.e., a clear, L-shaped recession; a vibrant, Check-mark-shaped ({\large\rotatebox[origin=c]{-120}{\sffamily 7}}) recovery; and emerging, V-shaped bounces) in terms of the \vvv of different provinces.

\begin{figure}
    \centering
    \subfloat[The \vvv in Beijing]{\label{fig:a-beijing-v3}\includegraphics[width=0.49\textwidth,trim={0.48cm 1.08cm 0.58cm 1.08cm},clip]{fig/f3-pro-v3-beijing}} \
    \subfloat[The \nvc for Beijing]{\label{fig:a-beijing-nvc}\includegraphics[width=0.49\textwidth,trim={0.48cm 1.08cm 0.58cm 1.08cm},clip]{fig/f3-pro-nvc-beijing}} \
    \subfloat[The \vvv in Hubei]{\label{fig:a-hubei-v3}\includegraphics[width=0.49\textwidth,trim={0.48cm 1.08cm 0.58cm 1.08cm},clip]{fig/f3-pro-v3-hubei}} \
    \subfloat[The \nvc for Hubei]{\label{fig:a-hubei-nvc}\includegraphics[width=0.49\textwidth,trim={0.48cm 1.08cm 0.58cm 1.08cm},clip]{fig/f3-pro-nvc-hubei}} \
    \subfloat[The \vvv in Shanghai]{\label{fig:a-shanghai-v3}\includegraphics[width=0.49\textwidth,trim={0.48cm 1.08cm 0.58cm 1.08cm},clip]{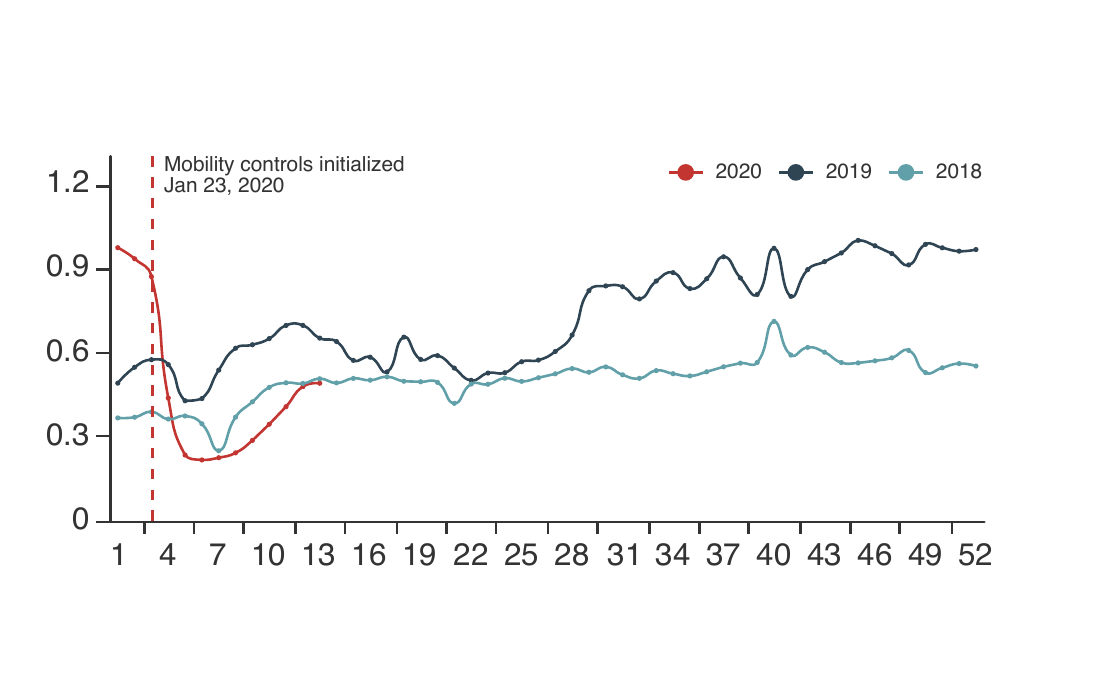}} \
    \subfloat[The \nvc for Shanghai]{\label{fig:a-shanghai-nvc}\includegraphics[width=0.49\textwidth,trim={0.48cm 1.08cm 0.58cm 1.08cm},clip]{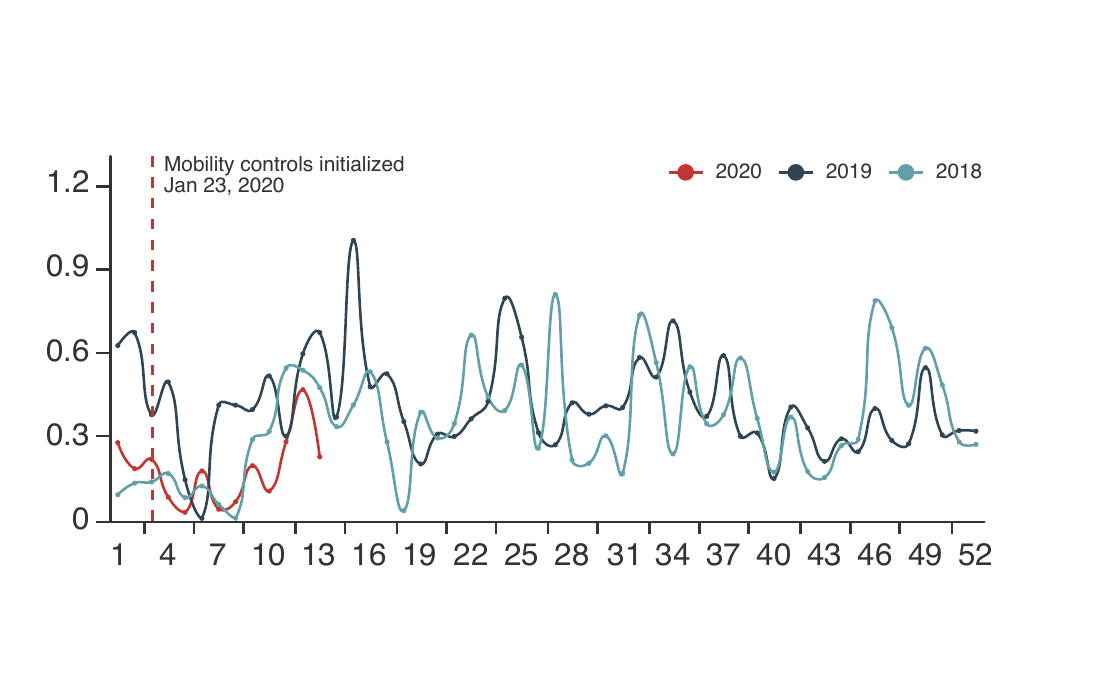}} \
    \subfloat[The \vvv in Tianjin]{\label{fig:a-tianjin-v3}\includegraphics[width=0.49\textwidth,trim={0.48cm 1.08cm 0.58cm 1.08cm},clip]{fig/f3-pro-v3-tianjin}} \
    \subfloat[The \nvc for Tianjin]{\label{fig:a-tianjin-nvc}\includegraphics[width=0.49\textwidth,trim={0.48cm 1.08cm 0.58cm 1.08cm},clip]{fig/f3-pro-nvc-tianjin}} \
    \caption{\label{fig:v3-nvc-31-provinces}The \vvv and the \nvc of 31 provinces. (1 of 8)}
\end{figure}

\begin{figure}
    \centering
    \ContinuedFloat
    \subfloat[The \vvv in Xizang]{\label{fig:a-xizang-v3}\includegraphics[width=0.49\textwidth,trim={0.48cm 1.08cm 0.58cm 1.08cm},clip]{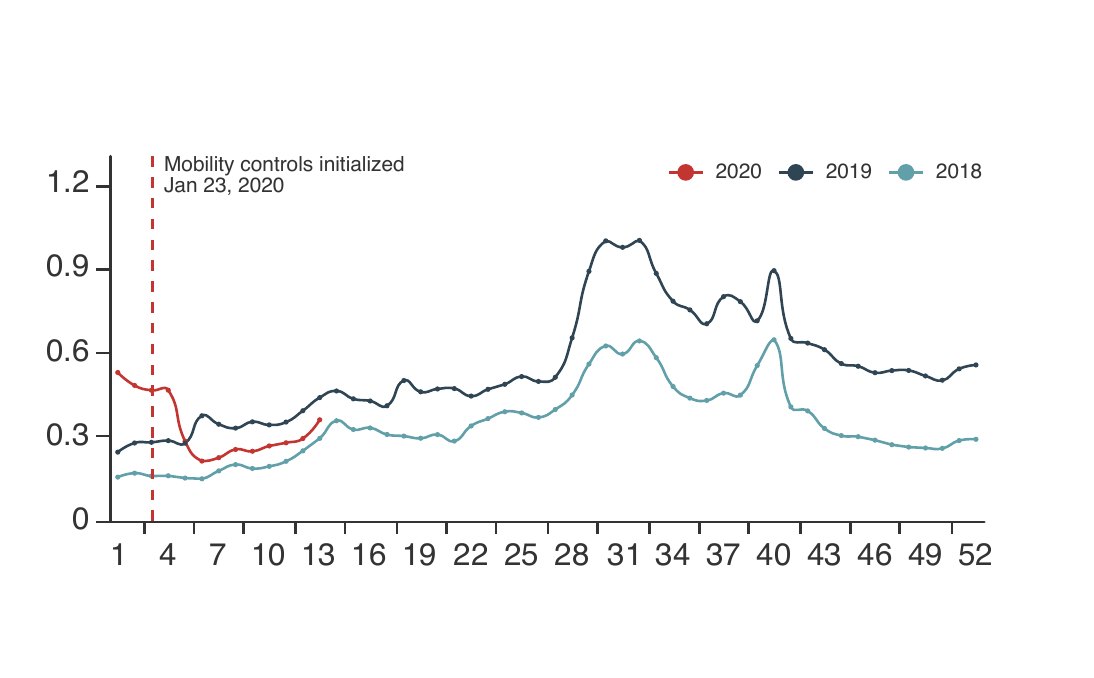}} \
    \subfloat[The \nvc for Xizang]{\label{fig:a-xizang-nvc}\includegraphics[width=0.49\textwidth,trim={0.48cm 1.08cm 0.58cm 1.08cm},clip]{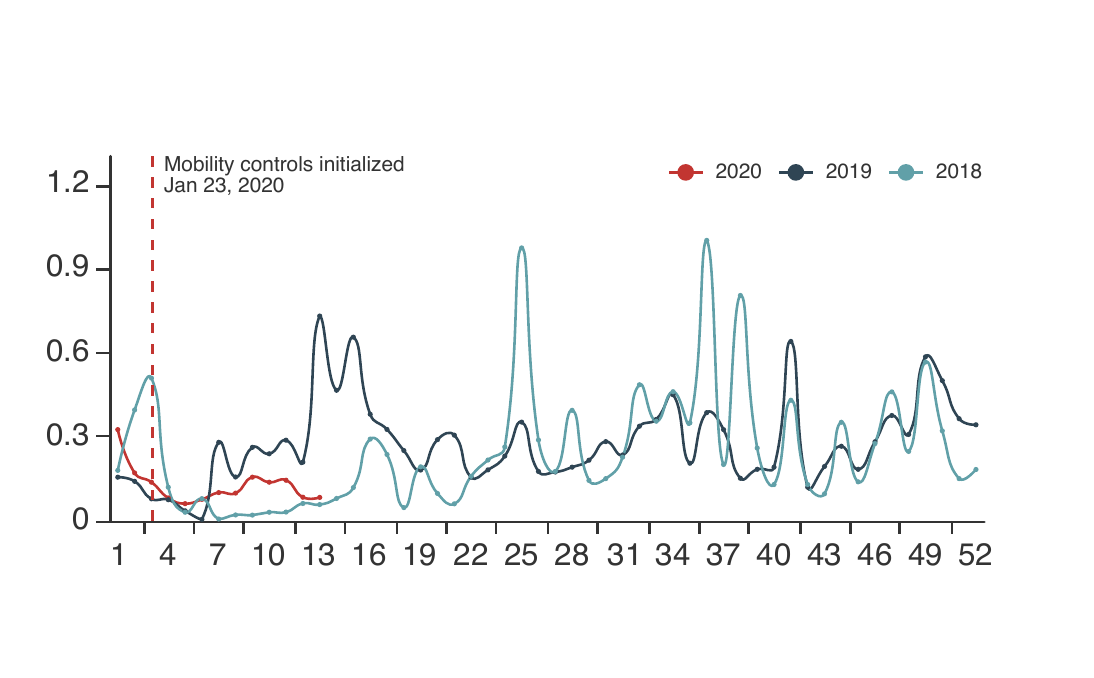}} \
    \subfloat[The \vvv in Shaanxi]{\label{fig:a-shaanxi-v3}\includegraphics[width=0.49\textwidth,trim={0.48cm 1.08cm 0.58cm 1.08cm},clip]{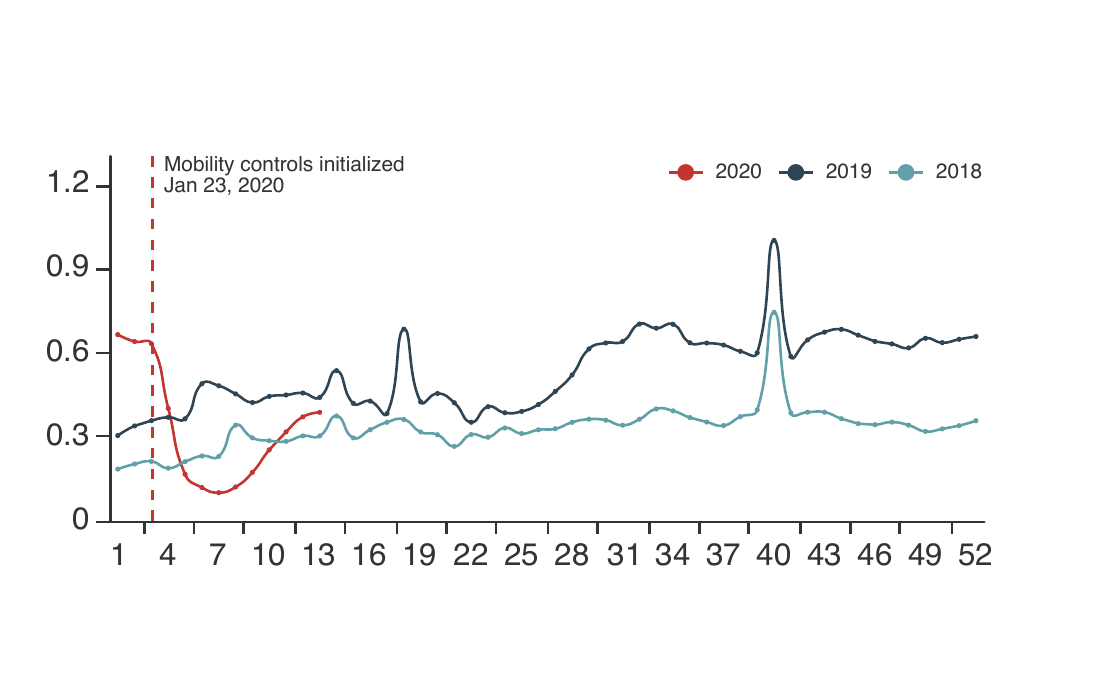}} \
    \subfloat[The \nvc for Shaanxi]{\label{fig:a-shaanxi-nvc}\includegraphics[width=0.49\textwidth,trim={0.48cm 1.08cm 0.58cm 1.08cm},clip]{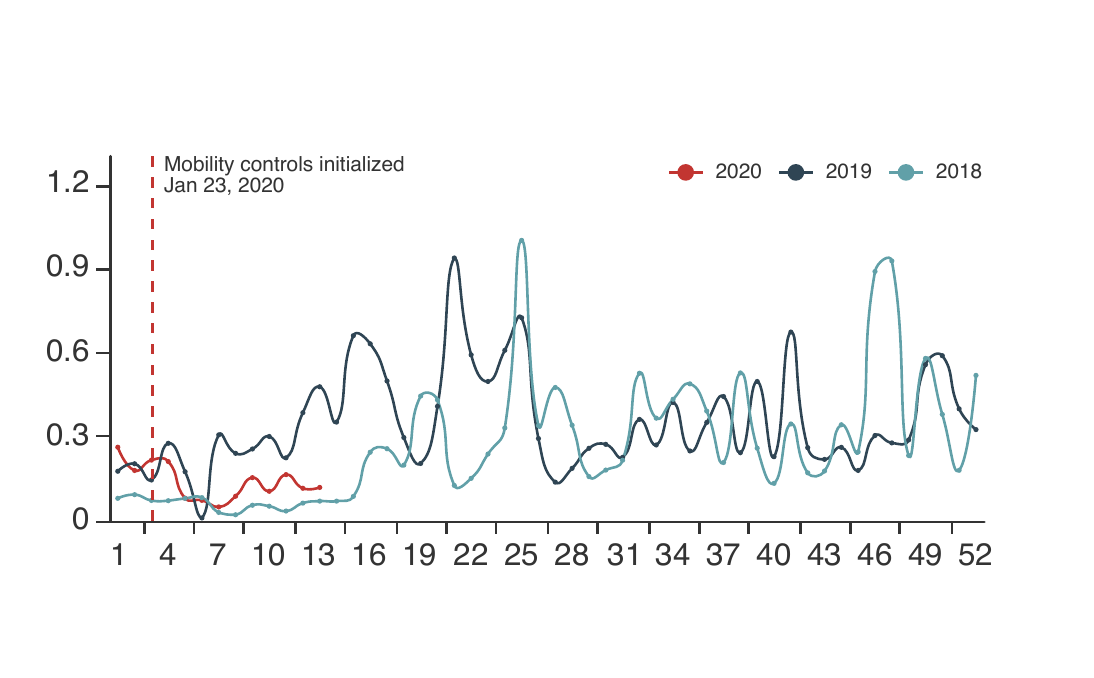}} \
    \subfloat[The \vvv in Jiangsu]{\label{fig:a-jiangsu-v3}\includegraphics[width=0.49\textwidth,trim={0.48cm 1.08cm 0.58cm 1.08cm},clip]{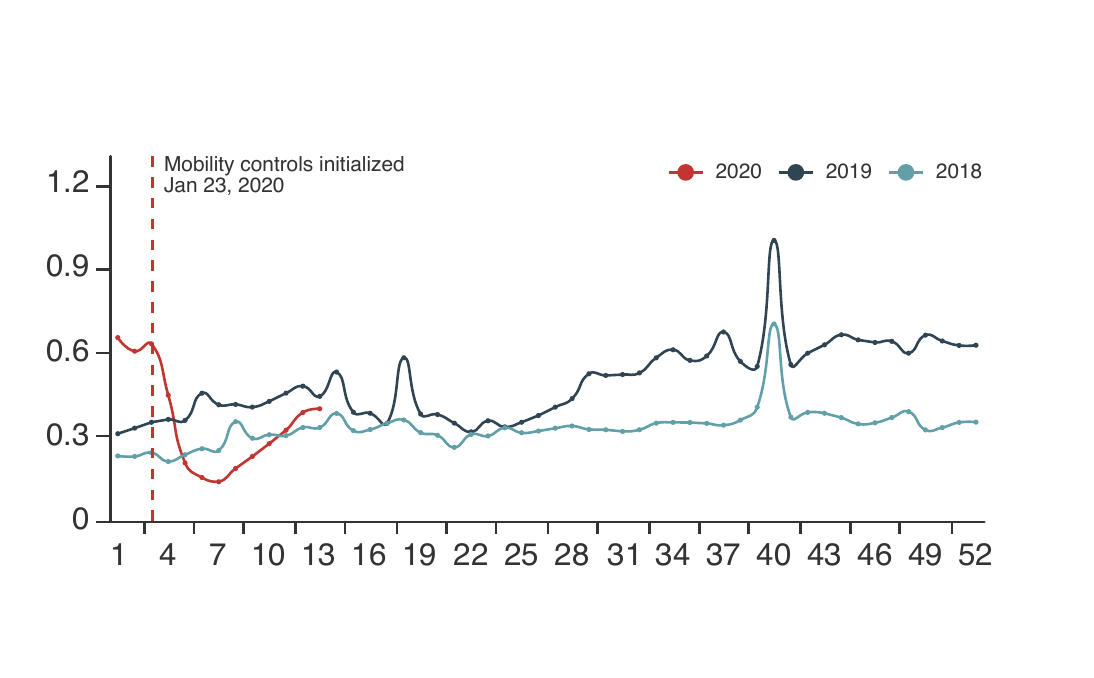}} \
    \subfloat[The \nvc for Jiangsu]{\label{fig:a-jiangsu-nvc}\includegraphics[width=0.49\textwidth,trim={0.48cm 1.08cm 0.58cm 1.08cm},clip]{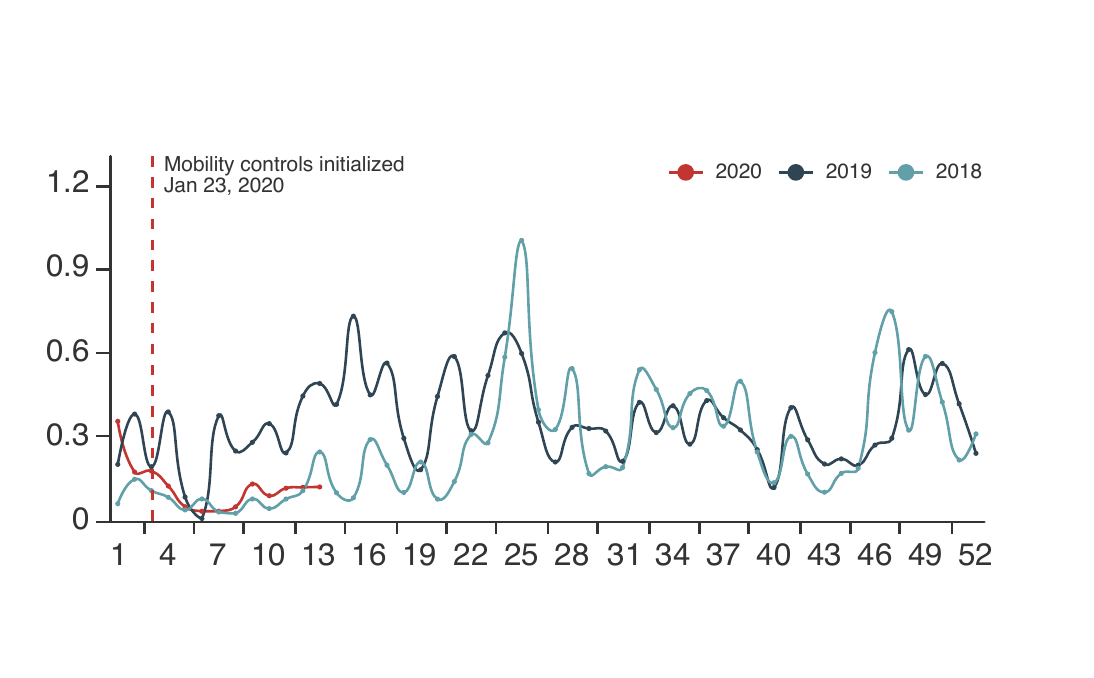}} \
    \subfloat[The \vvv in Hebei]{\label{fig:a-hebei-v3}\includegraphics[width=0.49\textwidth,trim={0.48cm 1.08cm 0.58cm 1.08cm},clip]{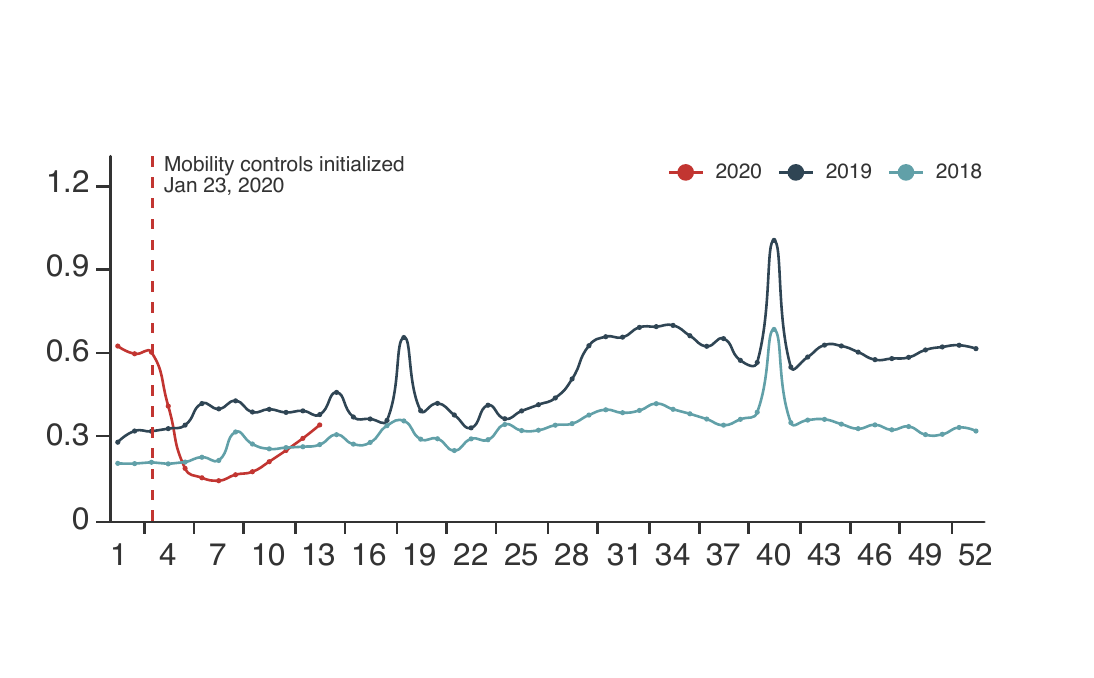}} \
    \subfloat[The \nvc for Hebei]{\label{fig:a-hebei-nvc}\includegraphics[width=0.49\textwidth,trim={0.48cm 1.08cm 0.58cm 1.08cm},clip]{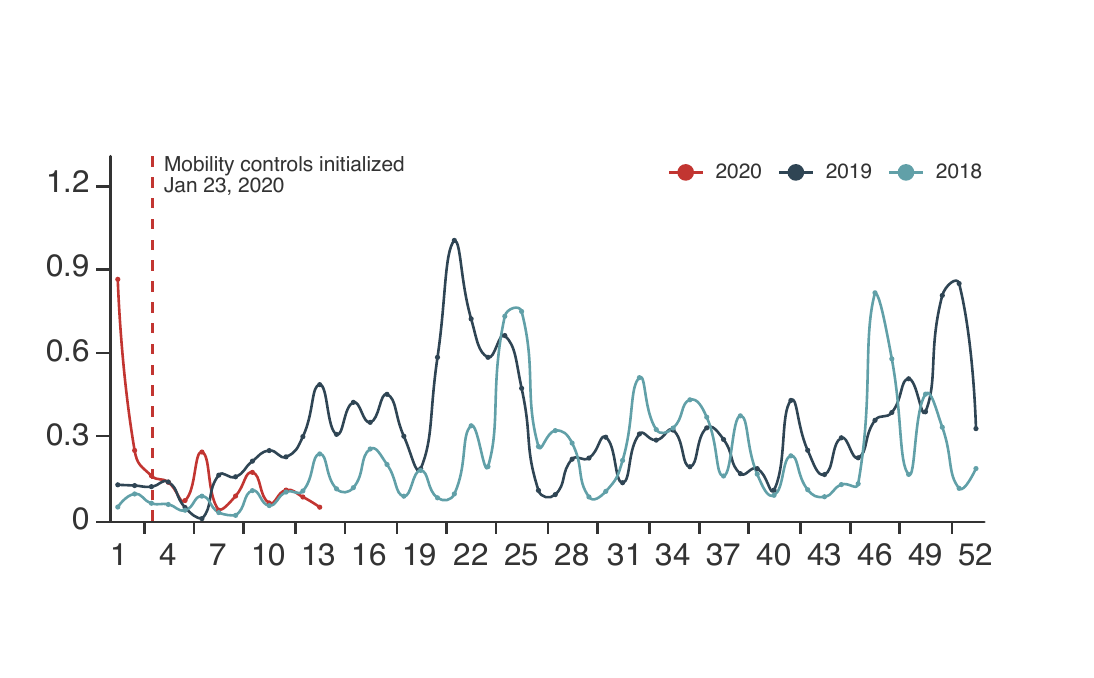}} \
    \caption{The \vvv and the \nvc of 31 provinces. (2 of 8)}
\end{figure}

\begin{figure}
    \centering
    \ContinuedFloat 
    \subfloat[The \vvv in Chongqing]{\label{fig:a-chongqing-v3}\includegraphics[width=0.49\textwidth,trim={0.48cm 1.08cm 0.58cm 1.08cm},clip]{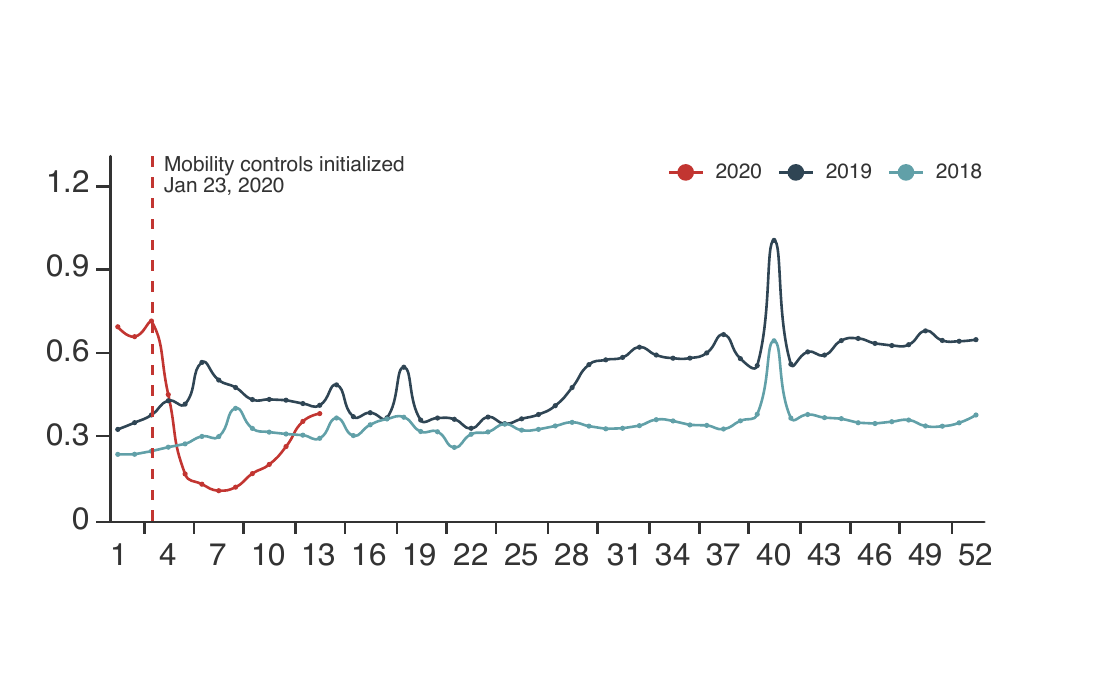}} \
    \subfloat[The \nvc for Chongqing]{\label{fig:a-chongqing-nvc}\includegraphics[width=0.49\textwidth,trim={0.48cm 1.08cm 0.58cm 1.08cm},clip]{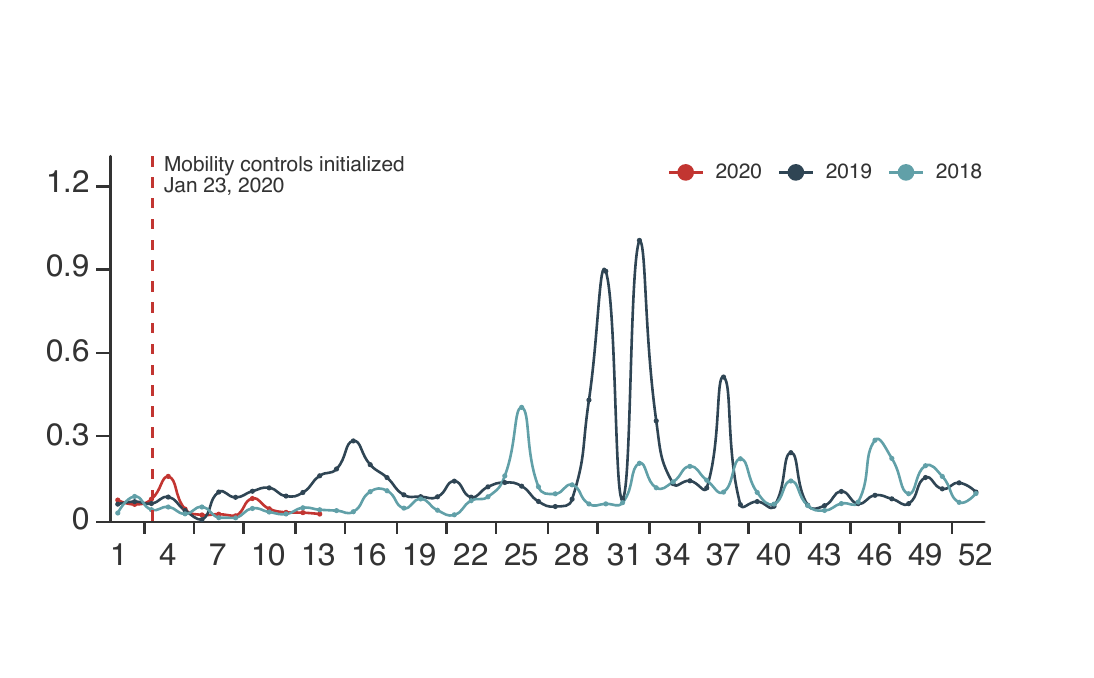}} \
    \subfloat[The \vvv in Zhejiang]{\label{fig:a-zhejiang-v3}\includegraphics[width=0.49\textwidth,trim={0.48cm 1.08cm 0.58cm 1.08cm},clip]{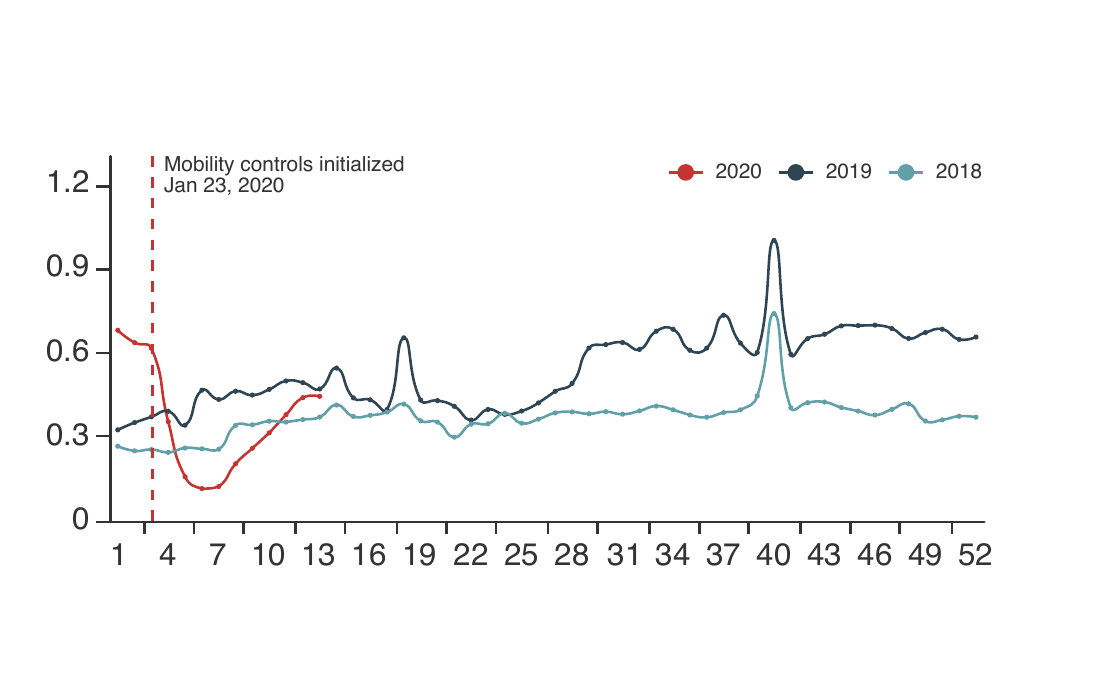}} \
    \subfloat[The \nvc for Zhejiang]{\label{fig:a-zhejiang-nvc}\includegraphics[width=0.49\textwidth,trim={0.48cm 1.08cm 0.58cm 1.08cm},clip]{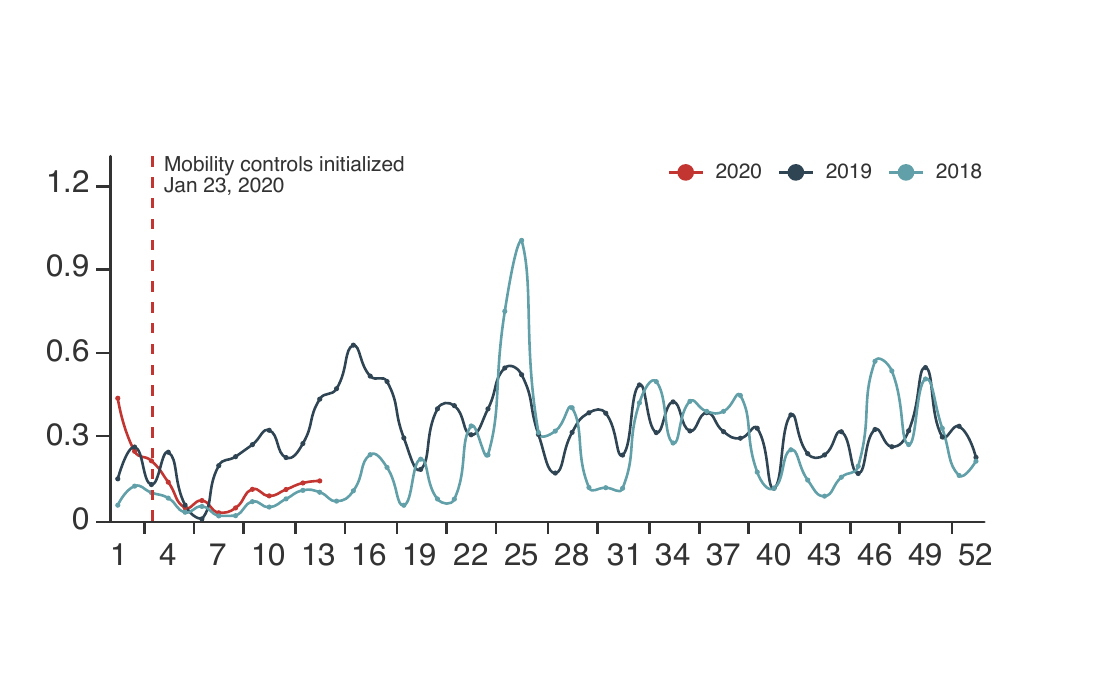}} \
    \subfloat[The \vvv in Heilongjiang]{\label{fig:a-heilongjiang-v3}\includegraphics[width=0.49\textwidth,trim={0.48cm 1.08cm 0.58cm 1.08cm},clip]{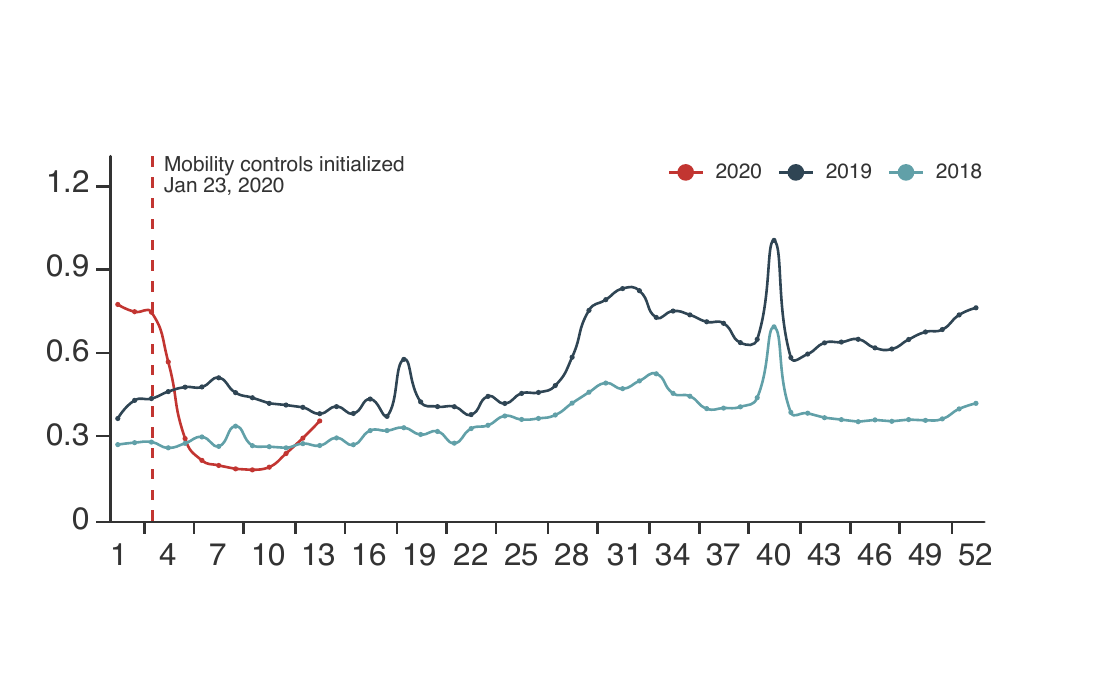}} \
    \subfloat[The \nvc for Heilongjiang]{\label{fig:a-heilongjiang-nvc}\includegraphics[width=0.49\textwidth,trim={0.48cm 1.08cm 0.58cm 1.08cm},clip]{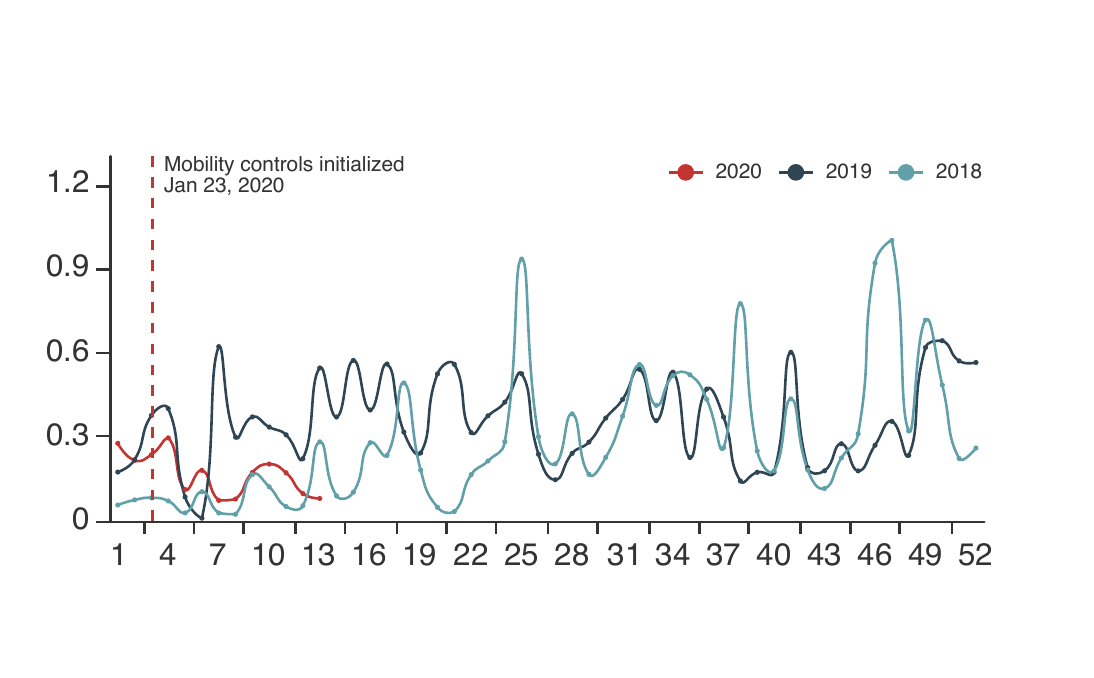}} \
    \subfloat[The \vvv in Hainan]{\label{fig:a-hainan-v3}\includegraphics[width=0.49\textwidth,trim={0.48cm 1.08cm 0.58cm 1.08cm},clip]{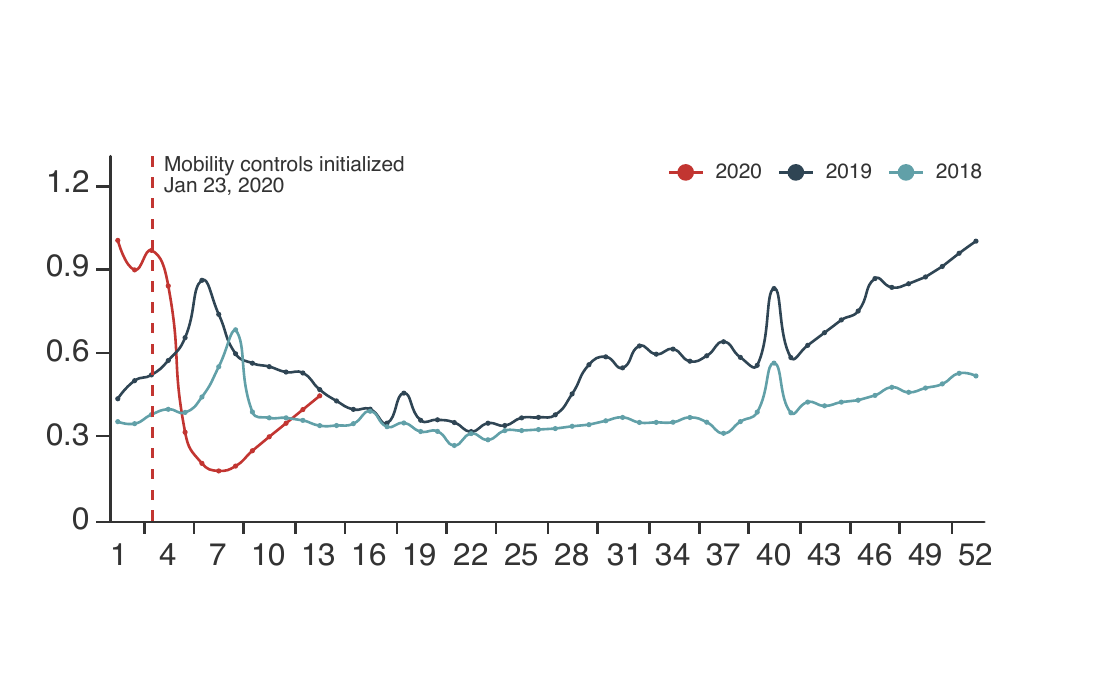}} \
    \subfloat[The \nvc for Hainan]{\label{fig:a-hainan-nvc}\includegraphics[width=0.49\textwidth,trim={0.48cm 1.08cm 0.58cm 1.08cm},clip]{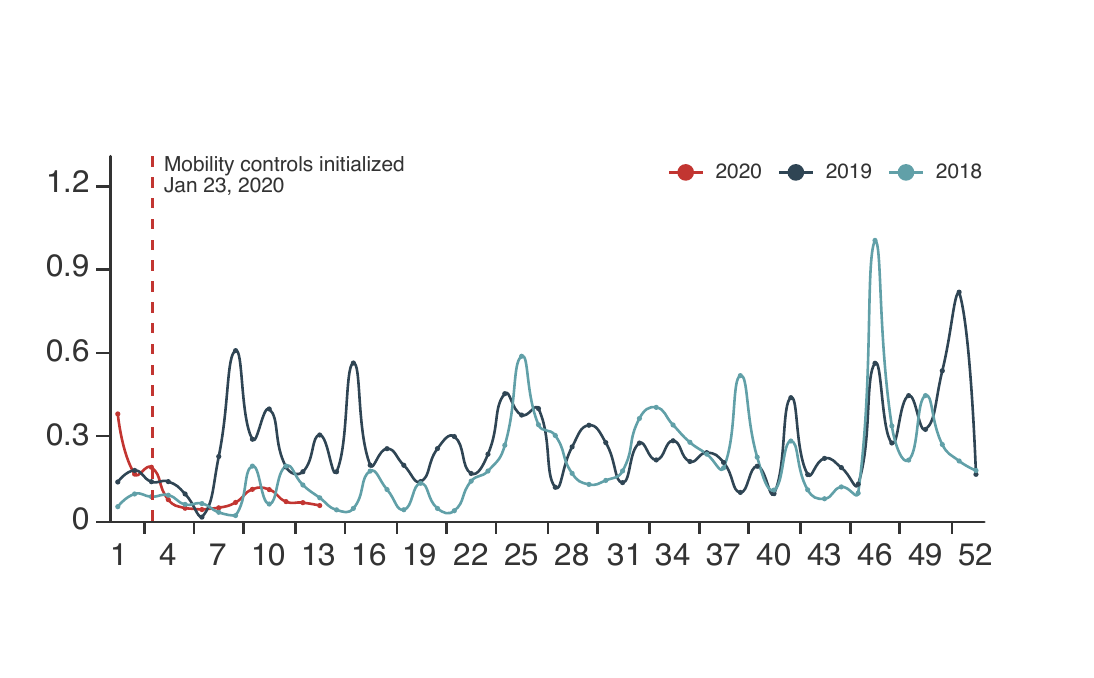}} \
    \caption{The \vvv and the \nvc of 31 provinces. (3 of 8)}
\end{figure}

\begin{figure}
    \centering
    \ContinuedFloat     
    \subfloat[The \vvv in Guangdong]{\label{fig:a-guangdong-v3}\includegraphics[width=0.49\textwidth,trim={0.48cm 1.08cm 0.58cm 1.08cm},clip]{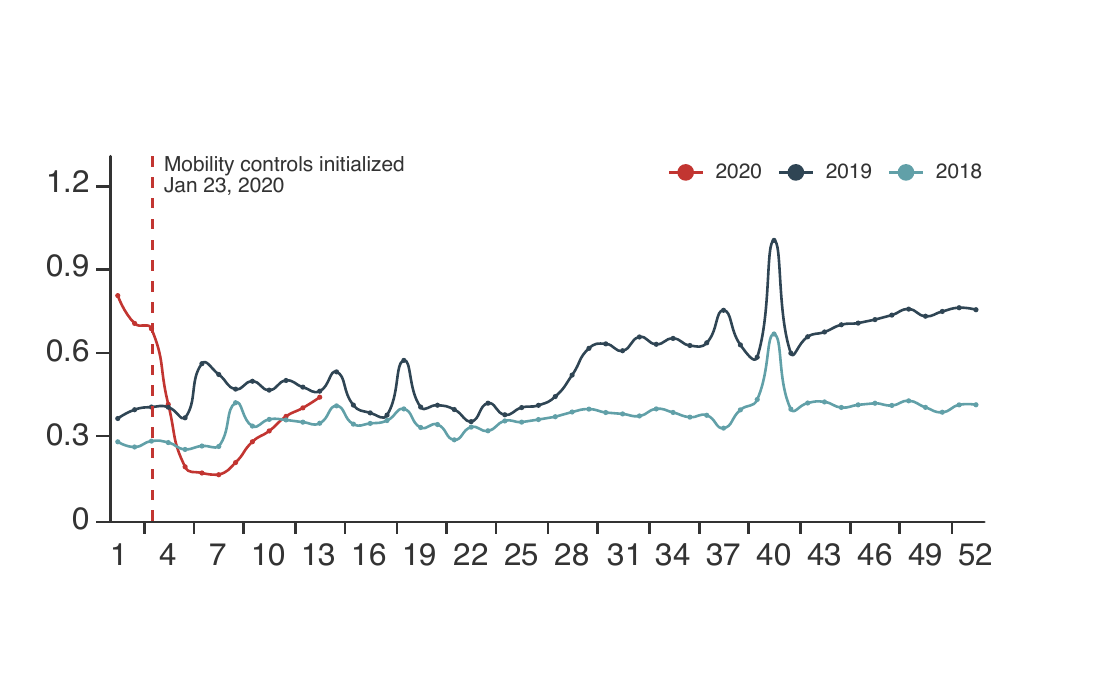}} \
    \subfloat[The \nvc for Guangdong]{\label{fig:a-guangdong-nvc}\includegraphics[width=0.49\textwidth,trim={0.48cm 1.08cm 0.58cm 1.08cm},clip]{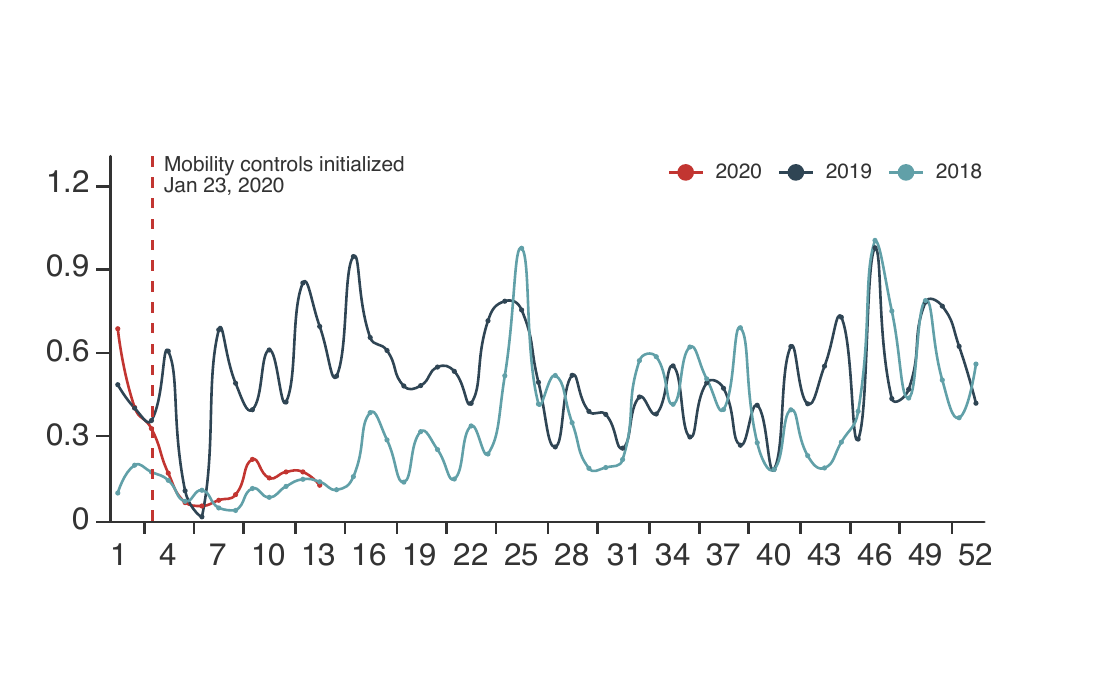}} \
    \subfloat[The \vvv in Henan]{\label{fig:a-henan-v3}\includegraphics[width=0.49\textwidth,trim={0.48cm 1.08cm 0.58cm 1.08cm},clip]{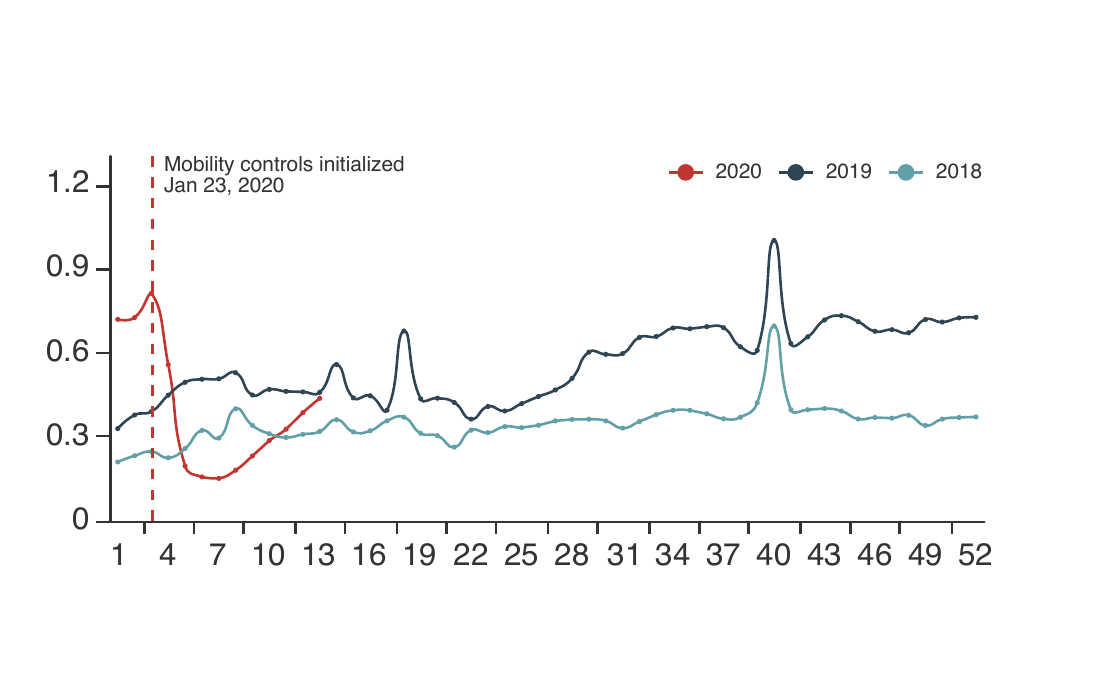}} \
    \subfloat[The \nvc for Henan]{\label{fig:a-henan-nvc}\includegraphics[width=0.49\textwidth,trim={0.48cm 1.08cm 0.58cm 1.08cm},clip]{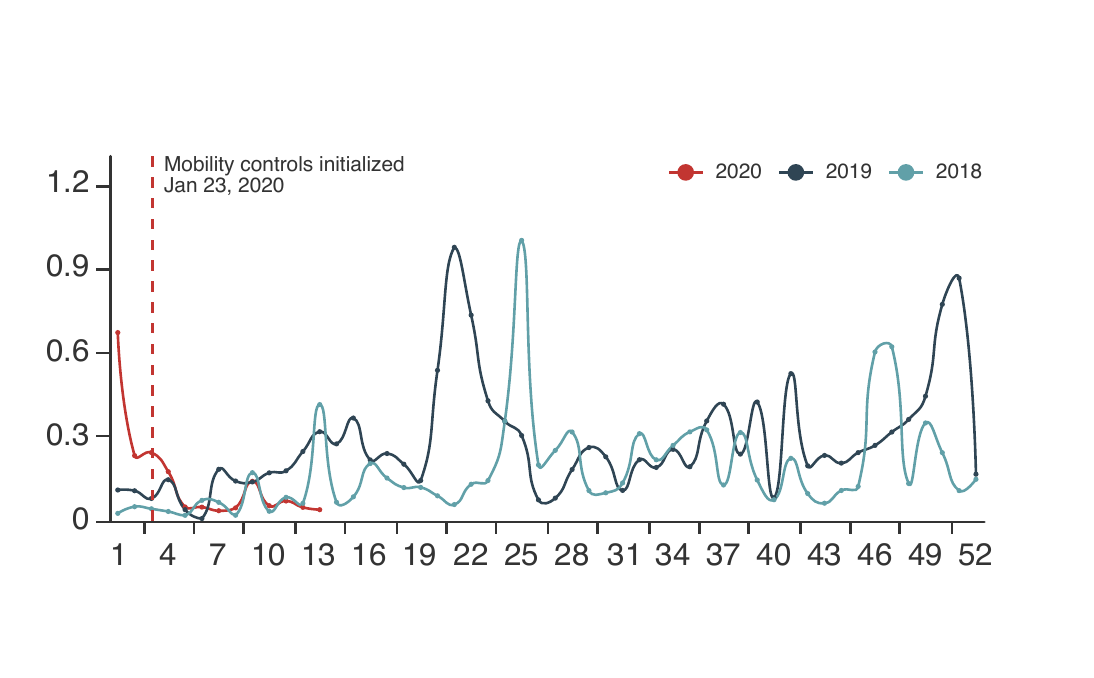}} \
    \subfloat[The \vvv in Hunan]{\label{fig:a-hunan-v3}\includegraphics[width=0.49\textwidth,trim={0.48cm 1.08cm 0.58cm 1.08cm},clip]{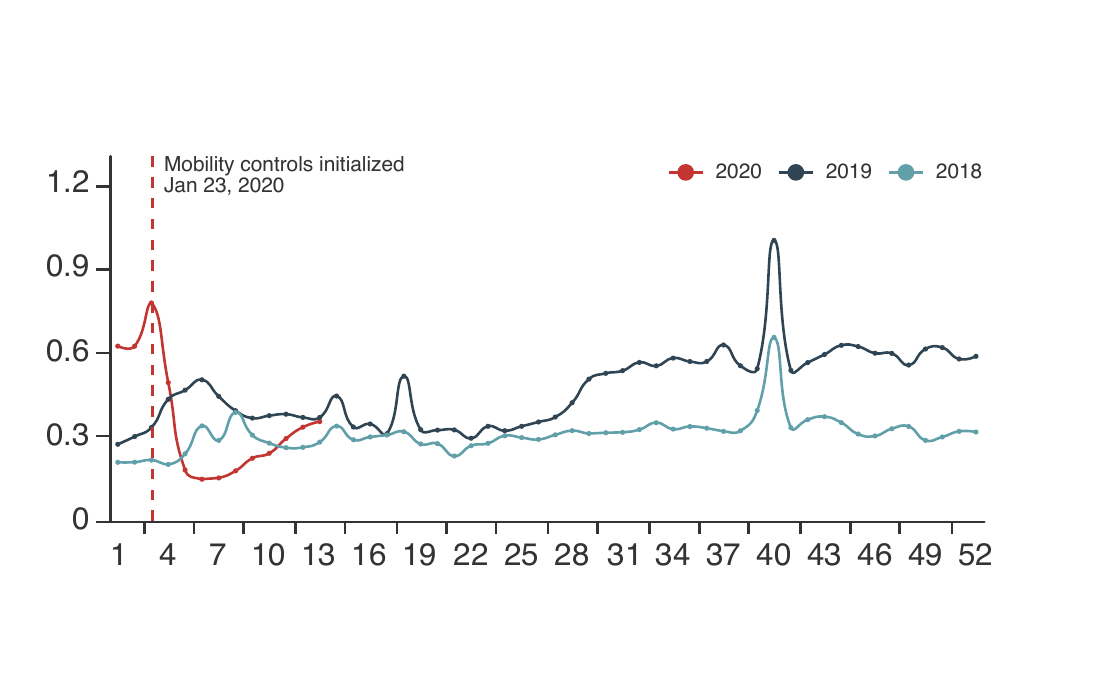}} \
    \subfloat[The \nvc for Hunan]{\label{fig:a-hunan-nvc}\includegraphics[width=0.49\textwidth,trim={0.48cm 1.08cm 0.58cm 1.08cm},clip]{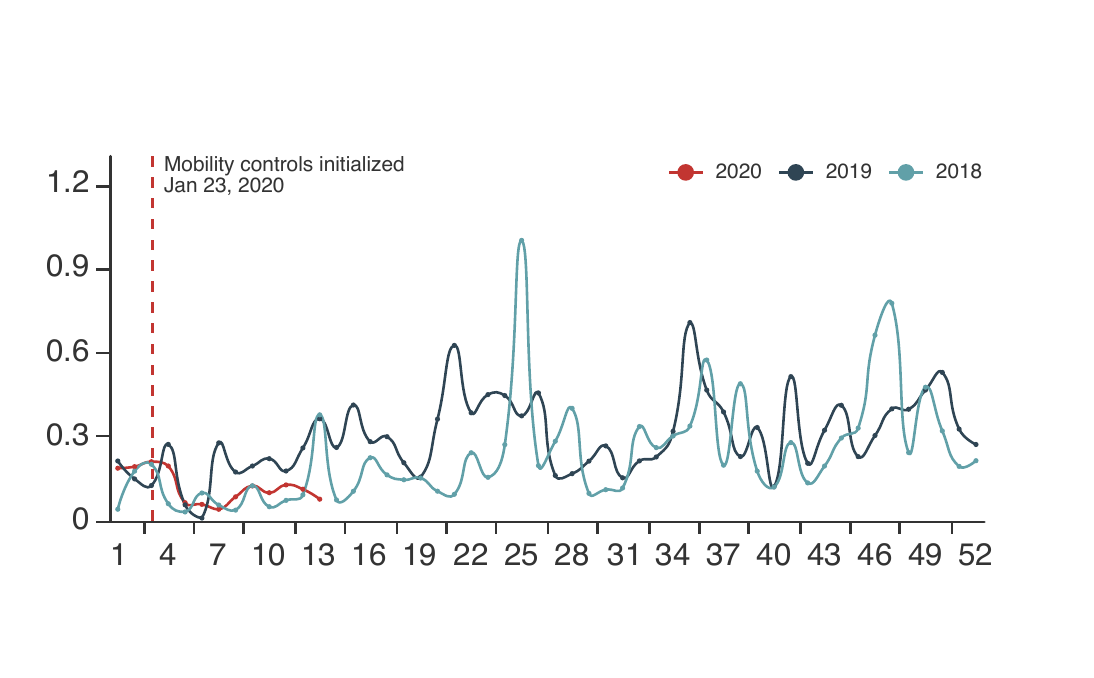}} \
    \subfloat[The \vvv in Jiangxi]{\label{fig:a-jiangxi-v3}\includegraphics[width=0.49\textwidth,trim={0.48cm 1.08cm 0.58cm 1.08cm},clip]{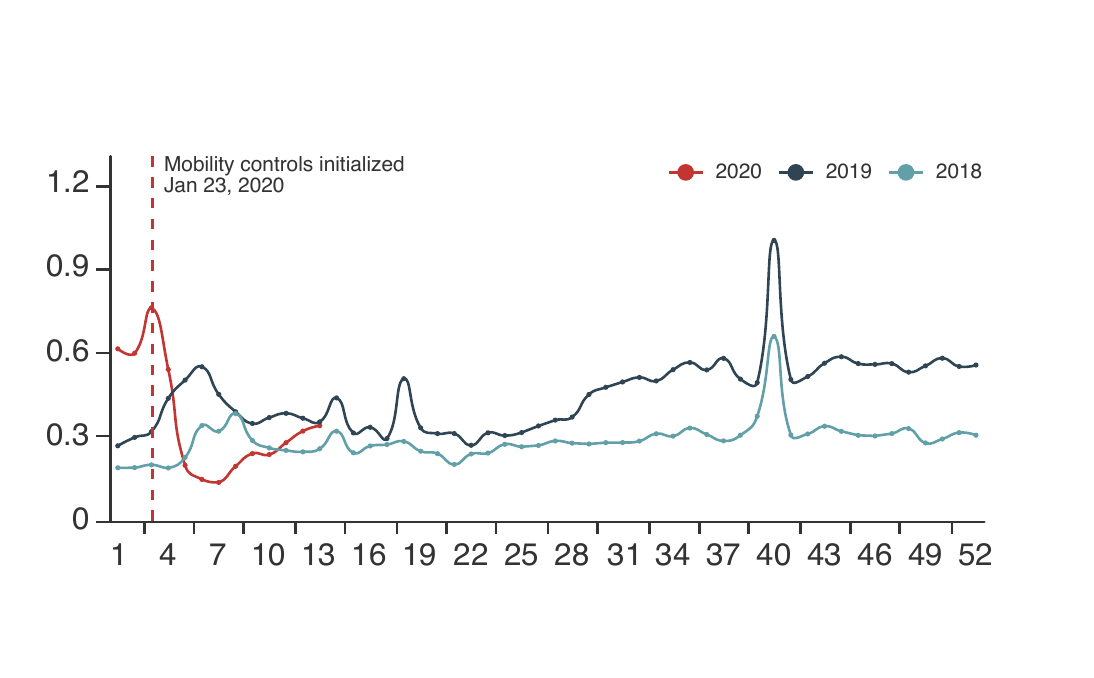}} \
    \subfloat[The \nvc for Jiangxi]{\label{fig:a-jiangxi-nvc}\includegraphics[width=0.49\textwidth,trim={0.48cm 1.08cm 0.58cm 1.08cm},clip]{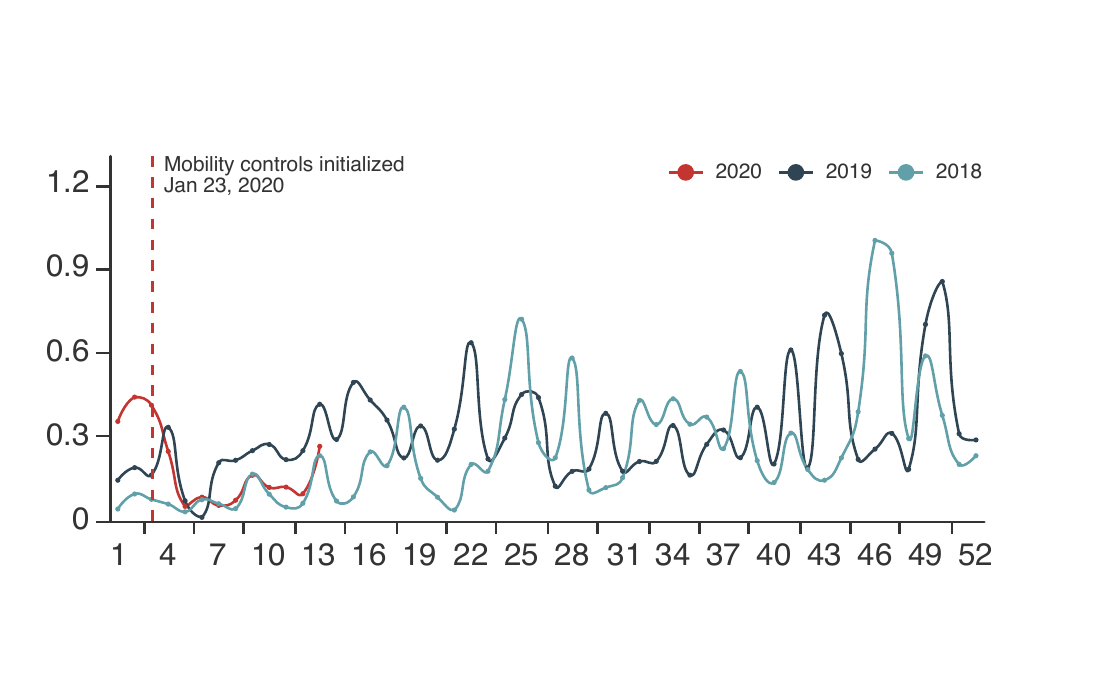}} \
    \caption{The \vvv and the \nvc of 31 provinces. (4 of 8)}
\end{figure}

\begin{figure}
    \centering
    \ContinuedFloat      
    \subfloat[The \vvv in Fujian]{\label{fig:a-fujian-v3}\includegraphics[width=0.49\textwidth,trim={0.48cm 1.08cm 0.58cm 1.08cm},clip]{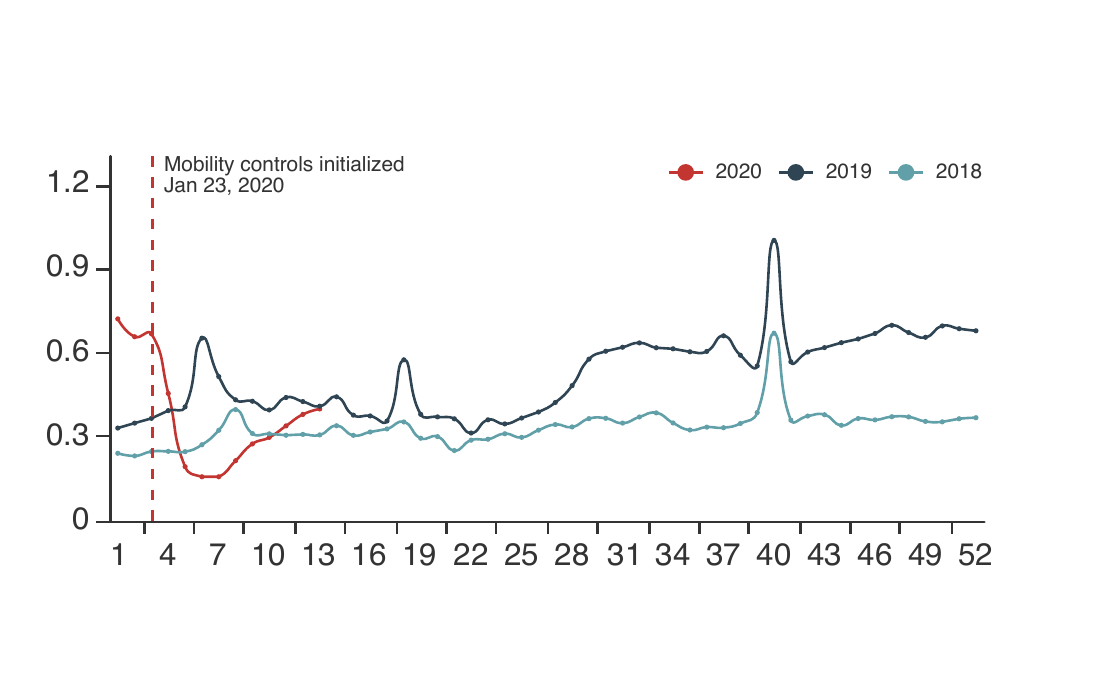}} \
    \subfloat[The \nvc for Fujian]{\label{fig:a-fujian-nvc}\includegraphics[width=0.49\textwidth,trim={0.48cm 1.08cm 0.58cm 1.08cm},clip]{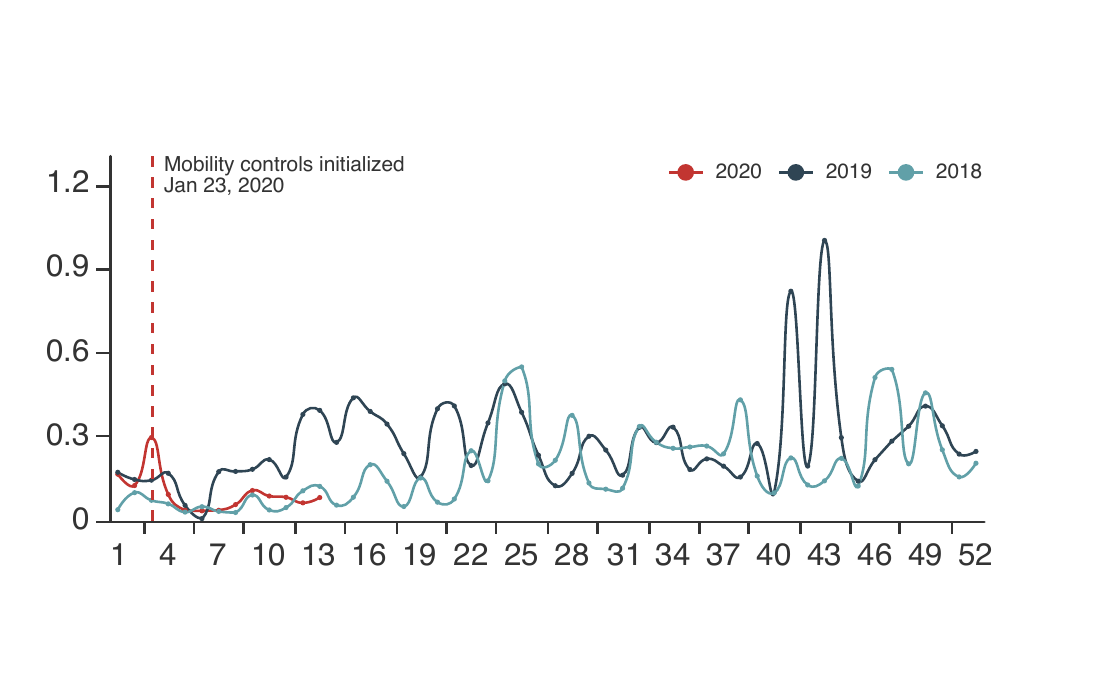}} \
    \subfloat[The \vvv in Gansu]{\label{fig:a-gansu-v3}\includegraphics[width=0.49\textwidth,trim={0.48cm 1.08cm 0.58cm 1.08cm},clip]{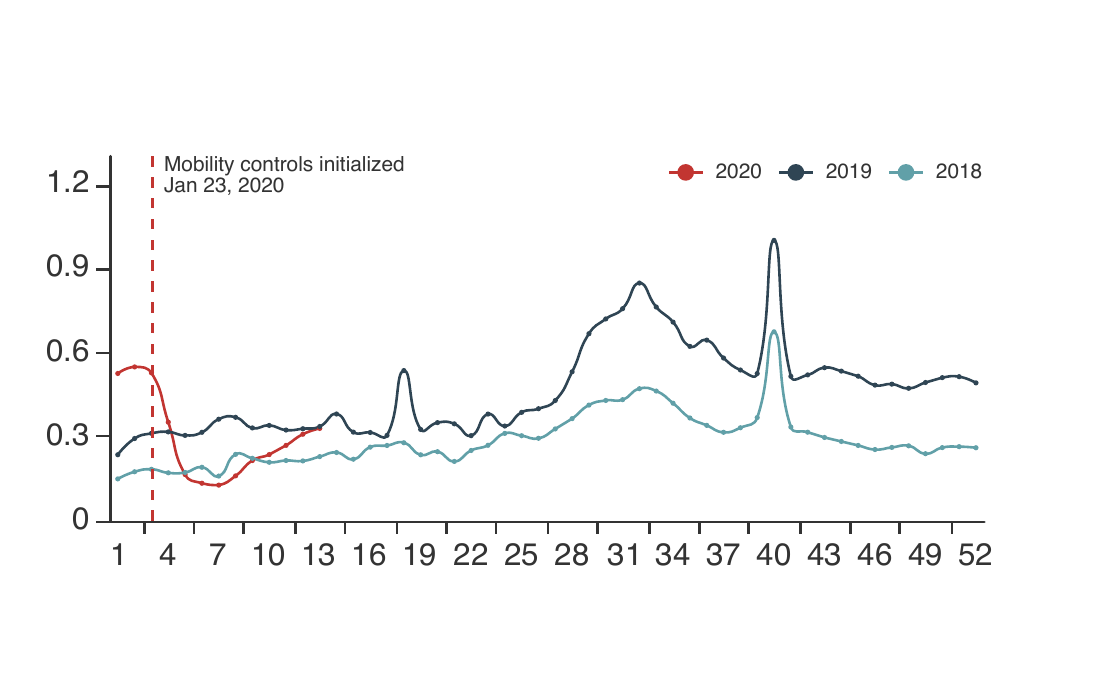}} \
    \subfloat[The \nvc for Gansu]{\label{fig:a-gansu-nvc}\includegraphics[width=0.49\textwidth,trim={0.48cm 1.08cm 0.58cm 1.08cm},clip]{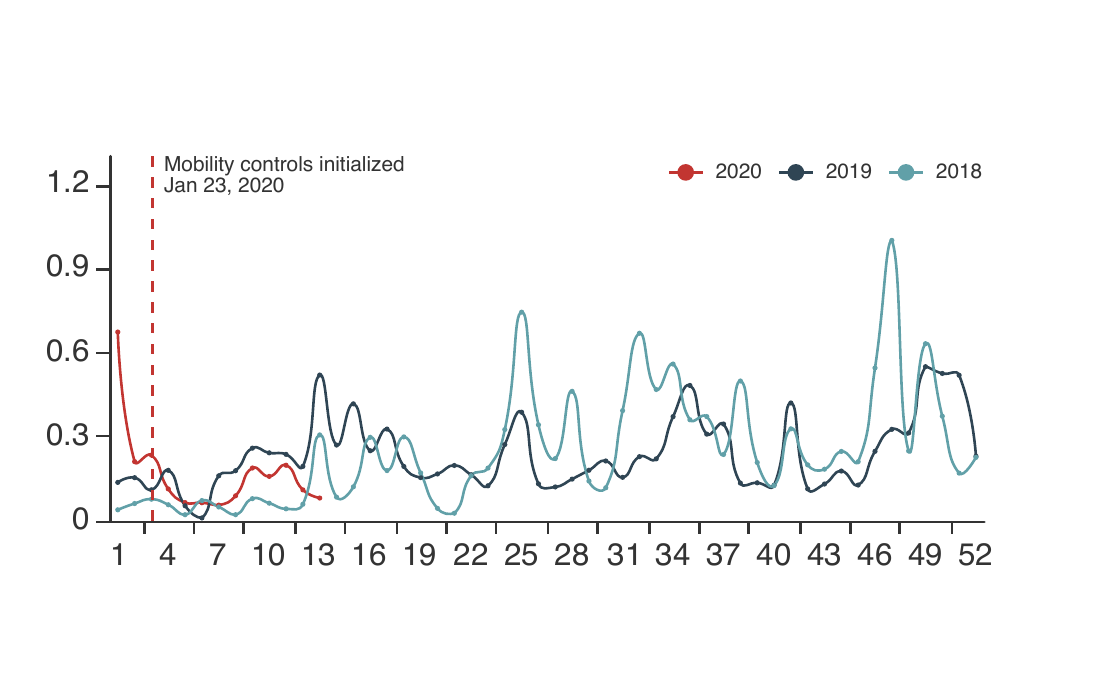}} \
    \subfloat[The \vvv in Sichuan]{\label{fig:a-sichuan-v3}\includegraphics[width=0.49\textwidth,trim={0.48cm 1.08cm 0.58cm 1.08cm},clip]{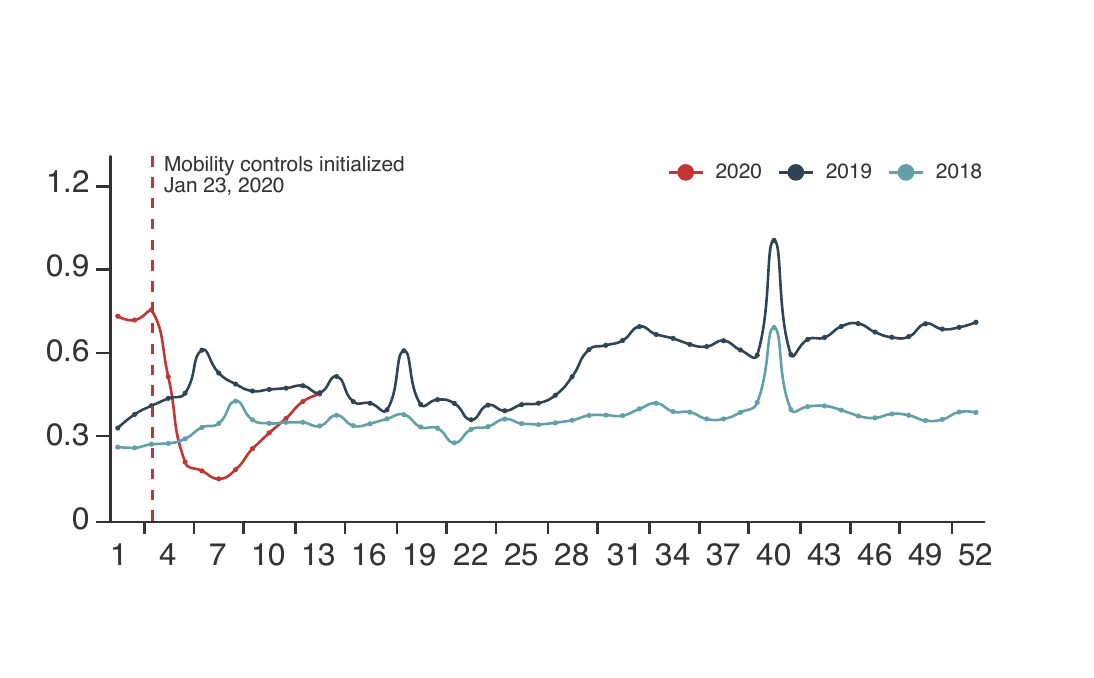}} \
    \subfloat[The \nvc for Sichuan]{\label{fig:a-sichuan-nvc}\includegraphics[width=0.49\textwidth,trim={0.48cm 1.08cm 0.58cm 1.08cm},clip]{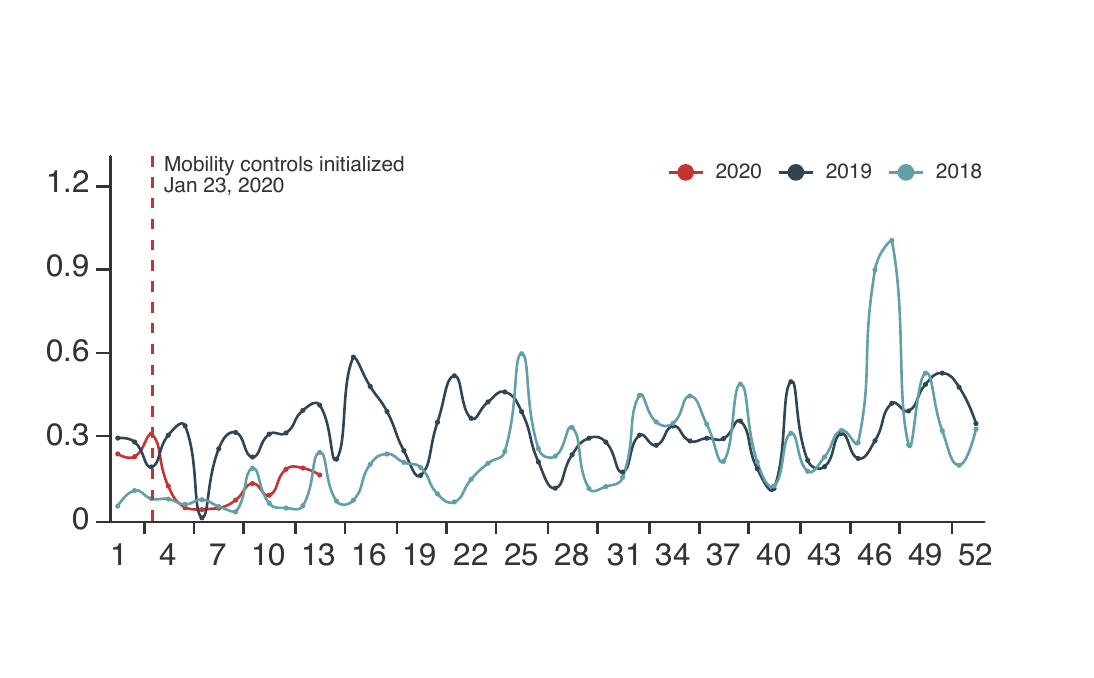}} \
    \subfloat[The \vvv in Anhui]{\label{fig:a-anhui-v3}\includegraphics[width=0.49\textwidth,trim={0.48cm 1.08cm 0.58cm 1.08cm},clip]{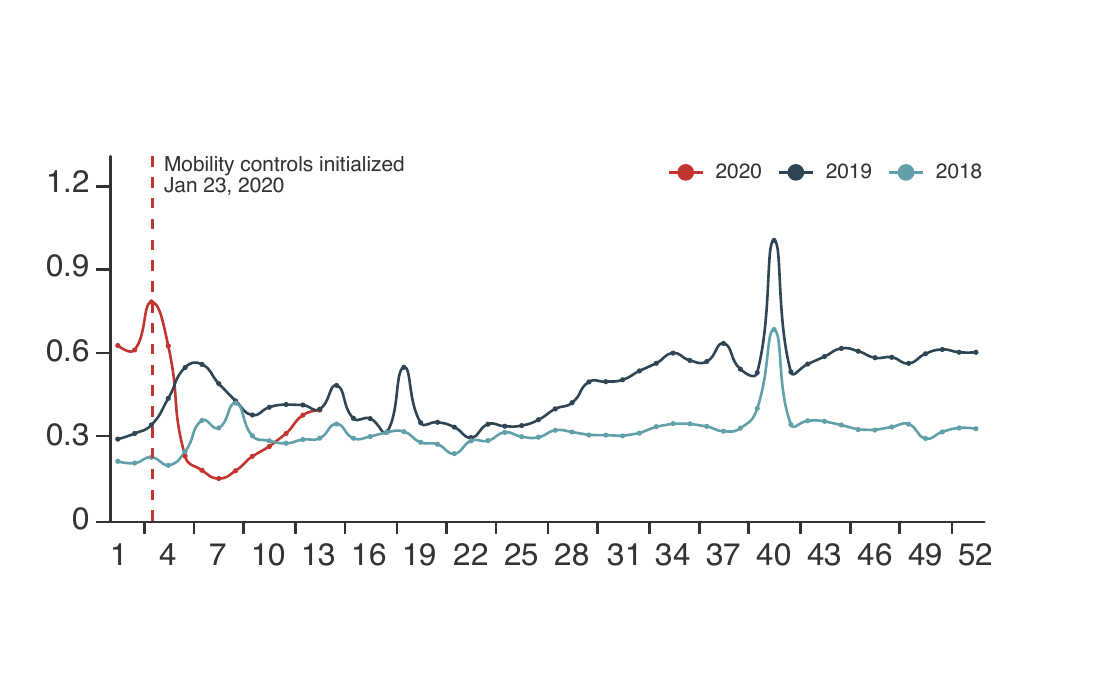}} \
    \subfloat[The \nvc for Anhui]{\label{fig:a-anhui-nvc}\includegraphics[width=0.49\textwidth,trim={0.48cm 1.08cm 0.58cm 1.08cm},clip]{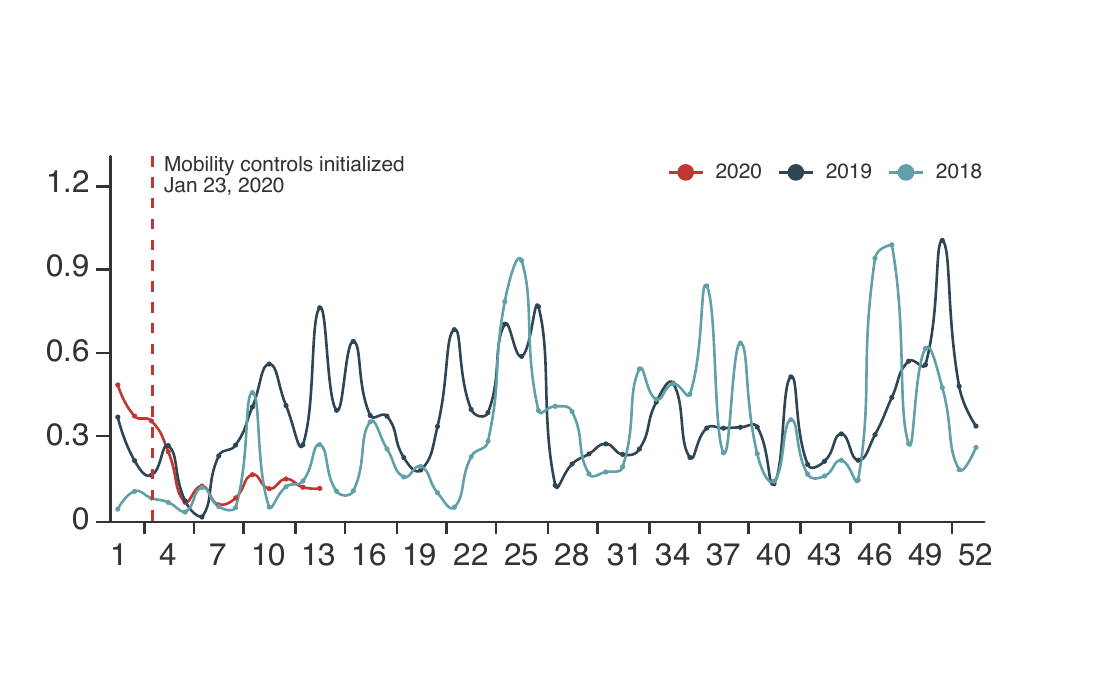}} \
    \caption{The \vvv and the \nvc of 31 provinces. (5 of 8)}
\end{figure}

\begin{figure}
    \centering
    \ContinuedFloat     
    \subfloat[The \vvv in Shanxi]{\label{fig:a-shanxi-v3}\includegraphics[width=0.49\textwidth,trim={0.48cm 1.08cm 0.58cm 1.08cm},clip]{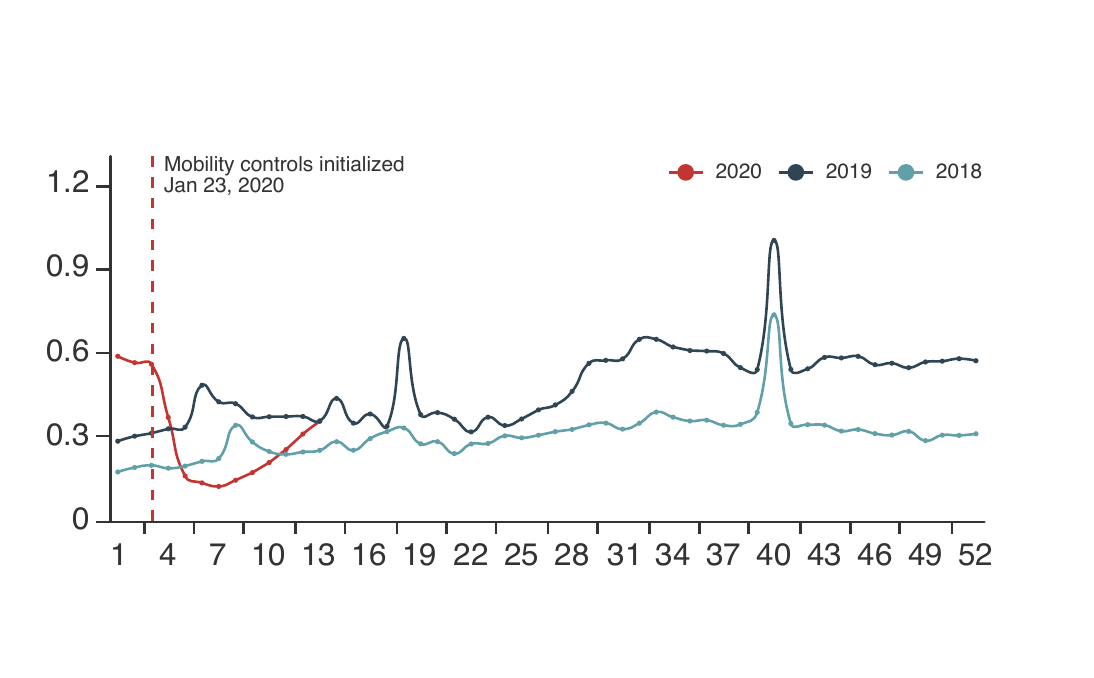}} \
    \subfloat[The \nvc for Shanxi]{\label{fig:a-shanxi-nvc}\includegraphics[width=0.49\textwidth,trim={0.48cm 1.08cm 0.58cm 1.08cm},clip]{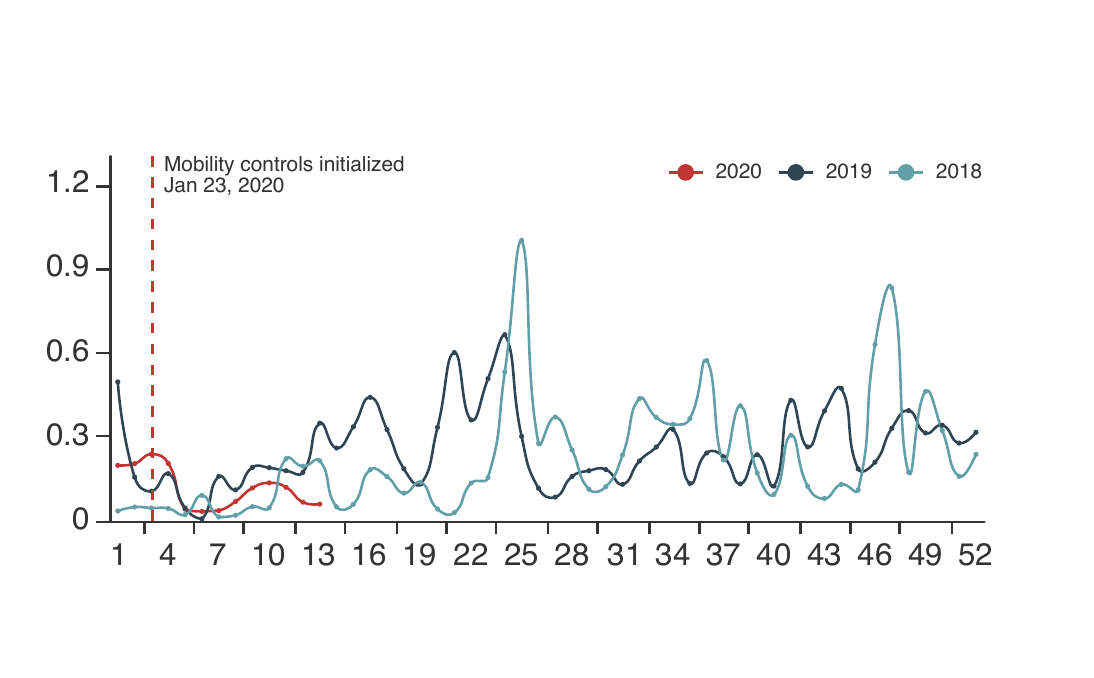}} \
    \subfloat[The \vvv in Liaoning]{\label{fig:a-liaoning-v3}\includegraphics[width=0.49\textwidth,trim={0.48cm 1.08cm 0.58cm 1.08cm},clip]{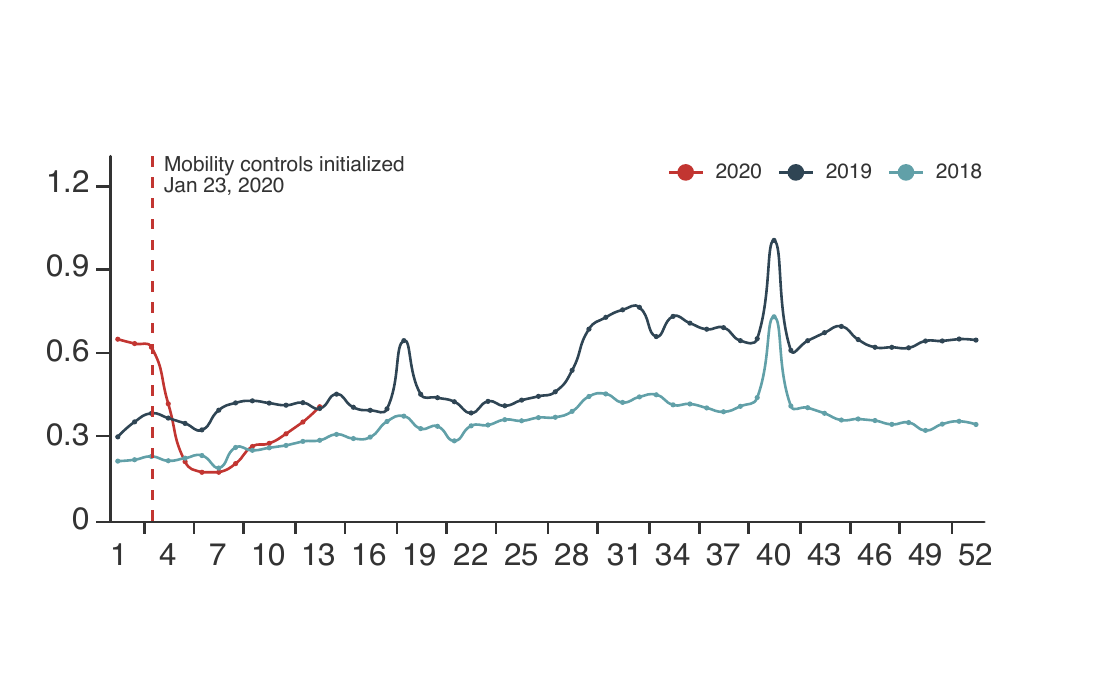}} \
    \subfloat[The \nvc for Liaoning]{\label{fig:a-liaoning-nvc}\includegraphics[width=0.49\textwidth,trim={0.48cm 1.08cm 0.58cm 1.08cm},clip]{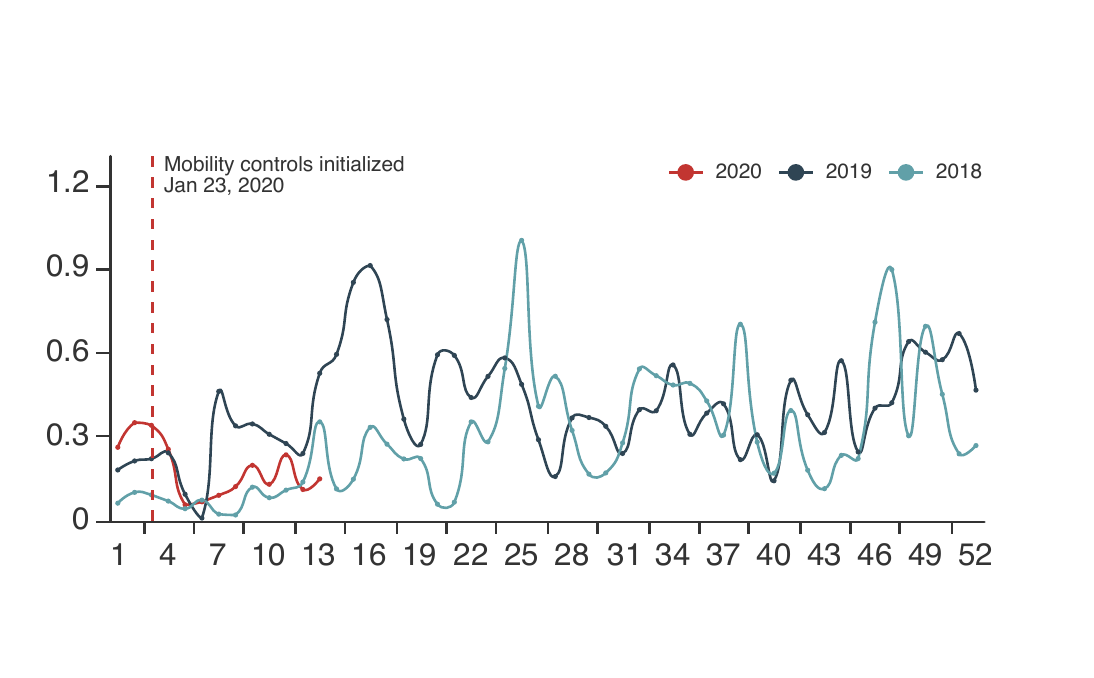}} \
    \subfloat[The \vvv in Shandong]{\label{fig:a-shandong-v3}\includegraphics[width=0.49\textwidth,trim={0.48cm 1.08cm 0.58cm 1.08cm},clip]{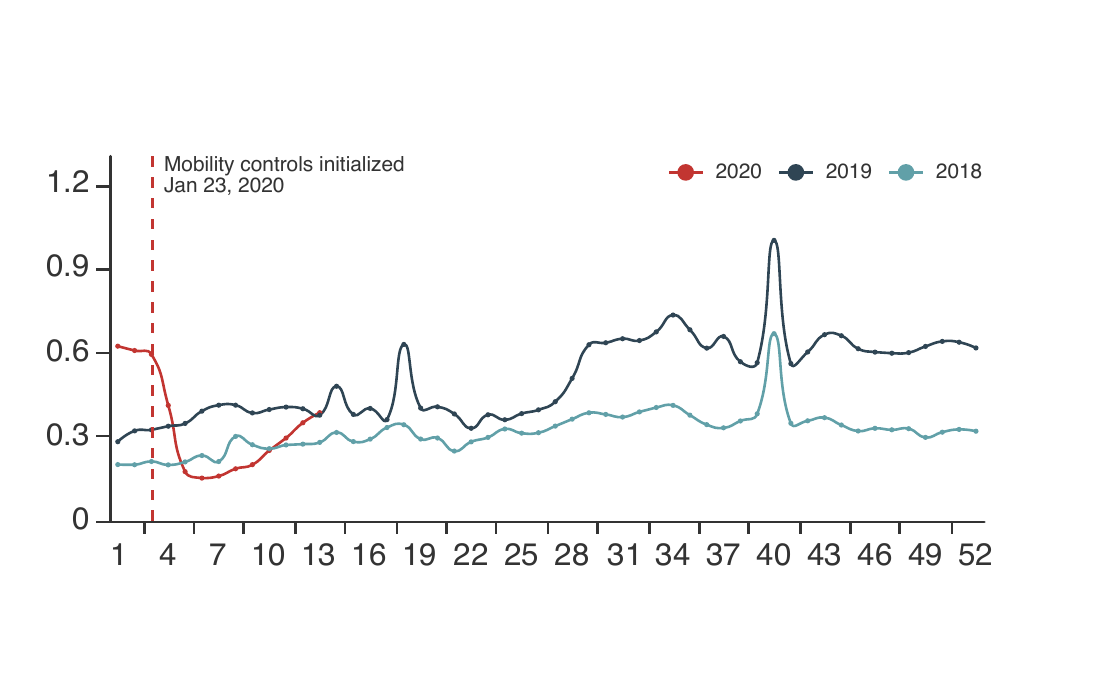}} \
    \subfloat[The \nvc for Shandong]{\label{fig:a-shandong-nvc}\includegraphics[width=0.49\textwidth,trim={0.48cm 1.08cm 0.58cm 1.08cm},clip]{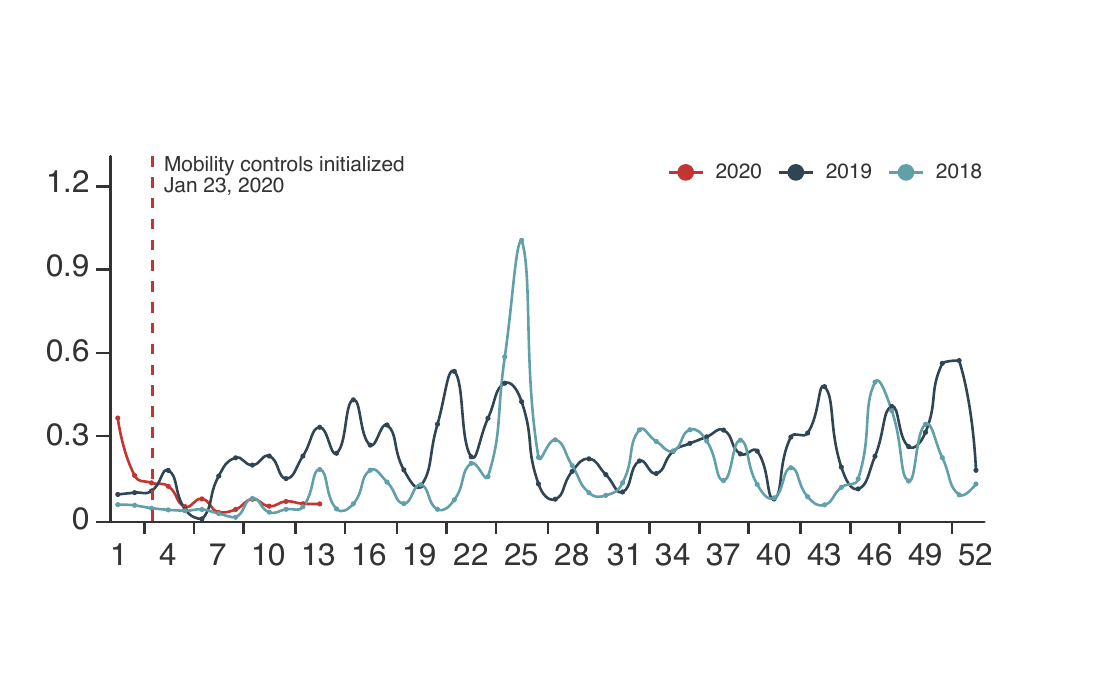}} \
    \subfloat[The \vvv in Ningxia]{\label{fig:a-ningxia-v3}\includegraphics[width=0.49\textwidth,trim={0.48cm 1.08cm 0.58cm 1.08cm},clip]{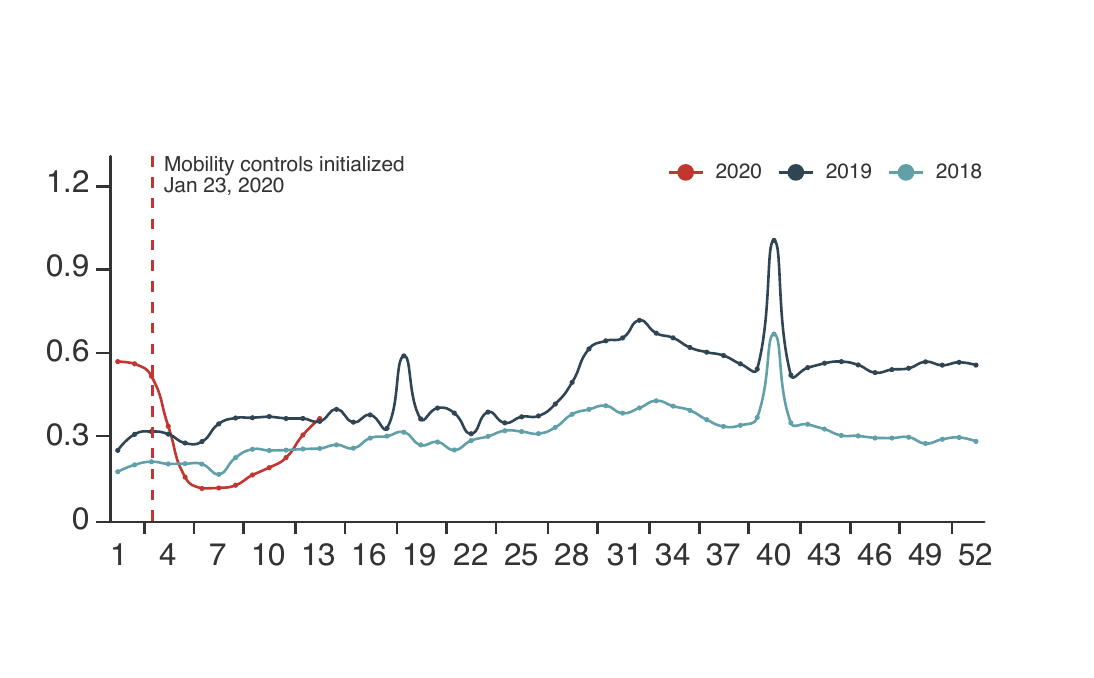}} \
    \subfloat[The \nvc for Ningxia]{\label{fig:a-ningxia-nvc}\includegraphics[width=0.49\textwidth,trim={0.48cm 1.08cm 0.58cm 1.08cm},clip]{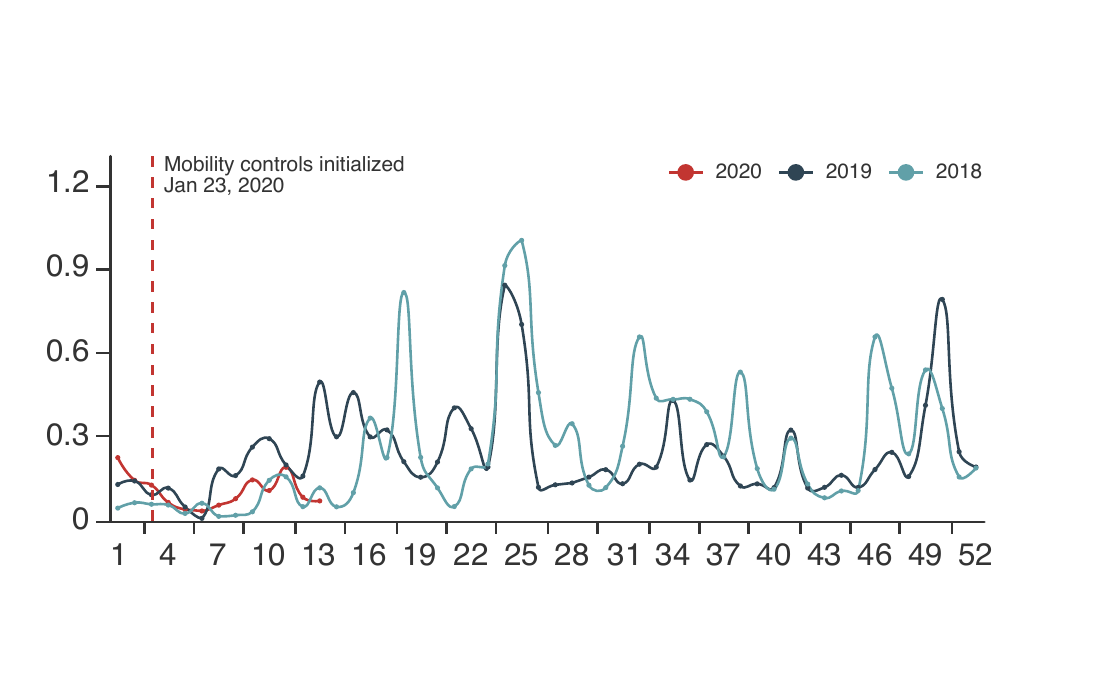}} \
    \caption{The \vvv and the \nvc of 31 provinces. (6 of 8)}
\end{figure}

\begin{figure}[t]
    \centering
    \ContinuedFloat     
    \subfloat[The \vvv in Inner Mongolia]{\label{fig:a-inner-mongolia-v3}\includegraphics[width=0.49\textwidth,trim={0.48cm 1.08cm 0.58cm 1.08cm},clip]{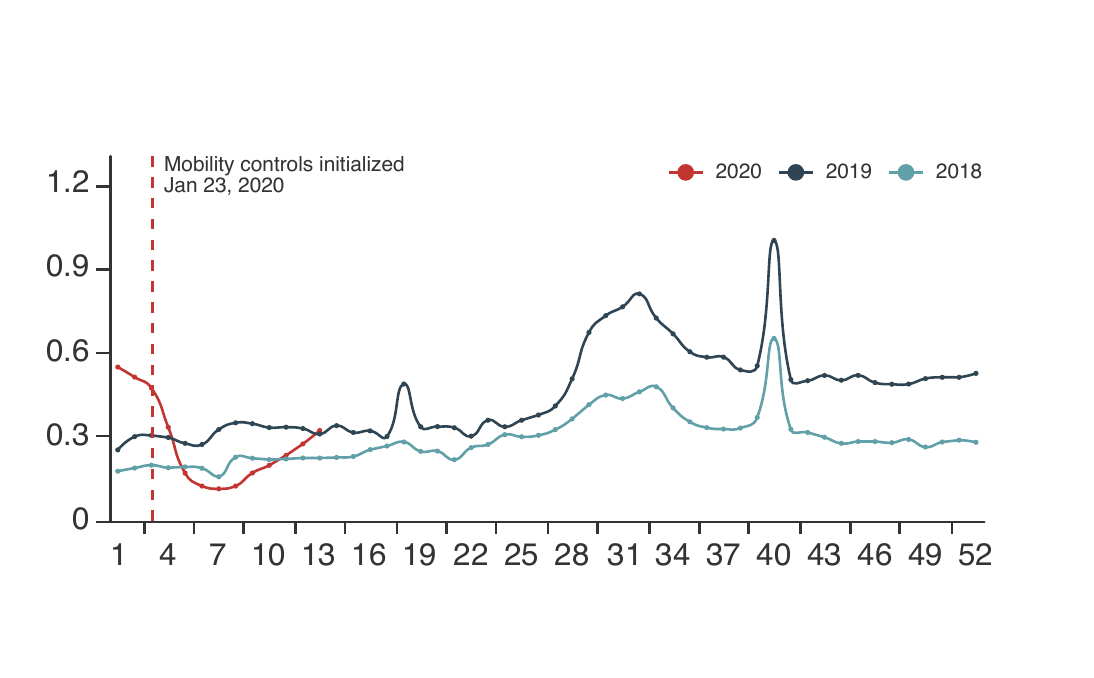}} \
    \subfloat[The \nvc for Inner Mongolia]{\label{fig:a-inner-mongolia-nvc}\includegraphics[width=0.49\textwidth,trim={0.48cm 1.08cm 0.58cm 1.08cm},clip]{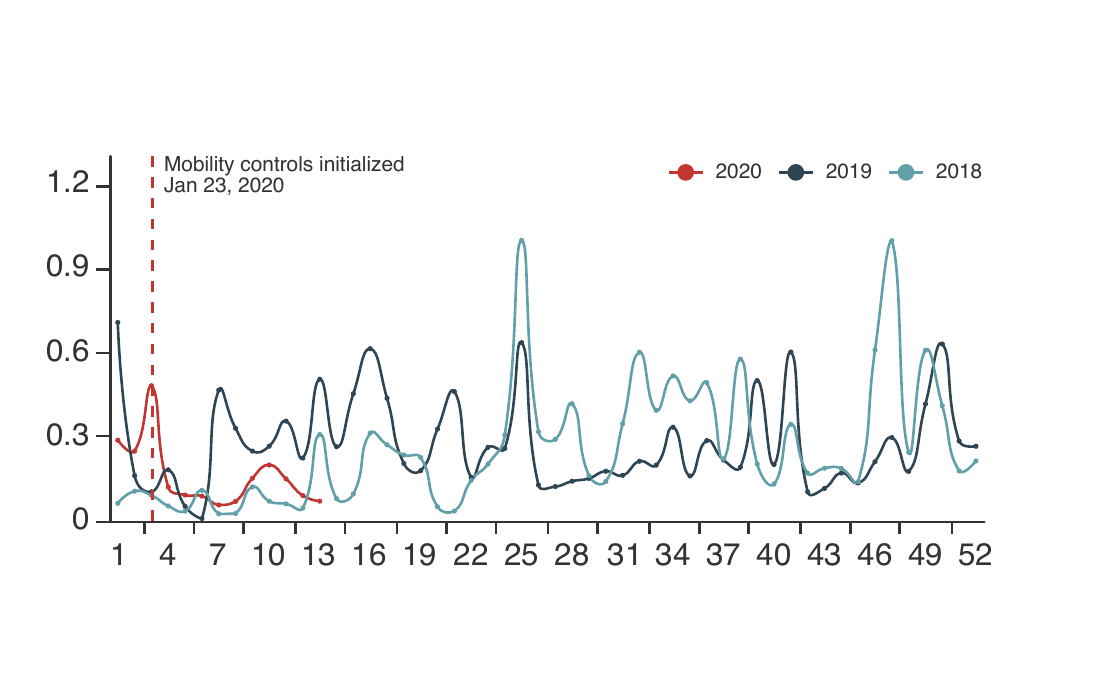}} \
    \subfloat[The \vvv in Guizhou]{\label{fig:a-guizhou-v3}\includegraphics[width=0.49\textwidth,trim={0.48cm 1.08cm 0.58cm 1.08cm},clip]{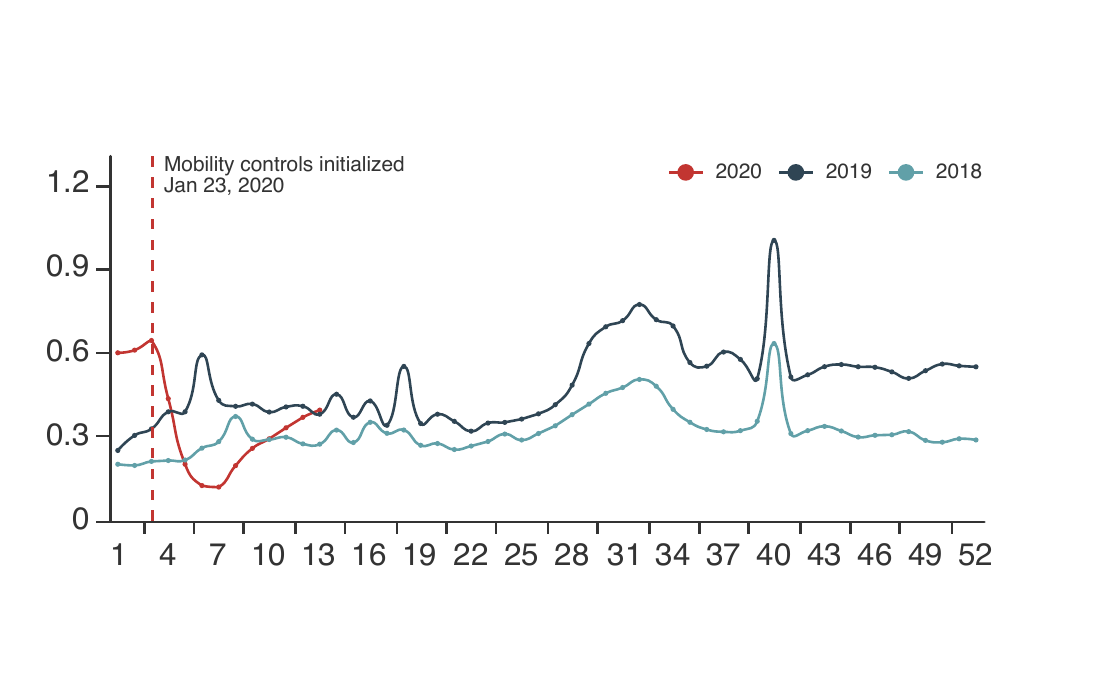}} \
    \subfloat[The \nvc for Guizhou]{\label{fig:a-guizhou-nvc}\includegraphics[width=0.49\textwidth,trim={0.48cm 1.08cm 0.58cm 1.08cm},clip]{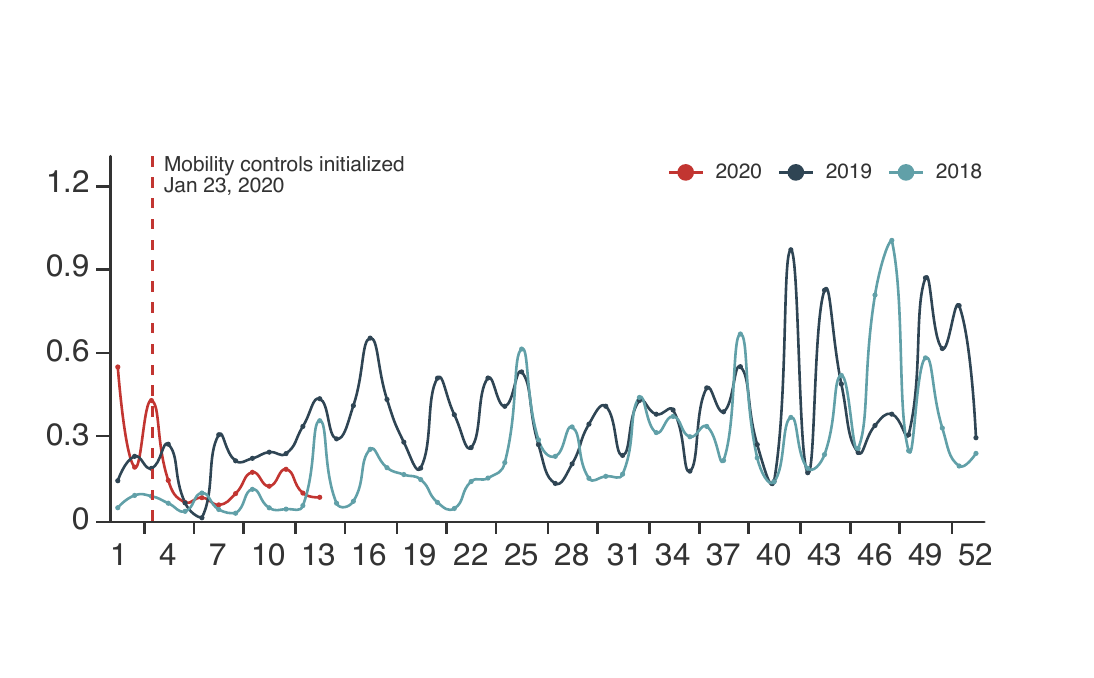}} \
    \subfloat[The \vvv in Guangxi]{\label{fig:a-guangxi-v3}\includegraphics[width=0.49\textwidth,trim={0.48cm 1.08cm 0.58cm 1.08cm},clip]{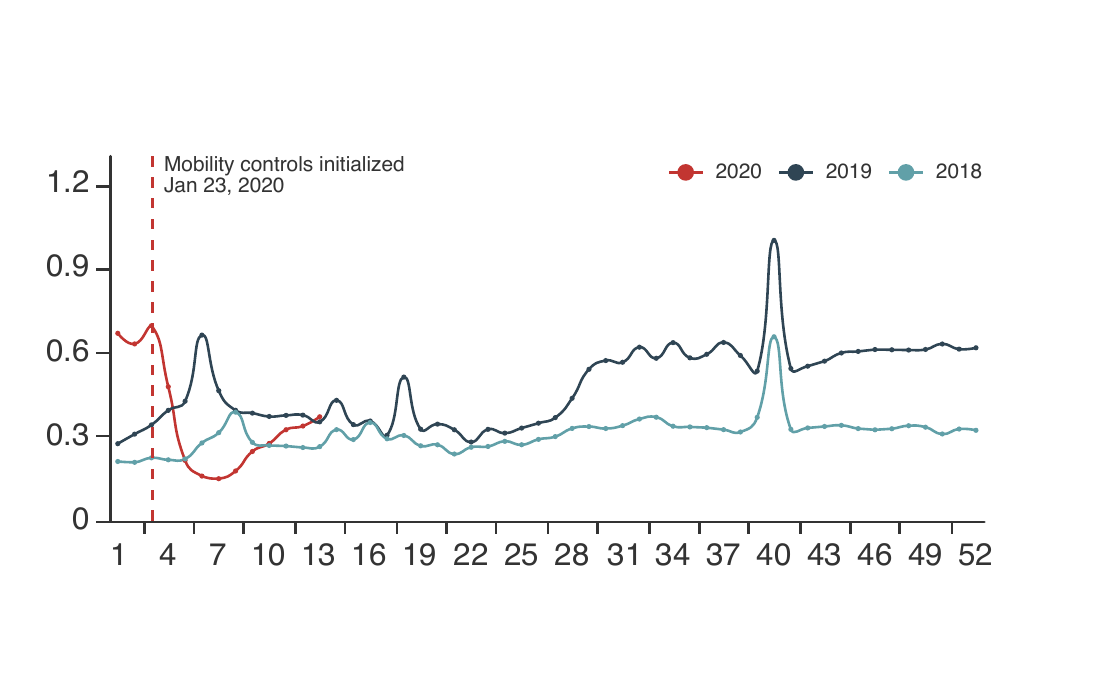}} \
    \subfloat[The \nvc for Guangxi]{\label{fig:a-guangxi-nvc}\includegraphics[width=0.49\textwidth,trim={0.48cm 1.08cm 0.58cm 1.08cm},clip]{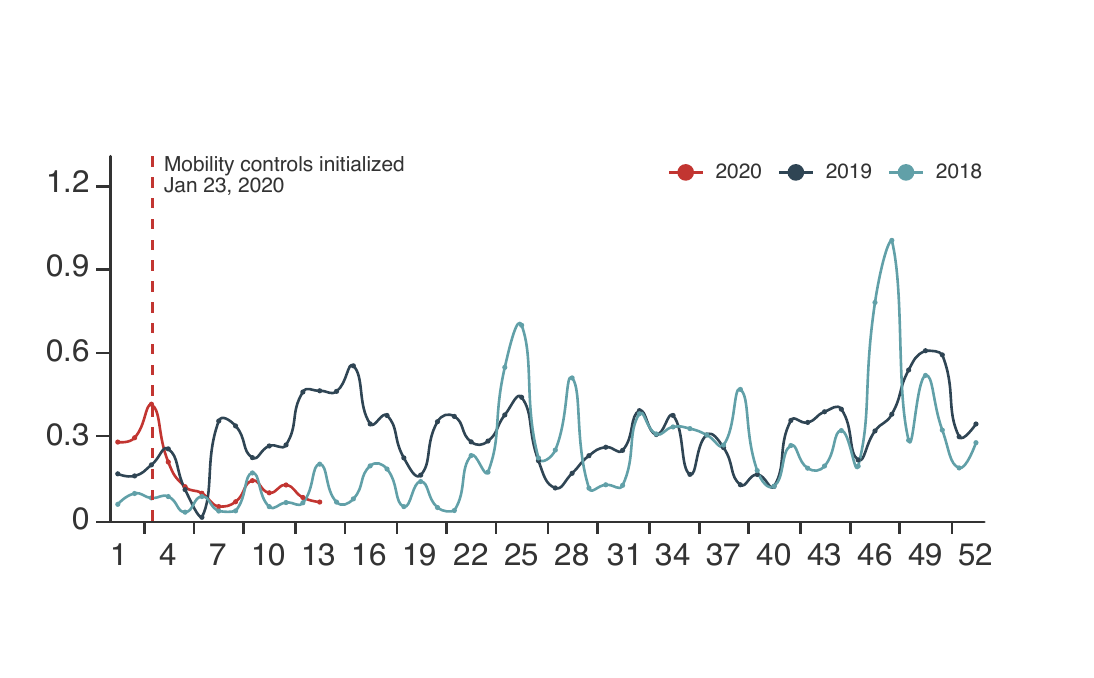}} \
    \subfloat[The \vvv in Yunnan]{\label{fig:a-yunnan-v3}\includegraphics[width=0.49\textwidth,trim={0.48cm 1.08cm 0.58cm 1.08cm},clip]{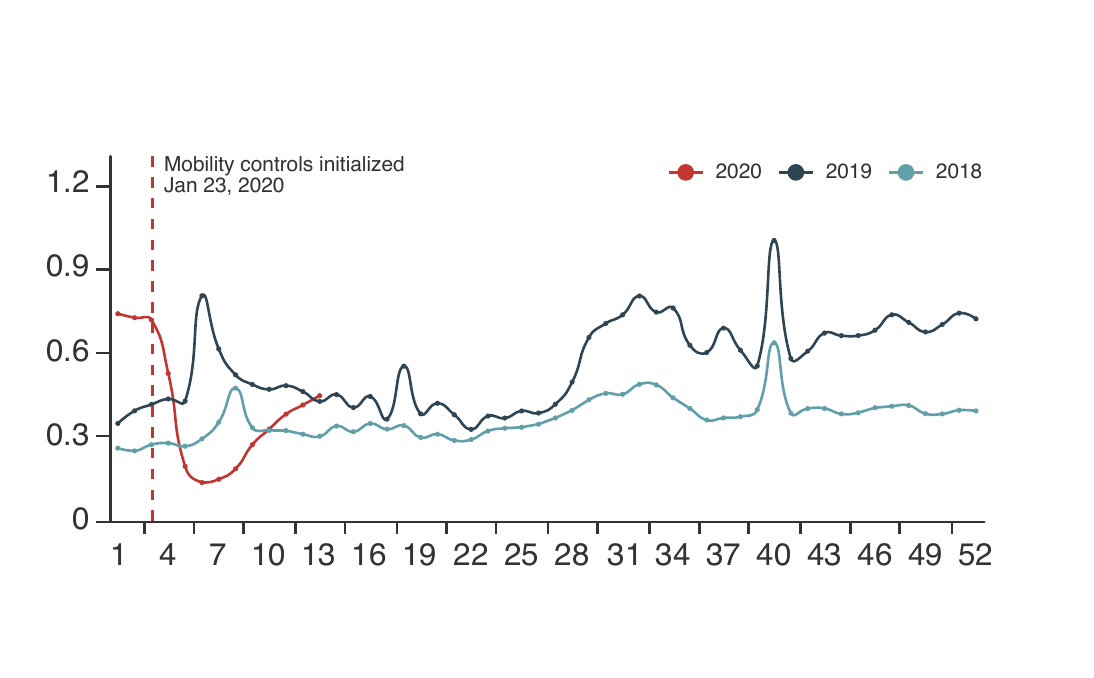}} \
    \subfloat[The \nvc for Yunnan]{\label{fig:a-yunnan-nvc}\includegraphics[width=0.49\textwidth,trim={0.48cm 1.08cm 0.58cm 1.08cm},clip]{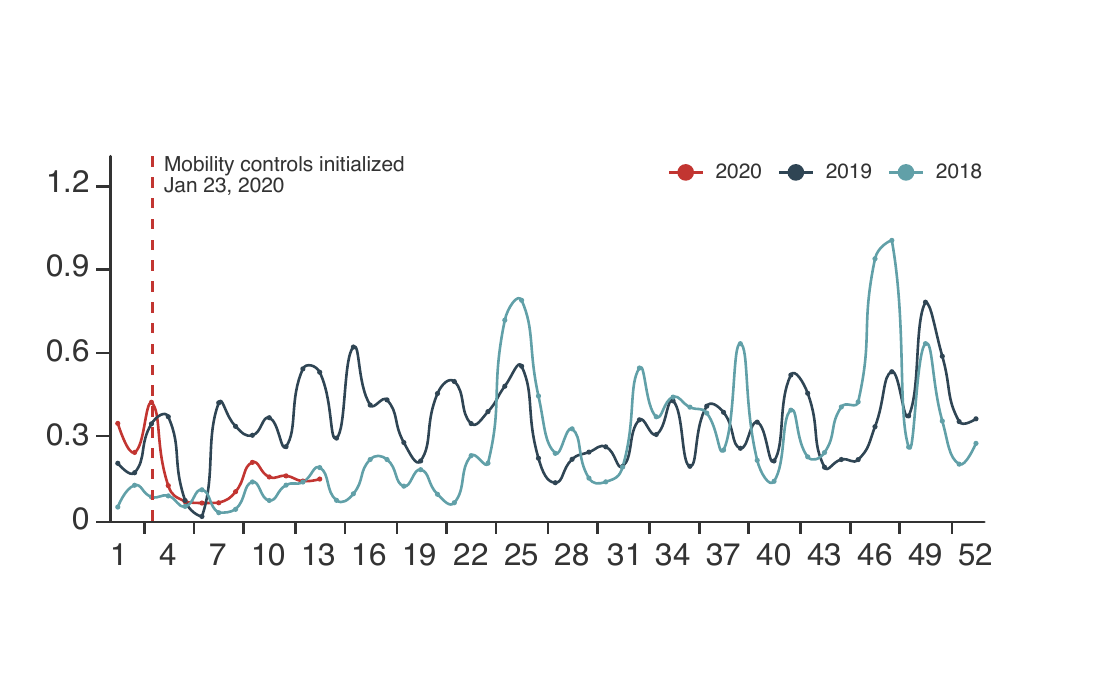}} \
    \caption{The \vvv and the \nvc of 31 provinces. (7 of 8)}
\end{figure}

\begin{figure}[t]
    \centering
    \ContinuedFloat     
    \subfloat[The \vvv in Jilin]{\label{fig:a-jilin-v3}\includegraphics[width=0.49\textwidth,trim={0.48cm 1.08cm 0.58cm 1.08cm},clip]{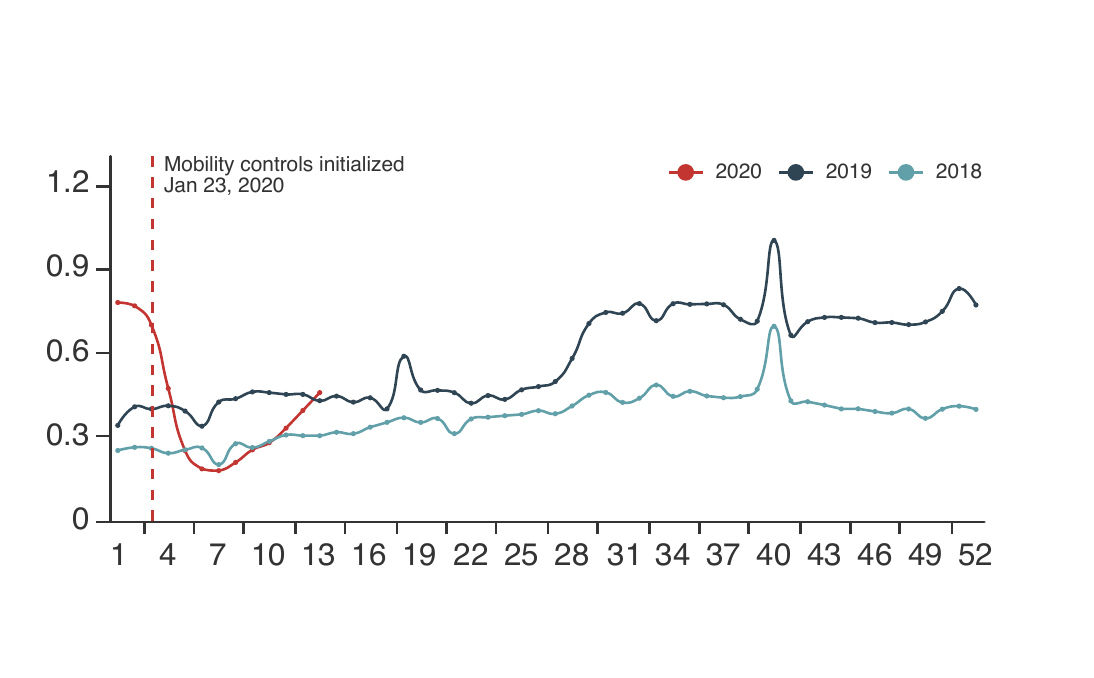}} \
    \subfloat[The \nvc for Jilin]{\label{fig:a-jilin-nvc}\includegraphics[width=0.49\textwidth,trim={0.48cm 1.08cm 0.58cm 1.08cm},clip]{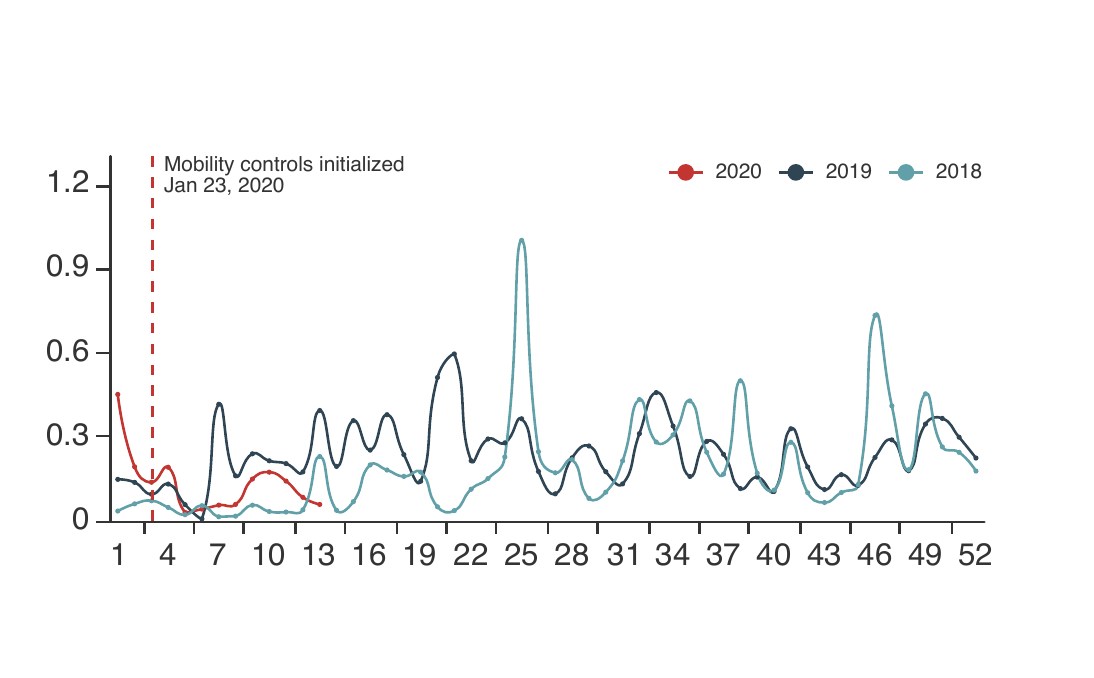}} \
    \subfloat[The \vvv in Qinghai]{\label{fig:a-qinghai-v3}\includegraphics[width=0.49\textwidth,trim={0.48cm 1.08cm 0.58cm 1.08cm},clip]{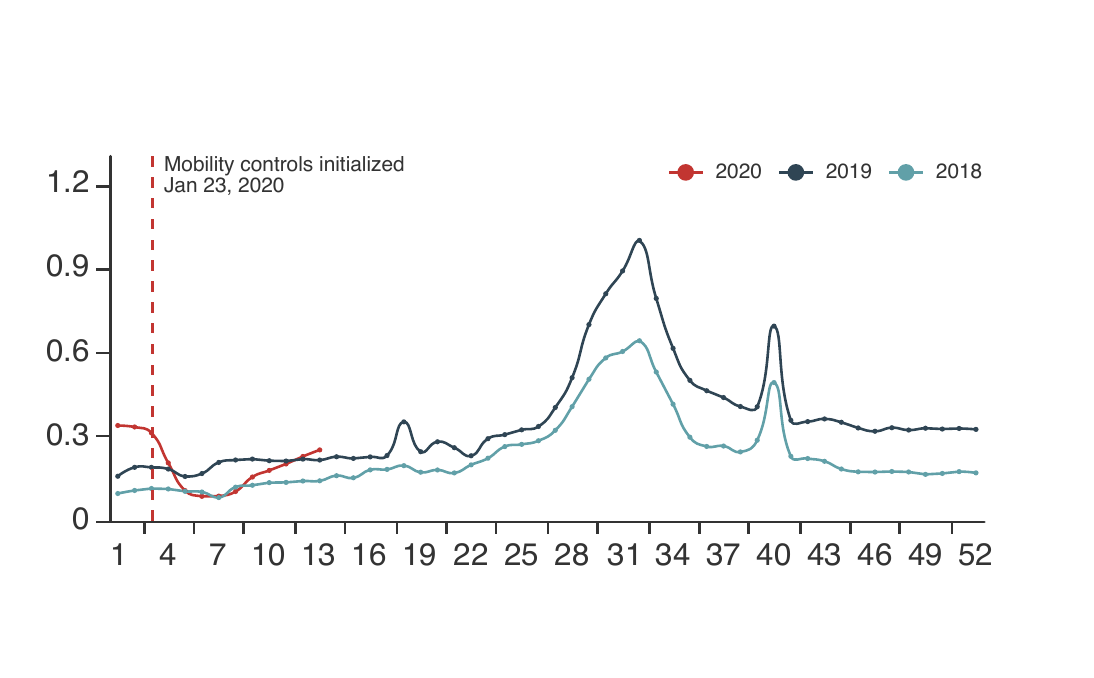}} \
    \subfloat[The \nvc for Qinghai]{\label{fig:a-qinghai-nvc}\includegraphics[width=0.49\textwidth,trim={0.48cm 1.08cm 0.58cm 1.08cm},clip]{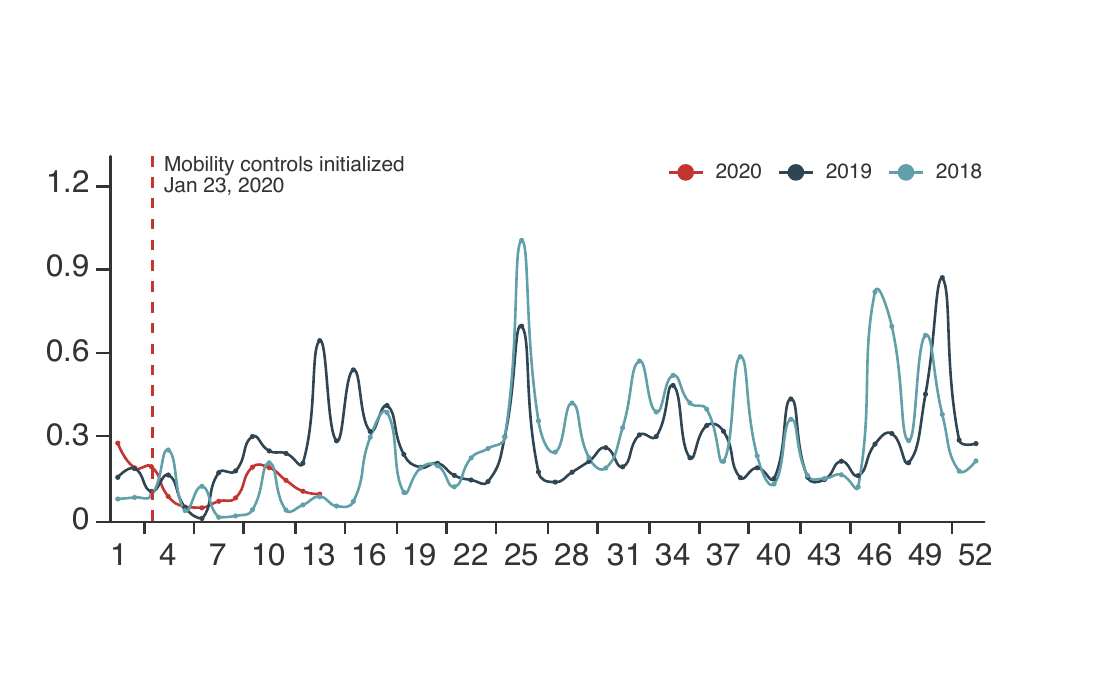}} \
    \subfloat[The \vvv in Xinjiang]{\label{fig:a-xinjiang-v3}\includegraphics[width=0.49\textwidth,trim={0.48cm 1.08cm 0.58cm 1.08cm},clip]{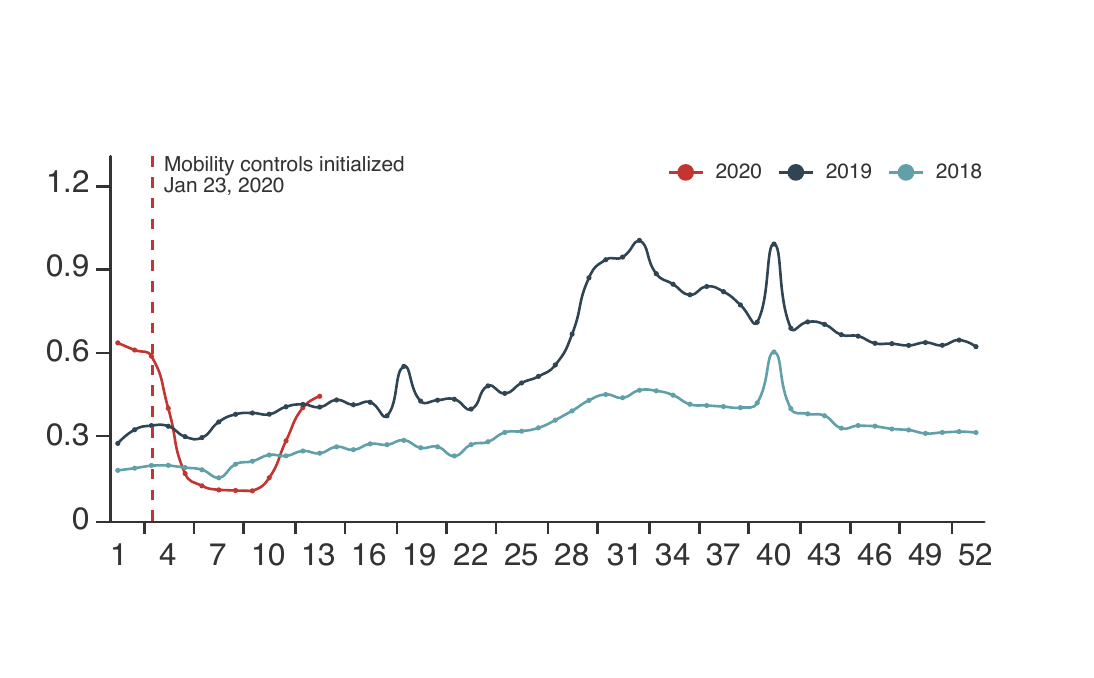}} \
    \subfloat[The \nvc for Xinjiang]{\label{fig:a-xinjiang-nvc}\includegraphics[width=0.49\textwidth,trim={0.48cm 1.08cm 0.58cm 1.08cm},clip]{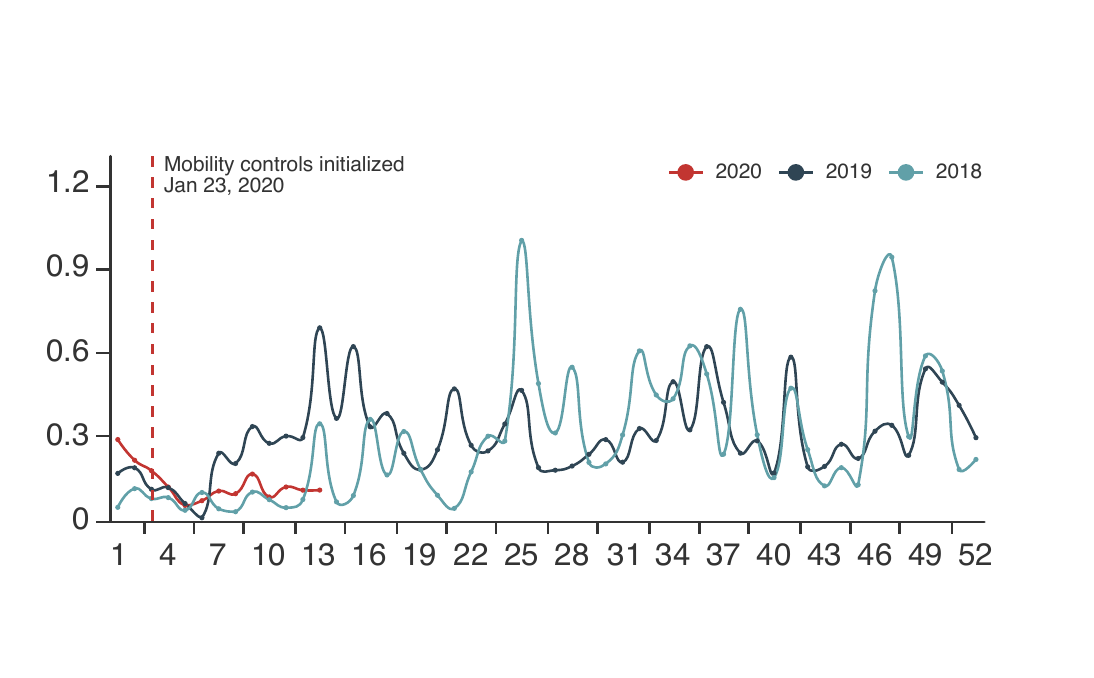}} \
    \caption{The \vvv and the \nvc of 31 provinces. (8 of 8)}
\end{figure}

\end{document}